\documentclass[aps,pra,superscriptaddress,onecolumn]{revtex4-2}

\usepackage{bbm}
\usepackage{graphicx}
\usepackage{booktabs}
\usepackage{longtable}
\usepackage{array}
\usepackage{ragged2e}

\usepackage{amsmath,amssymb,amsfonts}
\usepackage{textcomp}
\usepackage{braket} 
\usepackage{bm}
\usepackage{xspace}
\usepackage[table]{xcolor}

\usepackage{multirow}

\newcommand{\bmt}{\bm{\theta}}
\newcommand{\bx}{\bm{x}}

\newcommand{\Tr}{\text{Tr}}

\newcommand{\btheta}{\bm{\theta}}
\newcommand{\bomega}{\bm{\omega}}

\usepackage{fontawesome5}
 \definecolor{NatureTeal}{HTML}{0F6B68}
\definecolor{NatureBlue}{HTML}{1F4E79}
\definecolor{NatureRed}{HTML}{A23B3B}

\newcommand{\projectpage}[1]{%
  \vspace{0.3em}%
  \noindent\mbox{%
    \textsf{\textcolor{NatureTeal}{\faAtom\ Project page:}}~%

    \href{#1}{\textcolor{NatureTeal}{\texttt{#1}}}%

  }%
}

\newcommand{\omniQ}{\textsf{OmniQ}\xspace}
\newcommand{\RA}{\textsf{RA}\xspace}

\usepackage[colorlinks=true,linkcolor=red,citecolor=blue,urlcolor=blue,
            breaklinks=true,pdfstartview=FitH]{hyperref}

\begin{document}

\title{Reimagining computational quantum research with AI scientists}

\author{Yuxuan Du}
\thanks{These authors contributed equally to this work.}
\affiliation{College of Computing and Data Science,
Nanyang Technological University, Singapore 639798, Singapore}
\affiliation{School of Physical and Mathematical Sciences,
Nanyang Technological University, Singapore 639798, Singapore}

\author{Hanqing Zhao}
\thanks{These authors contributed equally to this work.}
\affiliation{College of Computing and Data Science,
Nanyang Technological University, Singapore 639798, Singapore}

\author{Xinbiao Wang}
\affiliation{College of Computing and Data Science,
Nanyang Technological University, Singapore 639798, Singapore}

\author{Yiming Lu}
\affiliation{College of Computing and Data Science,
Nanyang Technological University, Singapore 639798, Singapore}

\author{Kaining Zhang}
\affiliation{College of Computing and Data Science,
Nanyang Technological University, Singapore 639798, Singapore}

\author{Dacheng Tao}
\thanks{Correspondance to \url{duyuxuan123@gmail.com}; \url{dacheng.tao@ntu.edu.sg}}
\affiliation{College of Computing and Data Science,
Nanyang Technological University, Singapore 639798, Singapore}

\begin{abstract}
AI scientists are expanding automated research across scientific domains, yet their ability to turn the scientific record into a foundation for autonomous discovery remains elusive. Here we uncover a fundamental barrier to this transition and demonstrate a route to overcome it in computational quantum science, where advances shape both what we can learn from complex quantum systems and what quantum computing can ultimately deliver. To identify this barrier, we introduce \textsf{OmniQ}, an expert-audited virtual research environment that converts published studies into executable reproduction tasks. Across 59 papers spanning classical simulation, quantum algorithms and quantum-property characterization, we systematically evaluate five agentic systems powered by large language models and six third-party agent harnesses. Our findings reveal that current AI scientists often fail to reproduce published results, exposing a fundamental mismatch between what scientific papers communicate and what autonomous agents require. By contrast, human expert intervention can turn these failures into successful reproductions, showing that the missing information can be identified and converted into reusable research experience. Guided by this observation, we develop Residual Agents (\RA), a quantum-tailored self-improving AI scientist that distils externally verified workflow repairs into such experience. \RA substantially improves research performance under fixed base models and limited adaptation budgets. For example, \RA raises DeepSeek V4.1 Flash performance from 68\% to 85\%, matching Claude Code with Opus~5 at a third of its cost despite using a weaker base model. We integrate all reproduced results into an extensible platform built on the model context protocol, making verified research workflows directly accessible to AI scientists and allowing new scientific papers to be continuously incorporated. These results establish a new paradigm for computational quantum research, where published results become reusable scientific capabilities that support autonomous discovery.

\smallskip
\noindent\projectpage{https://omniq-qid-lab.github.io/}
\end{abstract}

\maketitle

\section*{Introduction}

Computational quantum science develops algorithms to understand and exploit complex quantum systems~\cite{georgescu2014quantum,mcardle2020quantum,du2026ai}, spanning classical simulation~\cite{schollwock2005dmrg,orus2019tensor,berezutskii2025tnqc}, quantum algorithms~\cite{childs2010algebraic,bharti2022nisq,morales2026quantum}, and property characterization of quantum systems~\cite{carleo2019machine,eisert2020quantum,gebhart2023learning}. Despite this promise, algorithm design across all three directions is becoming increasingly difficult as quantum systems and hardware continue to scale. For classical simulation, pushing the limits of classical tractability while establishing reliable baselines for practical utility increasingly relies on  sophisticated methods such as tensor networks~\cite{cirac2021mps,tindall2024eagle,tindall2026disordered} and Pauli-propagation techniques~\cite{shao2024noisy,begusic2024fast,begusic2025sparse,rudolph2026paulipropagation,zhang2026stochastic}. For both near-term and fault-tolerant quantum algorithms, achieving computational advantage remains challenging under realistic resource constraints, requiring careful co-design of algorithms and hardware capabilities~\cite{kim2023evidence,liu2025certified,googleqai2025echoes,liao2026sample}. For quantum property characterization, extracting informative properties from limited measurements becomes difficult as the Hilbert space grows exponentially, motivating efficient measurement and inference protocols~\cite{wang2024transition,du2025efficient,hu2025shallow,wu2024learning,liao2026demonstration,du2026shadownet}. As a result, these challenges create an urgent need for more autonomous approaches to computational quantum science.

Recent breakthroughs in AI scientists offer a potential route towards autonomous computational quantum science~\cite{lu2026end,gottweis2026coscientist,ghareeb2026robin}. By integrating frontier large language models (LLMs) with planning, tool use and iterative feedback \cite{vaswani2017attention,brown2020language,openai2023gpt4,deepseek2025r1}, these systems have begun to automate algorithm discovery in mathematics~\cite{romeraparedes2024funsearch,trinh2024alphageometry,hubert2026alphaproof} and software engineering~\cite{novikov2025alphaevolve,jimenez2024swebench,yang2024sweagent,wang2025openhands}. This trend is also extending to other disciplines, such as biology~\cite{gao2024biomedical,swanson2025virtuallab,huang2026biomni}, chemistry~\cite{boiko2023coscientist,mbran2024chemcrow,ruan2024llmrdf} and materials science~\cite{huang2026automat}. All of these advances point to a shift from AI-assisted research towards autonomous algorithm discovery. Computational quantum science is a natural next arena because progress in this field may reshape both what we can learn from quantum systems and what quantum computing can ultimately deliver. In this context, early efforts have demonstrated AI-driven discovery of quantum error-correcting codes~\cite{cruzbenito2026bbcodes,liu2026sce,yan2026omniqec}, automated design of quantum circuits~\cite{arlt2026metadesign,knipfer2026aiagents,sakka2026llmvqc} and algorithmic search for quantum-chemistry applications~\cite{zou2025elagente,zhang2026forjador,li2026autonomous,finger2026algorithmdiscovery}. However, the progress remains fragmented and largely confined to narrowly defined settings, such as a single subproblem, a human-tailored workflow or a search space strongly constrained by prior knowledge. This raises a central question: 

\begin{figure*}[htp]
  \centering
\includegraphics[width=0.88\textwidth]{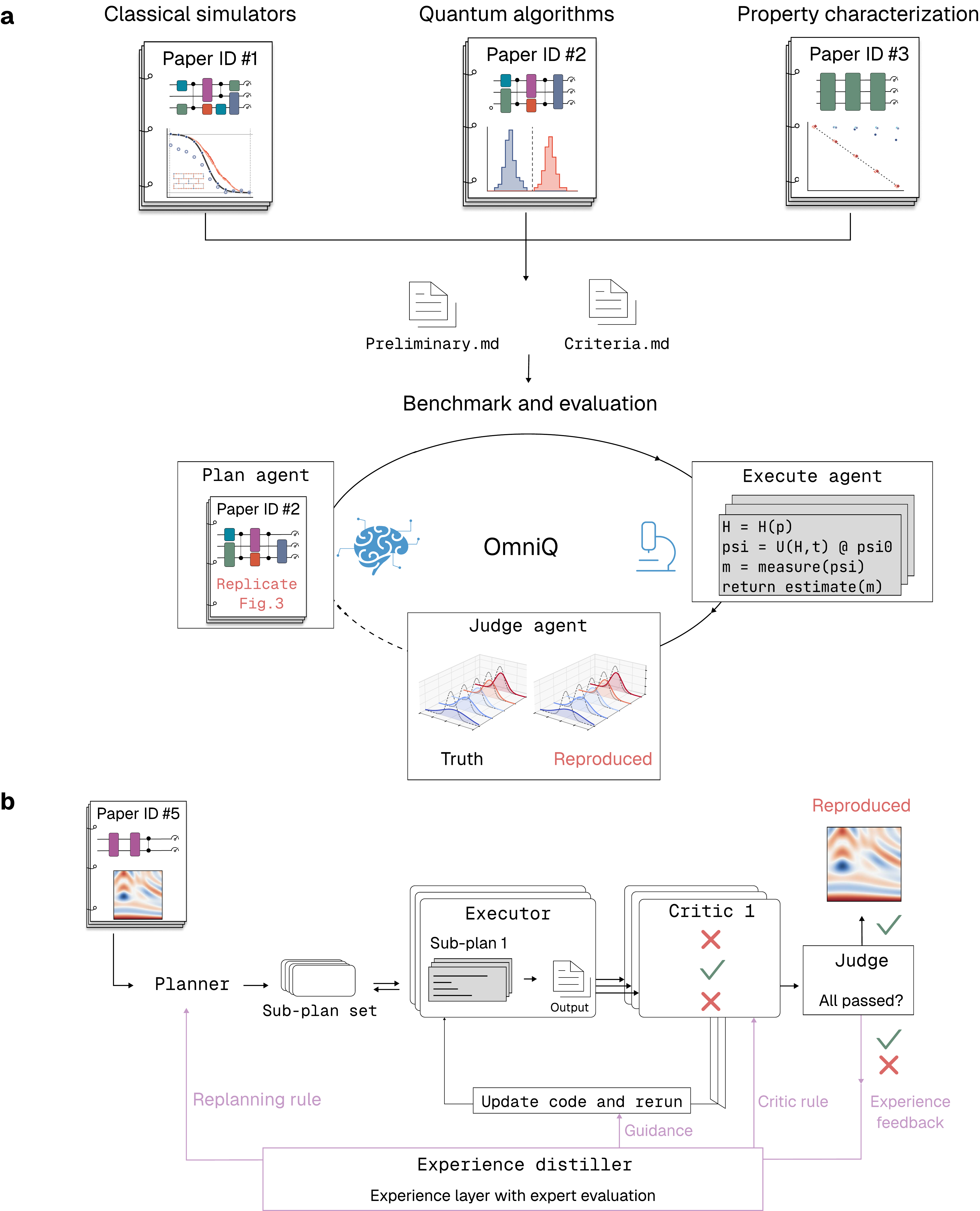}
  \caption{\small{\textbf{Paper-to-result reproduction in computational quantum science.}
  \textbf{a}, Scientific papers from three research directions in computational quantum science are converted into reproduction tasks centred on a target figure within \omniQ, an expert-audited virtual research environment. In \omniQ, a plan agent and an execute agent reproduce the source code and the target figure, and a judge agent compares the reproduced figure with the published one.
  \textbf{b}, The mechanism of the self-improving AI scientist \RA for computational quantum science. The planner decomposes a research plan into sub-plans, each executed by an executor and evaluated by a critic. Failed steps are revised and re-executed, with the planner replanning when necessary. A judge evaluates the assembled result. Upon failure, expert feedback is processed by the residual agent to refine executor guidance, critic criteria and planner strategies.}}
  \label{fig:framework}
\end{figure*}

\begin{center}
	\textit{Can AI scientists drive the next wave of discovery in computational quantum science?} 
\end{center}

\begin{figure*}[htp]
  \centering
  \includegraphics[width=0.90\textwidth]{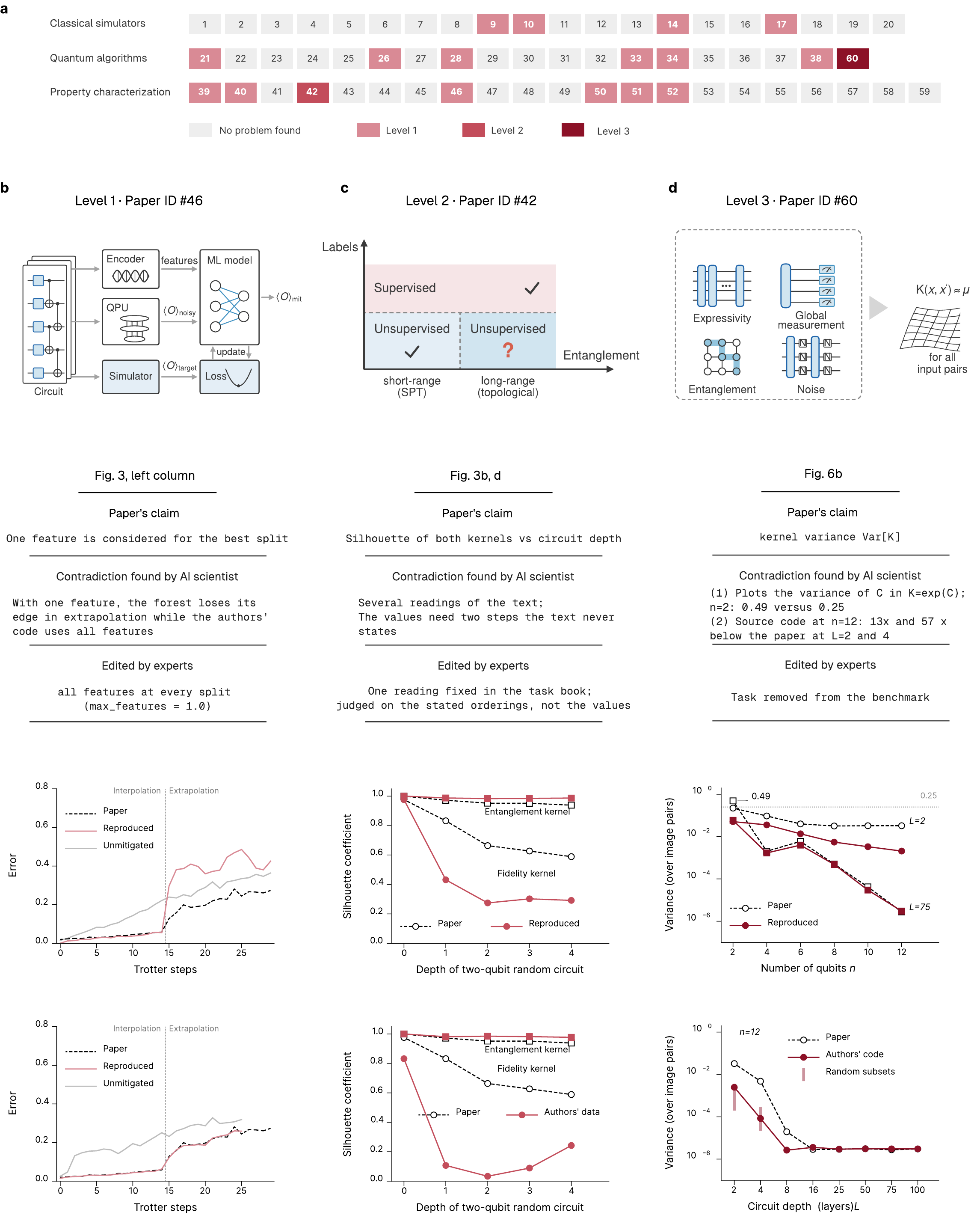}
  \caption{\small\textbf{AI-assisted reproduction reveals inconsistencies in the scientific record.} \textbf{a}, Reproduction assessment across three directions of computational quantum science. Grey denotes studies with no identified problem, whereas increasingly dark red indicates Level I–III issues according to their severity and resolvability. \textbf{b}–\textbf{d}, Representative Level I–III cases. From top to bottom, each case shows the scientific problem, the identified reproduction issue—including the paper’s claim, the contradiction found by the AI scientist and expert feedback—and the reproduced results before and after expert correction, each compared with the original result. For the panel \textbf{b}, a machine-learning configuration discrepancy resolved using the released code. For the panel \textbf{c}, an ambiguous experimental procedure requiring expert resolution. For the panel \textbf{d}, a fundamental inconsistency between the reported quantity, released code and theoretical constraints that cannot be resolved from the available scientific record. Dashed black and solid red curves denote the reported and reproduced results, respectively.}
  \label{fig:irreproducible}
\end{figure*}

In this study, we address this question by starting from paper-to-result reproduction, where verifiable outcomes allow AI scientists to accumulate scientific experience before moving to open-ended research. To this end, we introduce \omniQ, an expert-audited virtual research environment that transforms published studies in computational quantum science into executable reproduction tasks. More precisely, given a full paper and a target result, an AI scientist reconstructs the underlying method, implements and executes it under a comparable computational budget, and attempts to reproduce the reported outcome. The workflow is evaluated for consistency with the reported method and outcome, with expert intervention reserved for ambiguous or inconsistent cases. In doing so, \omniQ turns paper-to-result reproduction into a controlled and diagnosable research setting.

Using \omniQ, we systematically evaluate five LLM-powered AI scientists and six third-party agent harnesses across 60 papers spanning computational quantum science. Different from recent paper-to-agent approaches~\cite{miao2026paper2agent} that transform papers, code and data into executable agentic systems, here we consider a more direct reproduction setting in computational quantum science. That is, the agent must reconstruct the workflow underlying a reported result. Experimental results show that current AI scientists frequently fail at this task. Moreover, human expert review reveals a deeper problem. That is, in 18 studies, the paper description, released code and reported results are inconsistent. Such cases obscure the boundary between agent failure and limitations of the scientific record, posing a barrier to reliable autonomous discovery in computational quantum science.

After separating source-side inconsistencies from genuine agent failures, we find that human expert intervention can often turn failed attempts into successful reproductions, revealing domain knowledge that can be learned and reused. Building on this observation, we develop Residual Agents (\RA), an efficient self-improving AI scientist that distils externally verified workflow repairs into reusable quantum research experience. Unlike generic evolution mechanisms, \RA accumulates domain-specific lessons from successful repairs and applies them to subsequent research. Across base LLMs, \RA substantially improves performance under a limited adaptation budget, with gains that generalize to unseen tasks and transfer across models.  We further integrate all reproduced results into an extensible platform built on the model context protocol, enabling AI scientists and the broader research community to directly access, reuse, and extend verified research workflows. Collectively, these advances recast paper-to-result reproduction from a one-off translation problem into a foundation for reusable scientific capabilities and continual learning, opening a path towards autonomous discovery in computational quantum science and potentially beyond.

\section*{Results}

\subsection*{Failures of paper-to-result reproduction}

Autonomous computational quantum science envisions AI scientists that can independently understand and develop computational methods for quantum systems. Towards this goal, a crucial step is reproduction, which can accelerate the validation and reuse of existing methods while reducing the expert effort required to reconstruct complex computational workflows. We therefore ask whether an AI scientist can reconstruct the underlying workflow from the scientific record and faithfully reproduce a reported result. We refer to this setting as \textit{paper-to-result reproduction}.

We study paper-to-result reproduction using \omniQ, a virtual research environment that turns published computational quantum studies into executable and diagnosable reproduction tasks. As illustrated in Fig.~\ref{fig:framework}\textbf{a}, \omniQ is employed to curate 60 studies spanning classical simulation, quantum algorithms and quantum-property characterization. The studies are selected to balance scientific impact, publication in leading venues, methodological diversity, and reproducibility under realistic computational budgets, with approximately even coverage across the three research directions. More details are provided in Supplementary Information (SI)~A and B.

To faithfully assess reproduction quality, we classify reproduction quality into three levels according to how well the central scientific claim is preserved through reconstruction. Specifically, Level~I captures missing methodological details that can be recovered from the released code or expert interpretation. Level~II captures localized errors in the paper that can be resolved from the code, figures or broader scientific context, but otherwise lead to reproduction discrepancies. Level~III captures fundamental inconsistencies among the context, figures and released code that cannot be resolved through expert inspection, leaving the target result unreproducible from the available scientific record.

The reproduction experiment in \omniQ assigns eight agent configurations to each study, each tasked with independently reproducing a designated result under a computational budget comparable to that of the original work. Across representative frontier models and agentic systems, the reconstructed workflows and reported results are cross-checked across agents, compared against available source code, and independently verified by domain experts (See SI~B). As shown in Fig.~\ref{fig:irreproducible}\textbf{a}, this procedure identifies reproducibility issues in 18 of the 60 studies. Below, we examine three representative studies from each reproduction level to illustrate the distinct failure modes.

Level~I is exemplified by Paper~\#46~\cite{liao2024machine} on quantum error mitigation~\cite{cai2023quantum}, exploring whether learning-based mitigation can recover accurate observable expectation values beyond the training regime, enabling extrapolation to deeper quantum circuits without the full overhead of fault-tolerant error correction~\cite{roffe2019quantum}. Its Figure~3 directly probes this claim by tracking prediction errors across Trotter steps within and beyond the training region for Hamiltonian simulation tasks. The reproduction with \omniQ reveals a localized but consequential inconsistency. That is, the paper states that only one feature is considered when selecting the best split of the machine-learning model, whereas the released code uses all available features at each split (i.e., \texttt{max\_features=1.0}). As shown in the third panel of Fig.~\ref{fig:irreproducible}\textbf{b}, following the written description causes the reproduced model to lose its reported advantage in the extrapolation regime. In contrast, as shown in the bottom panel, following the released code resolves this discrepancy and recovers the reported extrapolation advantage.

\begin{figure*}[htp]
  \centering
  \includegraphics[width=0.85\textwidth]{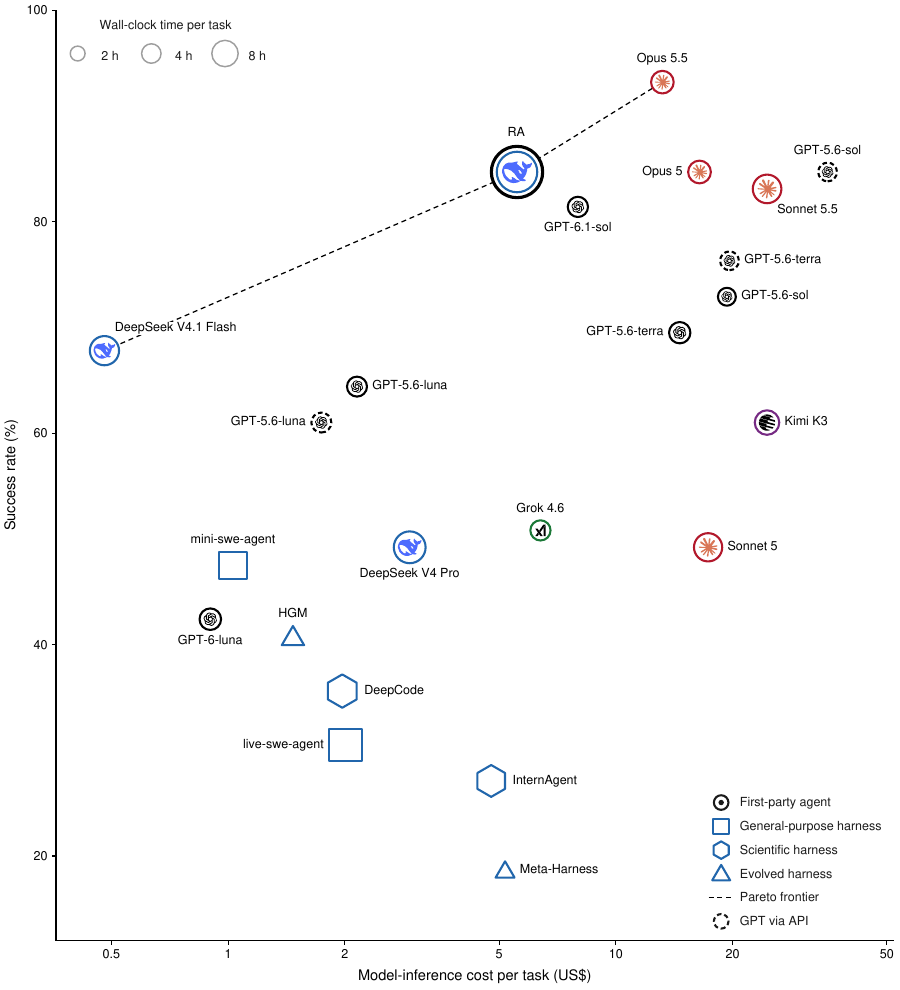}
  \caption{\small{\textbf{Performance of \RA for computational quantum science.} Reproduction success across 59 reproduction tasks is plotted against model-inference cost per task on a logarithmic scale. Marker shape denotes the agent system or harness type, marker color denotes the underlying model family or access mode, and marker size indicates wall-clock time per task; semi-transparent markers denote earlier model generations. Labels identify the corresponding model or harness. \RA, the proposed self-improving AI scientist, runs DSH as its executor and is drawn as the DSH marker inside a gear; the dashed line indicates the Pareto frontier between reproduction success and inference cost. A brown ring marks GPT models accessed through an application programming interface (API).}}
  \label{fig:benchmark}
\end{figure*}

Level~II is exemplified by Paper~\#42~\cite{che2026topological} on quantum-property characterization, where efficient learning methods are developed to identify topological order in many-body quantum systems without prior phase labels. Its Figures~3b and 3d quantify the robustness of the proposed fidelity and entanglement kernels under increasingly deep two-qubit random circuits using the silhouette coefficient, a standard metric of clustering quality \cite{rousseeuw1987silhouettes}. Reproduction with \omniQ fails to recover the reported quantitative values. As shown in the third panel of Fig.~\ref{fig:irreproducible}\textbf{c}, the silhouette coefficient, which measures clustering quality from $-1$ (poor) to $1$ (well separated), decreases from approximately $0.59$ in the paper to $0.29$ in the reproduction at circuit depth four. Expert inspection reveals that the experimental procedure is incompletely specified, as recovering the reported value requires an additional two-step procedure that is not described in the paper. Moreover, as shown in the last panel, applying the procedure described in the paper to the authors’ own data yields a silhouette coefficient of only $0.24$. After expert editing focuses the task on the reported scientific conclusion instead of unrecoverable numerical values, \RA reproduces the reported degradation of the fidelity kernel and the consistent advantage of the entanglement kernel.

Level~III is represented by Paper~\#60~\cite{thanasilp2024exponential} on quantum machine learning, which studies whether quantum kernels remain efficiently distinguishable as the number of qubits increases. This question is closely tied to whether quantum kernels can ultimately support quantum advantage in learning~\cite{huang2021power}. The scientific paper claims that highly expressive data encodings can cause quantum kernels to concentrate exponentially, making their estimation increasingly costly and ultimately limiting their usefulness for learning. 
Its Figure~6b provides numerical evidence for this claim by plotting the variance of the projected quantum kernel $\mathsf{K}^{\mathrm{PQ}}$ as a function of qubit number $n$ and circuit depth $L$. 
However, reproduction with \omniQ reveals a fundamental inconsistency: although the figure is described as reporting $\mathrm{Var}[\mathsf{K}^{\mathrm{PQ}}]$, the released code instead evaluates the variance of the distance $C$ entering the kernel $\mathsf{K}^{\mathrm{PQ}}=\exp(-\gamma C)$. As shown in the third panel of Fig.~\ref{fig:irreproducible}\textbf{d}, the discrepancy is already evident at $n=2$, where the reported value reaches $0.49$, exceeding the theoretical upper bound of $0.25$ on the variance of a kernel taking values in $[0,1]$. A separate discrepancy concerns the generation of the data underlying the figure. Even when $\mathrm{Var}[C]$ was used as the target, rerunning the released code fails to recover the published values. For the case of $n=12$ qubits, the published variances at both $L=2$ and $L=4$ are more than an order of magnitude larger than the recalculated values, as shown in the last panel. Expert inspection identified the statistical error but did not establish the configuration used to generate the archived data. Unlike Levels I and II, the discrepancy between the released code and the published figure remains unresolved, preventing faithful reproduction of the target result from the available scientific record.

\subsection*{Performance across frontier AI scientists}

We next explore how existing AI scientists perform on a cleaned computational quantum science dataset. Following expert review, resolvable Level I and II issues are corrected, whereas the unresolved Level III case (i.e., Paper \#60) is excluded, yielding a consistent dataset of 59 papers for agent evaluation. Using this cleaned dataset and \omniQ, we systematically evaluate leading base models from the GPT, Claude, DeepSeek, Kimi and Grok families~\cite{openai2026gpt56,anthropic2026opus55,deepseek2026v4,kimiteam2026k3,xai2026grok46}. We also consider first-party agentic systems, including Codex, Claude Code, Kimi Code, Grok Build and DeepSeek Harness (DSH), together with six widely used third-party agent harnesses, i.e., InternAgent, DeepCode, mini-swe-agent, live-swe-agent, Huxley--G\"odel Machine (HGM) and Meta-Harness~\cite{internagent2025,li2025deepcode,minisweagent2025,xia2025livesweagent,wang2025hgm,lee2026metaharness}. A reproduction task is counted as successful only when the reconstructed workflow satisfies the strict reproduction criterion defined above.

Figure~\ref{fig:benchmark} summarizes reproduction success together with inference cost per task and wall-clock time across different agents and harness mechanisms. A first observation is that performance varies markedly across frontier agentic systems. In particular, Claude Code with Opus~5.5 achieves the highest reproduction rate, recovering approximately $93\%$ of the computational quantum science dataset. Besides, Codex paired with several frontier base models exceeds $70\%$ reproduction success. At the other end of the spectrum, Codex with GPT-6-luna reaches only about $42\%$, and Meta-Harness with DeepSeek performs worst at $19\%$. These results indicate that once the scientific record is self-consistent, frontier AI scientists already possess the capability for autonomous computational quantum science.

The achieved results further reveal \textit{two unexpected findings}. First, reproduction success does not increase monotonically with inference cost. As shown in  Fig.~\ref{fig:benchmark}, Claude Code with Sonnet~5.5, Kimi Code with K3, and several Codex-based systems use larger inference budgets than Claude Code with Opus~5.5 yet perform worse. Meanwhile, DSH with DeepSeek~V4.1 Flash achieves approximately $68\%$ success at the lowest cost, outperforming every other system below \$7 per task and several AI scientists costing an order of magnitude more. This observation suggests that higher inference spending often translates into more deliberation but not more execution, a ``\textit{thinking more but doing less}’’ behavior that echoes recent observations of overthinking in agentic reasoning~\cite{chen2025not}. Collectively, this decoupling identifies the allocation of inference budgets between reasoning and scientific execution as a central determinant of autonomous computational science.

A second unexpected finding is that existing agent harnesses do not necessarily improve reproduction and can even degrade performance. As illustrated in Fig.~\ref{fig:benchmark}, third-party harnesses span a wide performance range, from about $19\%$ for Meta-Harness with DeepSeek to $48\%$ for mini-swe-agent with DeepSeek, all substantially below the strongest first-party systems. Even on the same base LLM model, none of the six exceeds the vendor's own harness. With DeepSeek V4 Pro, their reproduction success ranges from $19\%$ to $48\%$, compared with $49\%$ for DSH. Failure analysis reveals common behaviors, i.e., agents may declare success without validating the reproduced result against the paper, replace the required computation with a cheaper surrogate, or silently reduce the problem size. Therefore, harness mechanisms tuned on other tasks thus do not transfer to computational quantum science~\cite{yao2026harness,hu2026beyond}, which calls for scientific experience grounded in the computational quantum science domain.

 \subsection*{\RA: Self-improving AI scientists}

The performance gap between proprietary and open-source AI scientists in Fig.~\ref{fig:benchmark} underscores a dilemma. On the one hand, leading proprietary systems achieve strong reproduction performance, but their closed implementations limit the transparency and adaptability of scientific workflows. Conversely, existing open-source alternatives struggle to reproduce published results reliably, limiting their utility for computational quantum science. 

To bridge this gap, we develop Residual Agents (\RA), an open-source, self-improving AI scientist for autonomous computational quantum science that learns from verified failures to improve future reproductions. As shown in Fig.~\ref{fig:framework}\textbf{b}, \RA splits a reproduction task into sub-plans and coordinates three roles powered by the same base LLM. Specifically, a planner develops a resource-bounded plan, executors carry out each sub-plan, and a critic validates the results against the paper. When a step fails, it is repaired or replanned based on the critic’s assessment. Beyond resolving individual failures, \RA identifies the \textit{residual} between failed executions and their verified corrections, distilling these differences into generalizable failure signatures stored in a residual memory alongside the evolving memories of the planner and critic. As such, adaptation is confined to the experience layer and driven only by verified evidence. To ensure reliable evolution, a human expert reviews each proposed update before it is incorporated, while the underlying model, executors, and orchestration remain fixed. Further details and its associated research platform are provided in SI~C and SI~D, respectively.  

We next evaluate \RA across all 59 reproduction tasks. During development, \RA uses only the five most challenging tasks, whose verified failures form its experience. The remaining 54 tasks are unseen during development and evaluate how well this experience generalizes. For evaluation, \RA uses DeepSeek V4.1 Flash with DSH as the executor and is given 24 hours per task. We compare it directly with DSH using the same base model and computational budget, with all systems evaluated under the same criteria as in the preceding reproduction experiments.

Figure~\ref{fig:benchmark} compares \RA with existing frontier AI scientists, with further details provided in SI~E. In particular, \RA improves the reproduction success of DeepSeek V4.1 Flash from $68\%$ to approximately $85\%$, at an inference cost of only \$5.6 per task. This performance matches Claude Code with Opus~5 and Codex with GPT-5.6-sol API, which achieve approximately $85\%$ success at \$16 and \$35 per task, respectively. Although the best-performing system still achieves a higher success rate, \RA reaches competitive performance at substantially lower inference cost. 

The internal trajectories further reveal how \RA improves reproduction. Across the 59 tasks, the critic rejects more than half of intermediate results, triggering repair or replanning. Notably, $90\%$ of successful tasks require at least one correction, including all 13 tasks recovered beyond DSH. These results show that \RA gains its advantage primarily by identifying and correcting failures. More experimental results are deferred to SI~E.

  \section*{Discussion}

Our study establishes paper-to-result reproduction as a foundation for developing autonomous AI scientists. Supported by \omniQ, we show that reliable reproduction in computational quantum science depends jointly on the capabilities of AI scientists and the consistency of the scientific record. We further develop \RA, an open-source AI scientist that converts verified reproduction failures into reusable scientific experience, enabling improvement without modifying the underlying model. By making verified reproduction workflows accessible through an extensible research platform, we further lay the groundwork for a community-driven ecosystem of reusable scientific capabilities.

Three important directions remain for future research. The first is to extend \omniQ towards fully autonomous paper-to-result reproduction across scientific disciplines. Recent studies have explored AI-driven scientific reproduction in diverse domains, including Paper2Agent~\cite{miao2026paper2agent} for computational biology, AutoExperiment~\cite{kim2026reproduction} for machine learning, and SocSci-Repro-Bench~\cite{alizadeh2026reproduction} for social sciences. However, our study reveals a fundamental challenge. That is, inconsistencies in the scientific record can make faithful reproduction itself ambiguous, requiring expert intervention to resolve. In this regard, integrating \omniQ and \RA with these domain-specific approaches could enable more reliable and autonomous scientific reproduction across disciplines.

A second direction is to further improve the performance of paper-to-result reproduction in computational quantum science. Although \RA outperforms several more expensive models at lower cost, achieving fully reliable scientific reproduction with greater efficiency remains an open challenge. Toward this goal, there are three potential solutions. First, adaptive allocation of test-time computation could improve reproduction performance by directing greater computational resources toward challenging tasks~\cite{snell2025scaling,yang2025towards}. Second, integrating specialized scientific tools and reusable implementations of previously reproduced algorithms could improve the accuracy and reliability of complex scientific reproductions~\cite{ding2025scitoolagent}. Third, optimizing the performance--cost trade-off through adaptive model selection and efficient agent orchestration could further reduce the cost of scientific reproduction without compromising reliability~\cite{ong2025routellm}.

The last direction is to extend \RA beyond scientific reproduction towards autonomous algorithm discovery in computational quantum science. Recent studies have demonstrated the potential of AI scientists to discover more efficient quantum circuits through iterative search and independent verification~\cite{gu2026discovering,khattar2026leveraging,crichigno2026automating}. Building on these advances, \RA could strengthen autonomous discovery by first establishing reliable computational baselines and then learning from the differences between unsuccessful candidates and verified improvements. By accumulating these insights as reusable experience, \RA could guide subsequent algorithm design and optimization beyond individual search tasks. Extending its self-improvement mechanism from verified reproduction repairs to verified algorithmic advances offers a promising route towards more reliable and efficient autonomous quantum algorithm discovery.

%\bibliographystyle{unsrt}
%\bibliography{references}

\newpage
\clearpage
\appendix 

\renewcommand{\thefigure}{SI \arabic{figure}}
\setcounter{figure}{0}

\renewcommand{\thetable}{SI \arabic{table}}
\setcounter{table}{0}

\tableofcontents
\renewcommand{\appendixname}{SI}
\newcommand{\bi}{\boldsymbol{i}}
\newcommand{\bj}{\boldsymbol{j}}

\bigskip

\section{Preliminary and literature review}\label{append:sec:preliminary}
 
For completeness, this section provides background on computational quantum science and AI scientists. We first introduce three primary areas of computational quantum science: \textit{classical simulation of quantum systems} in SI.~\ref{SI:subsec:classical_simulator}, \textit{quantum algorithms} in SI.~\ref{SI:subsec:quantum_algorithm}, and \textit{quantum property characterization} in SI.~\ref{SI:subsec:property_characterization}. For each area, we review representative methods and computational procedures relevant to our reproduction tasks. Next, we recap representative AI agents and harnesses in SI.~\ref{append:sec:agents}, followed by a review of related work on AI scientists in computational quantum science in SI.~\ref{SI:subsec:literature_review}.

\subsection{Classical simulation of quantum systems}\label{SI:subsec:classical_simulator}
Classical simulation uses classical resources to predict quantum properties and assess potential quantum computational advantages. Here we explain how to estimate observable expectation values, which can be performed either by explicitly simulating quantum states or through methods that directly estimate the target quantities. Mathematically, given classical descriptions of an $N$-qubit initial state $\rho_0$, a quantum evolution channel $\mathcal{U}$, and an observable $O$, the goal is to estimate the expectation value
\begin{equation}\label{SI:A:eq-1}
    f_O(\rho)=\Tr \left[ \mathcal{U}(\rho_0) O \right],
\end{equation}
where $\rho=\mathcal{U}(\rho_0)$ denotes the evolved quantum state. For unitary dynamics described by a unitary operator $U$, the corresponding channel takes the form $\mathcal{U}(\rho_0)=U\rho_0U^{\dagger}$. 

A central challenge in classical simulation is the exponential growth of the Hilbert space with the number of qubits $N$, making general quantum-state simulation computationally expensive. To address this challenge, various classical simulation algorithms have been developed. In what follows, we introduce six representative classes of classical simulation methods: \textit{tensor networks}, \textit{neural quantum states}, \textit{Pauli propagation}, \textit{quantum Monte Carlo methods}, \textit{Lindbladian simulation for open quantum systems}, and \textit{free-fermion simulation}. For each class, we summarize its formulation, computational principles, and conditions for efficient simulation.

\medskip
\noindent\textbf{Tensor networks.}
Tensor-network methods simulate quantum dynamics by representing the evolving quantum state as a network of interconnected tensors~\cite{white1992density,vidal2004efficient,schollwock2011density}. Specifically, consider a decomposition of the evolution channel into elementary channels $\mathcal{U}=\mathcal{U}_L\circ\cdots\circ\mathcal{U}_1$. At each stage of the evolution, the $N$-qubit state, described by a density matrix $\rho=\sum_{\bi,\bj}\rho_{\bi\bj} |\bi\rangle\langle\bj|$, is approximated by a tensor network on a graph $G=(V,E)$ with $|V|=N$, i.e.,
\begin{equation}
    \rho_{\bi\bj}
    \approx
    \sum_{\alpha_1=1}^{\chi_1}\cdots
    \sum_{\alpha_{|E|}=1}^{\chi_{|E|}}
    \prod_{v\in V}
    A^{[v]}_{\bi_v,\bj_v,\{\alpha_e:e\ni v\}}, \nonumber
\end{equation}
where $\bi=(\bi_1,\ldots,\bi_N)$ and $\bj=(\bj_1,\ldots,\bj_N)$ are computational-basis indices, $A^{[v]}$ denotes the local tensor at vertex $v$, and each edge $e\in E$ carries a bond index $\alpha_e$ of dimension $\chi_e$. The state at the next stage, following the application of an $m$-qubit channel $\mathcal{U}_\ell$ to the qubit set $S_\ell$, can then be approximated by applying $\mathcal{U}_\ell$ to the contracted tensor $\mathcal{T}_\ell$ of the corresponding subgraph and decomposing it back into the original network structure with truncated bond dimensions, i.e.,
\begin{equation}
    \{A^{[v]}\}_{v\in S_\ell}
    \xrightarrow{\mathrm{contraction}}
    \mathcal{T}_\ell
    \xrightarrow{\mathcal{U}_\ell}
    \mathcal{T}_\ell'
    \xrightarrow{\mathrm{decomposition+truncation}}
    \{\widetilde A^{[v]}\}_{v\in S_\ell}.  \nonumber
\end{equation}

The efficiency of tensor-network simulation depends on both the entanglement structure of the quantum state and the geometry of the underlying network $G$. States with area-law entanglement often admit compact representations with small bond dimensions ${\chi_e}$, whereas volume-law entanglement can require exponentially large bond dimensions. Consequently, tensor-network methods are particularly effective for weakly entangled systems, enabling efficient exact simulation in certain structured cases and accurate approximate simulation more broadly through tensor contraction. Representative methods include matrix product states, matrix product operators, and projected entangled-pair states.

\medskip
\noindent\textbf{Neural quantum states.} According to Eq.~(\ref{SI:A:eq-1}), a neural quantum state (NQS) simulates quantum dynamics by approximating the evolving quantum state {$\mathcal{U}(\rho_0)$}. Different from tensor networks, NQS methods represent the wavefunction amplitudes using a parameterized neural network~\cite{carleo2017solving,schmitt2020quantum}. For pure-state dynamics with $\rho_0=|\psi_0\rangle\langle\psi_0|$ and $\mathcal U(\rho_0)=|\psi\rangle\langle\psi|$, the goal is to approximate $|\psi\rangle$ with high fidelity using an NQS $|\psi_{\btheta}\rangle$ of the form
\begin{equation}
    |\psi_{\btheta}\rangle
    :=
    \frac{1}{\sqrt{\sum_{\bi}|g_{\btheta}(\bi)|^2}}
    \sum_{\bi}g_{\btheta}(\bi)|\bi\rangle,  \nonumber
\end{equation}
where $g_{\btheta}(\cdot)$ is a complex-valued neural network that outputs an unnormalized amplitude for each configuration $\bi$.

A common strategy for simulating quantum dynamics with neural quantum states (NQS) is to sequentially optimize the network parameters after each elementary unitary evolution. Starting from an initial NQS $|\psi_{\btheta_0}\rangle=|\psi_0\rangle$, the parameters $\btheta_\ell$ at step $\ell$ are obtained by minimizing the infidelity between the normalized variational state $|\psi_{\btheta}\rangle$ and the target state $U_\ell|\psi_{\btheta_{\ell-1}}\rangle$, i.e., $\mathcal I_\ell(\btheta)
=1-|\langle\psi_{\btheta}|U_\ell|\psi_{\btheta_{\ell-1}}\rangle|^2$ with $\btheta_\ell\in\arg\min_{\btheta}\mathcal I_\ell(\btheta)$.  To enable efficient optimization, the infidelity gradient can be expressed in terms of expectations over the Born distribution $p_{\btheta}(\bi)=|\langle\bi|\psi_{\btheta}\rangle|^2$ as
\begin{equation}\label{SI:A:eqn2}
    \frac{\partial\mathcal I_k}{\partial\btheta_a}
    =-2\operatorname{Re}\left[
    \mathbb E_{\bi\sim p_{\btheta}}
    \left[
    \frac{\langle \bi|U_\ell|\psi_{\btheta_{\ell-1}}\rangle}
    {\langle \bi|\psi_{\btheta}\rangle}
    \right]^*
    \mathbb E_{\bi\sim p_{\btheta}}
    \left[
    \frac{\langle \bi|U_\ell|\psi_{\btheta_{\ell-1}}\rangle}
    {\langle \bi|\psi_{\btheta}\rangle}
    \left(
    O_a^*(\bi)-\mathbb E_{p_{\btheta}}[O_a^*]
    \right)
    \right]
    \right],
\end{equation}
where $O_a(\bi)=\partial_{\bm{\theta}_a}\log g_{\btheta}(\bi)$ is the logarithmic derivative of the neural-network amplitude.

The infidelity gradient in Eq.~(\ref{SI:A:eqn2}) can be estimated using Monte Carlo sampling, allowing NQS parameters to be optimized without explicitly representing the full quantum state. However, gradient estimation can suffer from numerical instability and sampling inefficiency, making optimization challenging~\cite{bagrov2021kinetic}. Improved neural-network architectures, sampling strategies, and gradient estimators have been developed to address these limitations~\cite{sinibaldi2023unbiasing}, enabling NQS to simulate increasingly complex quantum states and dynamics.

\medskip
\noindent\textbf{Pauli propagation.} Pauli propagation simulates quantum dynamics by evolving observables
in the Heisenberg picture and is particularly efficient for circuits
with limited non-Clifford resources~\cite{rall2019simulation,beguvsic2024fast}.
Following Eq.~(\ref{SI:A:eq-1}), for an $N$-qubit channel
$\mathcal{U}$ and an observable $O$, the expectation value is given by
$f_O(\rho_0)=\Tr[O\mathcal{U}(\rho_0)]
=\Tr[\mathcal{U}^{\dagger}(O)\rho_0]$,
where $\mathcal{U}^{\dagger}
=\mathcal{U}_1^\dagger\circ\cdots\circ\mathcal{U}_L^\dagger$
is the adjoint channel.
The evolved observable is represented as a Pauli expansion, i.e.,
\begin{equation}
\mathcal{U}^{\dagger}(O)
=\sum_{\bm{a}\in A}c_{\bm{a}}P_{\bm{a}},  \nonumber
\end{equation}
where $A\subseteq\mathbb{F}_2^{2N}$ indexes the nonzero terms,
$P_{\bm{a}}\in\{\mathbb{I},X,Y,Z\}^{\otimes N}$ denotes a Pauli
string, and $c_{\bm{a}}$ is its coefficient. 
The expectation value is then evaluated as
$f_O(\rho_0)=\sum_{\bm{a}\in A}c_{\bm{a}}\Tr(P_{\bm{a}}\rho_0)$. As a result, the computational efficiency depends on the growth of the Pauli expansion
during backward propagation.  

To reduce this computational cost, approximate Pauli-propagation algorithms truncate the Pauli expansion during the propagation process by retaining nontrivial Pauli strings~\cite{beguvsic2024fast}. Denote the retained Pauli strings by $A'\subseteq\mathbb{F}_2^{2N}$. The resulting approximation of the evolved observable takes the form $\hat{\mathcal{U}}^{\dagger}(O)=\sum_{\bm{a}\in A'}\hat{\bm{c}}_{\bm{a}}P_{\bm{a}}$, and the corresponding estimate of $f_{O}(\rho)$ is given by
\begin{equation}
\hat{f}_O(\rho)
=\Tr[\hat{\mathcal{U}}^{\dagger}(O)\rho_0]=\sum_{\bm{a}\in A'}
\hat{\bm{c}}_{\bm{a}}\Tr(P_{\bm{a}}\rho_0),  \nonumber
\end{equation}
where $\hat{\bm{c}}_{\bm{a}}$ denotes the coefficient obtained from the truncated propagation procedure. An effective Pauli-propagation algorithm should satisfy $|f_O(\rho)-\hat f_O(\rho)|\leq\epsilon$ for a prescribed accuracy $\epsilon$, while keeping the number of retained Pauli terms and the computational cost manageable.  Typical approaches in this regime include Pauli-weight truncation, coefficient-based pruning, and Monte Carlo sampling of Pauli paths~\cite{angrisani2025classically,beguvsic2024fast,rall2019simulation}.

\medskip
\noindent\textbf{Quantum Monte Carlo methods.} Quantum Monte Carlo methods directly estimate observable expectations $\Tr[O\mathcal U(\rho_0)]$ through stochastic sampling of suitable representations of quantum systems~\cite{blankenbecler1981monte,ceperley1995path, troyer2005computational,cohen2015taming}. Depending on the formulation, these methods can address equilibrium properties, quantum dynamics, and quantum-circuit simulations, while their efficiency may be limited by sign or phase problems. As a representative example, consider the estimation of thermal expectation values for a Hamiltonian $H$ at inverse temperature $\beta$, given by
\begin{equation}
    \Tr[O\mathcal U_\beta(\rho_0)]
    =
    \frac{\Tr(Oe^{-\beta H})}
    {\Tr(e^{-\beta H})} =
    \sum_{\bi} \frac{\langle \bi|e^{-\beta H}|\bi\rangle}{
    \Tr(e^{-\beta H})}
    \frac{\langle \bi|Oe^{-\beta H}|\bi\rangle}{
    \langle \bi|e^{-\beta H}|\bi\rangle}, \nonumber
\end{equation}
where $\mathcal U_\beta(\rho_0) =\Tr(\rho_0)e^{-\beta H}/\Tr(e^{-\beta H})$ refers to a thermal-state preparation channel.

To exploit Hamiltonian locality, imaginary-time QMC decomposes
$e^{-\beta H}=(e^{-\Delta\tau H})^L$, where
$\Delta\tau=\beta/L$, and inserts computational-basis
resolutions of the identity between successive time slices.
This yields a path-integral representation of the thermal
expectation value 
$\langle O\rangle_\beta
=\sum_{\mathcal C}p(\mathcal C)O_{\mathrm{loc}}(\mathcal C)$,
where $\mathcal C=(\bi_1,\ldots,\bi_L)$ is a
configuration path with $\bi_{L+1}=\bi_1$, and
\begin{equation} 
p(\mathcal C)=\frac{w(\mathcal C)}{\sum_{\mathcal C'}w(\mathcal C')},\quad w(\mathcal C)=\prod_{\ell=1}^{L}\langle\bi_{\ell+1}|e^{-\Delta\tau H}|\bi_\ell\rangle,\quad O_{\mathrm{loc}}(\mathcal C)=\frac{\langle\bi_2|e^{-\Delta\tau H}O|\bi_1\rangle}{\langle\bi_2|e^{-\Delta\tau H}|\bi_1\rangle}. \nonumber
\end{equation}
When $w(\mathcal C)\geq 0$, the normalized weights define
a sampling distribution, allowing the expectation value
to be estimated without explicitly summing over
exponentially many paths, i.e.,
$\langle O\rangle_\beta
\approx
\frac{1}{M}\sum_{m=1}^{M}
O_{\mathrm{loc}}(\mathcal C^{(m)})$ with $\mathcal C^{(m)}\sim p$.

\medskip
\noindent\textbf{Lindbladian simulation for open quantum systems.}
Lindbladian simulation provides a framework for simulating the dynamics of an open quantum system interacting with an environment without explicitly representing the environmental degrees of freedom~\cite{lindblad1976generators}.
Under the assumption of time-homogeneous Markovian dynamics, the evolution channel over time $T$ takes the Lindblad form $\mathcal U=e^{T\mathcal L}$, where the Lindblad generator $\mathcal L$ is defined by
\begin{equation}\label{SI:A:eqn:3}
    \mathcal L(\rho_t)
    =
    -\imath[H,\rho_t]
    +
    \sum_\mu
    \left(
    L_\mu\rho_t L_\mu^\dagger
    -
    \frac12\{L_\mu^\dagger L_\mu,\rho_t\}
    \right),
\end{equation}
where $H$ is the system Hamiltonian, $\rho_t$ is the system density matrix at time $t\leq T$, $L_\mu$ are the jump operators describing dissipative processes, and $[A,B]=AB-BA$ and $\{A,B\}=AB+BA$ denote the commutator and anticommutator, respectively.

A standard approach to evaluating $f_O(\rho)$ in Eq.~\eqref{SI:A:eq-1}
under Lindblad dynamics is the quantum-trajectory method~\cite{dalibard1992wave,molmer1993monte},
which unravels the mixed-state evolution into an ensemble of stochastic
pure-state trajectories, i.e.,
$\rho_t=\mathbb E_{\xi\sim p}[|\psi_t^{(\xi)}\rangle\langle\psi_t^{(\xi)}|]$.
Each trajectory $|\psi_t^{(\xi)}\rangle$ undergoes non-unitary evolution
governed by the effective non-Hermitian Hamiltonian
$H_{\mathrm{eff}}=H-\frac{\imath}{2}\sum_\mu L_\mu^\dagger L_\mu$,
derived from the Lindblad generator in Eq.~\eqref{SI:A:eqn:3},
interrupted by stochastic quantum jumps associated with the operators $L_\mu$.
The observable expectation is then estimated by averaging over
$M$ independently sampled trajectories:
\begin{equation}
\Tr(O\rho_T)
\approx
\frac{1}{M}\sum_{m=1}^{M}
\langle\psi_T^{(m)}|O|\psi_T^{(m)}\rangle. \nonumber
\end{equation}

\medskip
\noindent\textbf{Free-fermion simulation.}
Free-fermion methods enable efficient classical simulation of fermionic
Gaussian states evolving under quadratic Hamiltonians, without explicitly
representing the exponentially large quantum state~\cite{terhal2002classical,valiant2002quantum}. To evaluate the observable expectation $f_O(\rho)$ in
Eq.~(\ref{SI:A:eq-1}), we first introduce suitable representations
of the initial state, its evolution, and the target observable.

For an $N$-mode fermionic system, define the Majorana operators
$\gamma_{2j-1}=\bm{a}_j+\bm{a}_j^\dagger$ and
$\gamma_{2j}=-\imath(\bm{a}_j-\bm{a}_j^\dagger)$,
where $\bm{a}_j^\dagger$ and $\bm{a}_j$ denote fermionic creation
and annihilation operators, respectively.
A full-rank fermionic Gaussian state can be expressed as
$\rho_0=e^{-K_0}/\Tr(e^{-K_0})$, where
$K_0=(\imath/4)\boldsymbol{\gamma}^\top B_0\boldsymbol{\gamma}$
and $B_0$ is a real antisymmetric $2N\times 2N$ matrix.
Consider the evolution
$\mathcal U=\mathcal U_L\circ\cdots\circ\mathcal U_1$,
where each unitary channel is generated by a quadratic Hamiltonian
$H_\ell=(\imath/4)\boldsymbol{\gamma}^\top M_\ell\boldsymbol{\gamma}$,
with $M_\ell$ real and antisymmetric.
The evolved state remains Gaussian, and its coefficient matrix
is updated as
\begin{equation}
\label{eq:free-fermion-update}
B_\ell=Q_\ell B_{\ell-1}Q_\ell^\top
\qquad \text{with}~Q_\ell=e^{M_\ell x_\ell}, \nonumber
\end{equation}
where $x_\ell$ denotes the evolution time at step $\ell$.
The final state $\rho=\mathcal U(\rho_0)$ is thus characterized
by $B_L$, from which its covariance matrix
$\Gamma_{ab}=\frac{\imath}{2}\Tr(\rho[\gamma_a,\gamma_b])$
can be obtained as $\Gamma=-\tan(B_L/2)$.
In addition, we consider Hermitian observables of the form
$O=\imath^k\gamma_{a_1}\cdots\gamma_{a_{2k}}$, where $a_1,\ldots,a_{2k}$ denote distinct Majorana indices.

With the above notations, the expectation value $f_O(\rho)$ under Free-fermion system can then be
efficiently evaluated using Wick's theorem~\cite{wick1950evaluation}. Mathematically, we have
\begin{equation}
\label{eq:free-fermion-observable}
f_O(\rho)=\mathsf{Pf}(\Gamma_A), \nonumber
\end{equation}
where $\mathsf{Pf}(\cdot)$ denotes the Pfaffian of an
antisymmetric matrix, and $\Gamma_A$ is the
$2k\times 2k$ submatrix of $\Gamma$ obtained by selecting
the rows and columns indexed by
$A=(a_1,\ldots,a_{2k})$, as specified by the Majorana
operators appearing in $O$.

\subsection{Quantum algorithms}\label{SI:subsec:quantum_algorithm}

Quantum algorithms address computational problems by processing information encoded in quantum states and extracting results through measurement. We focus on five representative categories spanning both fault-tolerant and near-term quantum computing:  \textit{non-variational quantum simulation}, \textit{quantum phase estimation}, \textit{variational quantum simulation}, and \textit{quantum machine learning}, \textit{quantum-enhanced Markov chain Monte Carlo}. The diversity of computational approaches and implementation strategies within these categories further complicates autonomous reproduction, requiring AI scientists to reconstruct the specific methods adopted in individual studies. For each category, we introduce the problem formulation and basic workflows of representative algorithms.

\medskip
\noindent\textbf{Non-variational quantum simulation.}
Quantum simulation aims to understand the properties and dynamics of quantum systems, whose exponentially growing state spaces often make classical computation intractable. Non-variational approaches address this challenge by directly constructing quantum operations from the system Hamiltonian to simulate time evolution or prepare ground states~\cite{fauseweh2024quantum}. Mathematically, for a Hamiltonian $H(t)$ and an initial state $|\psi(0)\rangle$, the evolving state obeys the Schr\"odinger equation
\begin{equation} \label{SIA_qa_schrodinger_eq}
\partial_t|\psi(t)\rangle=-\imath H(t)|\psi(t)\rangle.
\end{equation}
Below, we discuss this approach for time-independent and time-dependent Hamiltonians, respectively.

For time-independent $H$ in Eq.~(\ref{SIA_qa_schrodinger_eq}), its solution is $|\psi(t)\rangle=e^{-\imath Ht}|\psi(0)\rangle$.
Trotterization approximates this evolution by composing operations generated by individual Hamiltonian terms~\cite{childs2021theory}. Let the explored Hamiltonian be  $H=\sum_{\ell=1}^{m}h_\ell$, where each Hermitian term $h_\ell$ acts on a small number of qubits, and dividing the total time $t>0$ into $L$ steps of duration $\Delta t=t/L$ gives the first-order approximation
\begin{equation*}
U_{\mathrm{Trot}}(t)=\left(e^{-\imath h_m\Delta t}\cdots e^{-\imath h_1\Delta t}\right)^L\approx e^{-\imath Ht}.
\end{equation*}
Each local unitary can be decomposed into elementary gates, and applying the full sequence prepares an approximation to $|\psi(t)\rangle$.
For fixed $H$, the error satisfies $\|U_{\mathrm{Trot}}(t)-\exp({-\imath Ht})\|= \mathcal{O}({t^2}/{L})$~\cite{childs2021theory}.
Increasing $L$ reduces the error bound at the cost of greater circuit depth.
Higher-order Trotter formulas~\cite{suzuki1992general} improve this tradeoff by canceling lower-order errors through local evolutions with suitably chosen ordering and durations.
Further refinements exploit separated energy scales~\cite{bosse2025efficient} or introduce auxiliary correctors~\cite{bagherimehrab2026faster} to reduce product-formula errors.

For time-dependent Hamiltonians $H(t)$ in Eq.~(\ref{SIA_qa_schrodinger_eq}), $\ket{\psi(t)}$ composes short evolutions with the Hamiltonian held fixed within each step~\cite{poulin2011quantum}, enabling digital adiabatic ground-state preparation~\cite{lu2026digital}. More precisely, 
starting from an easily prepared ground state of $H_{\mathrm{init}}$, one varies $H(s)=(1-s)H_{\mathrm{init}}+sH$ from $s=0$ to $s=1$.
For a smooth path with a nondegenerate ground state and a nonzero spectral gap, sufficiently slow evolution follows the instantaneous ground state toward the target ground state~\cite{albash2018adiabatic}.
Beyond adiabatic preparation, quantum imaginary-time evolution~\cite{motta2020determining} and adaptive spectral filtering~\cite{karacan2025enhancing} can prepare approximate ground states by suppressing excited-state components through unitary updates inferred from measurements and postselected filters built from controlled evolution, respectively.

\medskip
\noindent\textbf{Quantum phase estimation.} Quantum phase estimation (QPE)~\cite{cleve1998quantum} extracts the eigenphases of a unitary operator $U$, defined by $U|\phi_j\rangle=e^{\imath\varphi_j}|\phi_j\rangle$ with $\varphi_j\in[0,2\pi)$.  QPE offers rigorous precision guarantees, achieving additive phase-estimation error $\varepsilon$ with $\mathcal{O}(1/\varepsilon)$ controlled applications of $U$ under standard assumptions. As a fundamental primitive underlying quantum algorithms such as Shor's factoring algorithm and Hamiltonian eigenvalue estimation, QPE plays a central role in the pursuit of quantum computational advantages.

The mechanism of QPE is as follows. Given an initial state $|\psi_{\mathrm{init}}\rangle=\sum_jc_j|\phi_j\rangle$ and $m$ ancillary qubits prepared in $|+\rangle^{\otimes m}$, the circuit performs
\begin{equation*}
|+\rangle^{\otimes m}|\psi_{\mathrm{init}}\rangle
\xrightarrow{\prod_{r=0}^{m-1}\mathrm{C}_rU^{2^r}}
\frac{1}{2^{m/2}}\sum_{j,s}c_j e^{\imath s\varphi_j}|s\rangle|\phi_j\rangle
\xrightarrow{\text{inverse QFT}}
\sum_{j,y}c_j a_{jy}|y\rangle|\phi_j\rangle.
\end{equation*}
Here, $\mathrm{C}_rU^{2^r}$ denotes $U^{2^r}$ controlled by ancillary qubit $r$, and the inverse quantum Fourier Transform (QFT) acts on the ancillary register.
The indices $s,y\in\{0,\ldots,2^m-1\}$ label its computational-basis states, with $a_{jy}=2^{-m}\sum_{s=0}^{2^m-1}\exp({\imath s(\varphi_j-2\pi y/2^m)})$.
Measuring this register yields $y$ with probability $p(y)=\sum_j|c_j|^2|a_{jy}|^2$, giving the estimate $\hat\varphi=2\pi y/2^m$.
For each eigenstate component, $|a_{jy}|^2$ is concentrated near its eigenphase modulo $2\pi$, with resolution $\mathcal{O}(2^{-m})$.
Improving the resolution requires higher controlled powers of $U$, up to $U^{2^{m-1}}$.

Alternative implementations reduce ancillary-qubit requirements through iterative measurements~\cite{dobsicek2007arbitrary}, guide measurements through Bayesian updates~\cite{wiebe2016efficient}, or accommodate inexact initial states with sufficient target-state overlap~\cite{ni2023low}. Beyond extracting spectral information, QPE also serves as a subroutine in quantum algorithms such as quantum amplitude estimation~\cite{brassard2002quantum}, which estimates the probability $a$ of a target measurement outcome for a state prepared by a unitary $W$.
Combining a reflection about $W|0\rangle$ with a phase flip on the target outcomes defines an amplification operator $Q$ whose relevant eigenphases $\pm2\theta$ satisfy $a=\sin^2\theta$.
Applying QPE to $Q$ gives an estimate $\hat\theta$ of $\theta$ and hence $\hat a=\sin^2\hat\theta$, achieving $|\hat a-a|\leq\epsilon$ with $\mathcal{O}(\epsilon^{-1})$ calls to $W$ and $W^\dagger$, compared with $\mathcal{O}(\epsilon^{-2})$ for direct sampling.
To reduce circuit depth, classical inference can replace QPE by using measurements at different amplification powers~\cite{giurgica2022low}, with Bayesian schemes adapting these powers while accounting for device noise~\cite{ramoa2025bayesian}.

\medskip
\noindent\textbf{Variational quantum simulation.} Complementing non-variational approaches, variational quantum simulation aims to investigate quantum states and dynamics using variational quantum circuits (VQCs) tailored to the limited coherence times and circuit depths of near-term quantum devices~\cite{cerezo2021variational}. It approximates quantum states or time evolution through hybrid quantum-classical optimization, offering a hardware-efficient alternative to direct Hamiltonian-based simulation. Specifically,
the variational quantum eigensolver (VQE)~\cite{peruzzo2014variational} targets ground-state preparation through energy minimization, while variational quantum Hamiltonian evolution~\cite{barison2021efficient} simulates dynamics by fitting evolved states or time-evolution operators. In what follows, we present these two standard approaches separately.

\smallskip
\noindent\textit{Variational quantum eigensolver.} VQE approximates the ground state within the circuit family $|\psi(\bmt)\rangle=U(\bmt)|\psi_{\mathrm{init}}\rangle$, where $|\psi_{\mathrm{init}}\rangle$ is an easily prepared reference state and $\bmt$ denotes the parameters of  $U(\bmt)$, i.e., the employed VQC with a tailored circuit architecture. For the target Hamiltonian $H$, the objective is to minimize
\begin{equation}\label{SIA_qa_variational_energy_eq}
E(\bmt)=\langle\psi(\bmt)|H|\psi(\bmt)\rangle\geq E_0,
\end{equation}
where $E_0$ is the ground-state energy. A classical optimizer updates $\bmt$ to minimize Eq.~(\ref{SIA_qa_variational_energy_eq}).
Gradients used in the optimization can be estimated by evaluating the energy value $E(\bmt)$ at shifted circuit parameters~\cite{schuld2019evaluating}.
The optimized VQC (i.e., $U(\bmt^*)$) prepares a ground-state approximation, whose accuracy depends on the circuit family and the quality of the optimization~\cite{mcclean2016theory}.
Further developments address circuit construction through adaptive operator selection based on energy gradients~\cite{grimsley2019adaptvqe} and parameter initialization through tensor-network pretraining~\cite{rudolph2023synergistic}.

Beyond ground-state preparation, the VQE framework can be extended to combinatorial optimization with a cost function $C(\bm{s})$ over binary configurations $\bm{s}\in\{0,1\}^N$, as in the quantum approximate optimization algorithm (QAOA)~\cite{farhi2014quantum,farhi2022quantum}. Here, the VQC alternates evolution under \[H_C=\sum_{\bm{s}}C(\bm{s})|\bm{s}\rangle\langle\bm{s}| \quad\text{and}\quad H_B=\sum_{j=1}^{N}X_j.\]
The evolution times are optimized to minimize Eq.~(\ref{SIA_qa_variational_energy_eq}) with $H=H_C$, after which low-cost solutions are selected from configurations sampled in the computational basis.

\smallskip
\noindent\textit{Variational quantum Hamiltonian evolution.} Variational approaches simulate Hamiltonian dynamics by optimizing $U(\bmt)$ that prepare approximate evolved states $|\psi(\bmt)\rangle=U(\bmt)|\psi_{\mathrm{init}}\rangle$.
For a time-independent Hamiltonian $H$, one approach starts from $|\psi(\bmt(0))\rangle\approx|\psi(0)\rangle$ and fits each short-time evolved state $e^{-\imath H\Delta t}|\psi(\bmt(j))\rangle$ with $|\psi(\bmt)\rangle$ by minimizing their infidelity
\begin{equation}\label{SIA_qa_variational_update_eq}
\bmt(j+1)=\operatorname*{arg\,min}_{\bmt}\left[1-\left|\bra{\psi(\bmt)}e^{-\imath H\Delta t}\ket{\psi(\bmt(j))}\right|^2\right],
\end{equation}
where $\bmt(j)$ denotes the parameters at time $j\Delta t$~\cite{barison2021efficient}. The short-time evolution $\exp({-\imath H\Delta t})$ can be approximated by Trotterization.
Repeating the update approximates the state $|\psi(t)\rangle$ at successive times.

Subsequent studies have explored complementary strategies to improve the efficiency and scalability of variational quantum Hamiltonian evolution. To reduce the required quantum resources, circuit knitting estimates the fidelity in Eq.~(\ref{SIA_qa_variational_update_eq}) using smaller subcircuits while constraining the sampling overhead~\cite{gentinetta2024overhead}.
Other approaches approximate the time-evolution operator $\exp({-\imath H\Delta t})$ directly by learning a diagonal factorization to extend simulation times~\cite{cirstoiu2020variational} or fitting its action on subsystems to reduce the number of qubits required for training~\cite{mizuta2022local}.

\medskip
\noindent\textbf{Quantum machine learning.} Quantum machine learning (QML) aims to leverage quantum computing to learn patterns from classical or quantum data, with the potential to enhance the expressivity and computational efficiency of learning models~\cite{Du:2025amm}.
Two representative formulations are quantum neural networks (QNNs)~\cite{schuld2020circuit}, which optimize circuits to learn predictive models, and quantum kernel methods~\cite{schuld2019quantum}, which use quantum circuits to estimate kernel values for classical model fitting.

\smallskip
\noindent\textit{Quantum neural networks.} A typical QNN encodes an input $\bm{x}$ as $|\phi(\bm{x})\rangle=V(\bm{x})|0\rangle$ and processes it with a VQC $U(\bmt)$.
The expectation value of an observable $O$ defines the model output, i.e.,
\begin{equation*}
f_{\bmt}(\bm{x})=\langle\phi(\bm{x})|U^\dagger(\bmt)O U(\bmt)|\phi(\bm{x})\rangle.
\end{equation*}
For supervised learning, consider a training set $\mathcal{D}=\{(\bm{x}^{(i)},y^{(i)})\}_{i=1}^{M}$, where $y^{(i)}$ is the target associated with input $\bm{x}^{(i)}$.
The parameters in $f_{\bmt}(\bm{x})$ are optimized to minimize the empirical loss $\mathcal{L}(\bmt)=M^{-1}\sum_{i=1}^{M}\ell(f_{\bmt}(\bm{x}^{(i)}),y^{(i)})$, where $\ell(z,y)$ measures the discrepancy between a model output $z$ and its target $y$. For regression, one can use the squared-error loss $\ell(z,y)=(z-y)^2$, and for binary classification with $y\in\{-1,1\}$, the logistic loss $\ell(z,y)=\log(1+\exp({-yz}))$. Once trained, the circuit predicts outputs for new inputs.
 
Subsequent studies have explored both the applications and theoretical properties of QNNs. Beyond predictive learning, QNNs have been extended to generative modeling of probability distributions~\cite{lloyd2018quantum,tian2023recent} and reinforcement learning of decision policies~\cite{chen2020variational}. Meanwhile, theoretical studies have investigated the expressivity, trainability, and generalization of QNNs. Expressivity analyses characterize the functions and quantum states representable by parameterized circuits~\cite{du2022efficient}. Regarding trainability, global cost functions and scrambling dynamics can cause gradients to vanish exponentially~\cite{mcclean2018barren}. Generalization analyses further investigate how model complexity, training data, and quantum measurements affect predictive performance on unseen inputs~\cite{du2023problem}.

\smallskip
\noindent\textit{Quantum kernel methods.} Quantum kernel methods construct predictors from a kernel $\kappa(\bm{x},\bm{x}')$, defined as an inner product between input representations in a feature space~\cite{havlivcek2019supervised,schuld2019quantum}.
For the training set $\mathcal{D}$, the predictor takes the form $f(\bm{x})=\sum_{i=1}^{M}\bm{\alpha}_i\kappa(\bm{x},\bm{x}^{(i)})$, where $\bm{\alpha}\in\mathbb{R}^M$ contains the learned coefficients. A typical example of quantum kernels is the fidelity kernel $\kappa(\bm{x},\bm{x}')=\operatorname{Tr}[\rho(\bm{x})\rho(\bm{x}')]$, where $\rho(\bm{x})=|\phi(\bm{x})\rangle\langle\phi(\bm{x})|$. The value of this kernel equals the probability of the all-zero outcome after applying $V^\dagger(\bm{x})V(\bm{x}')$ to $|0\rangle$ and measuring in the computational basis. 
The coefficients $\bm{\alpha}=(\bm{\alpha}_1, \cdots, \bm{\alpha}_M)^\top$ can then be fitted classically by minimizing the regularized least-squares objective of kernel ridge regression,
\begin{equation*}
\bm{\alpha}^{*}=\arg\min_{\bm{\alpha}\in\mathbb{R}^{M}}
\left\{\|K\bm{\alpha}-\bm{y}\|_2^2+\lambda\bm{\alpha}^{\top}K\bm{\alpha}\right\}, i.e.,
\end{equation*}
where $K$ is the kernel matrix with entries $K_{ij}=\kappa(\bm{x}^{(i)},\bm{x}^{(j)})$, $\bm{y}=(y^{(1)},\ldots,y^{(M)})^\top$, and $\lambda>0$ is a regularization parameter~\cite{saunders1998ridge}.
Solving $(K+\lambda\mathbb{I}_M)\bm{\alpha}^{*}=\bm{y}$ gives the fitted coefficients $\bm{\alpha}^{*}$.
The prediction for a new input $\bm{x}$ is then computed using the fitted coefficients $\bm{\alpha}^{*}$ and the kernel values $\kappa(\bm{x},\bm{x}^{(i)})$ between this input and the training samples.
Following this framework, projected kernels select observable features to improve generalization~\cite{huang2021power}, kernel alignment adapts data encodings to the learning task~\cite{hubregtsen2022training,lei2024neural}, and covariant kernels exploit group structure in the data~\cite{glick2024covariant}.

\medskip
\noindent\textbf{Quantum-enhanced Markov chain Monte Carlo.}
Quantum-enhanced Markov chain Monte Carlo (MCMC)~\cite{layden2023quantum} combines quantum proposals with classical acceptance rules to sample equilibrium distributions and estimate thermal averages. By potentially accelerating the exploration of complex energy landscapes, it offers a promising approach to studying many-body systems and addressing sampling bottlenecks in statistical physics and computational chemistry.

A representative setting of Quantum-enhanced MCMC is a classical Ising model with configurations $\bm{s}\in\{-1,1\}^N$ and energy $E(\bm{s})$.
The target is the Boltzmann distribution $\pi(\bm{s})=\exp({-\beta E(\bm{s})})/\mathcal Z$, where $\beta$ is the inverse temperature and $\mathcal Z=\sum_{\bm{s}}\exp({-\beta E(\bm{s})})$ is the partition function. At each step of the Markov chain, the current configuration is encoded in a computational basis state $|\bm{s}\rangle$ and evolved under the Hamiltonian $H_{\mathrm{prop}}=\sum_{\bm{s}}E(\bm{s})|\bm{s}\rangle\langle\bm{s}|+\eta\sum_{j=1}^{N}X_j$, where $\eta\in\mathbb R$ controls the spin-flip strength.
After evolution for a time $t$, a computational-basis measurement yields a candidate $\bm{s}'$ with probability $p(\bm{s}'|\bm{s})=|\langle\bm{s}'|\exp({-\imath H_{\mathrm{prop}}t})|\bm{s}\rangle|^2$.
Since $H_{\mathrm{prop}}$ is real and symmetric, $p(\bm{s}'|\bm{s})=p(\bm{s}|\bm{s}')$, 
so the Metropolis--Hastings rule used to update the Markov chain accepts the proposed configuration $\bm{s}'$ with a probability that can be computed classically from the energy difference, i.e.,
\begin{equation*}
{\rm Pr}_{\mathrm{acc}}(\bm{s}'|\bm{s})=\min\left\{1,\frac{\pi(\bm{s}')p(\bm{s}|\bm{s}')}{\pi(\bm{s})p(\bm{s}'|\bm{s})}\right\}=\min\left\{1,e^{-\beta[E(\bm{s}')-E(\bm{s})]}\right\}.
\end{equation*}
This rule preserves $\pi$ as the stationary distribution, so averages along an ergodic chain converge to the corresponding thermal expectations.
The spin-flip strength $\eta$ and evolution time $t$ can be chosen to favor transitions between configurations that differ in many spins but have similar energies, which can accelerate equilibration compared with single-spin updates.

 Subsequent studies have further explored improved quantum proposal strategies~\cite{nakano2024markov}, faster mixing~\cite{hsieh2026adiabatic}, and reduced quantum resource requirements~\cite{ferguson2025quantum}, extending its applicability to larger and more complex systems~\cite{marshall2026quantum,sajjan2026polynomially}.

\subsection{Property characterization of quantum systems}\label{SI:subsec:property_characterization}

Quantum property characterization protocols aim to infer physical properties of quantum systems from finite measurement results. Following the framework in Ref.~\cite{du2026ai}, we review five representative directions: \textit{linear property estimation, nonlinear property estimation, Hamiltonian learning, quantum error mitigation, and quantum process characterization}. For each direction, we outline representative algorithms, focusing on their problem formulations, key assumptions, and computational procedures.

\medskip
\noindent\textbf{Linear properties of quantum states.}
The linear property of an $N$-qubit quantum state $\rho$ refers to the expectation value of a fixed observable $O$, i.e., $f_O(\rho)$ in Eq.~(\ref{SI:A:eq-1}). Applying different observables $O$ can characterize various physical quantities of quantum systems, such as magnetization, energy, and correlation functions. Two complementary approaches to estimating $f_O(\rho)$ are \textit{classical shadows}~\cite{huang2020predicting} and \textit{classical surrogates}~\cite{du2025efficient}. That is, classical shadows estimate multiple observable expectation values from measurements of a fixed quantum state, whereas classical surrogates learn to predict these values across a parametrized family of quantum states. Their mechanisms are summarized below.

\smallskip

\noindent \emph{Classical shadows.} Classical shadows enable the prediction of multiple quantum properties from shared measurement data without additional quantum measurements. Its mechanism is as follows. For an unknown $N$-qubit state $\rho$, the classical-shadow protocol performs $T$ independent measurement rounds. In each round, a unitary $U$ is sampled from a prescribed ensemble $\mathcal{U}$ and applied to a fresh copy of $\rho$, followed by a computational-basis measurement. The outcome is an $N$-bit string $\bm{b}\in\{0,1\}^N$, obtained with probability $\mathbb{P}(\bm{b}|U)=\bra{\bm{b}}U\rho U^\dagger\ket{\bm{b}}$. This procedure defines a linear map $\mathcal{M}$, known as the measurement channel, i.e., $\mathcal{M}(\rho) =\mathbb{E}_{U\sim \mathcal{U}}\mathbb{E}_{\bm{b}\sim \mathbb{P}(\bm{b})} U^{\dagger}\ket{\bm{b}}\bra{\bm{b}} U    = \sum_{\bm{b}} \int \mathsf{d} U U^{\dagger}\ket{\bm{b}}\bra{\bm{b}} U\bra{\bm{b}}U\rho U^{\dagger}\ket{\bm{b}}$.

When the measurement channel $\mathcal{M}$ is invertible,
each measurement outcome yields a classical-shadow snapshot
$\hat{\rho}_t=\mathcal{M}^{-1}
(U_t^\dagger|\bm{b}_t\rangle\langle\bm{b}_t|U_t)$.
For randomized single-qubit Pauli measurements,
where $U_t\sim\mathrm{Cl}(2)^{\otimes N}$,
the snapshot takes the explicit form
\begin{equation}\label{append:SM-A-CS-form}
\hat{\rho}_t
=\bigotimes_{j=1}^N
\left(
3U_{j,t}^\dagger|b_{j,t}\rangle\langle b_{j,t}|U_{j,t}
-\mathbb{I}_2
\right). \nonumber
\end{equation}
Averaging $T$ independent snapshots gives
$\hat{\rho}_T=T^{-1}\sum_{t=1}^T\hat{\rho}_t$,
which provides an unbiased estimator of $\rho$.
The same snapshots can then estimate multiple observables
$\{O_i\}_{i=1}^M$ through
$\hat{f}_{O_i}(\rho)=T^{-1}\sum_{t=1}^T
\Tr(O_i\hat{\rho}_t)$.

Subsequent advances, including derandomized measurements~\cite{huang2021derandomization} and measurement grouping~\cite{gresch2025shadowgrouping}, further reduce measurement costs by exploiting the structure of target observables.

\smallskip

\noindent \emph{Classical surrogates.} Unlike classical shadows for a fixed quantum state, classical surrogates aim to learn how a linear property $f_O(\rho(\boldsymbol{x}))=\Tr(\rho(\boldsymbol{x})O)$ varies across a parametrized family of quantum states $\rho(\boldsymbol{x})$, where $\boldsymbol{x}\in\mathbb{R}^d$ specifies circuit parameters or Hamiltonian couplings. The goal of a classical surrogate is to learn a predictive model $h(\bx)$ that approximates the observable expectation value $f_O(\bx)$ across a parametrized family of quantum states, with prediction error $\mathbb{E}_{\bx\sim\mathbb{D}}|f_O(\bx)-h(\bx)|^2\leq\epsilon$.

A common machine-learning-based approach represents the target function using a prescribed family of basis functions $\{\Phi_{\bomega}\}_{\bomega\in\Lambda}$. The construction involves two steps. First, a training dataset $\mathcal{T}=\{(\bx^{(i)},y^{(i)})\}_{i=1}^n$ is collected, where each label $y^{(i)}$ estimates $f_O(\bx^{(i)})$ through quantum measurements. Second, a subset of basis functions $\Lambda'\subseteq\Lambda$ is selected to construct the surrogate
\begin{equation}
h(\bx)=\sum_{\bomega\in\Lambda'}\hat{w}_{\bomega}\Phi_{\bomega}(\bx), \nonumber
\end{equation}
where the coefficients $\hat{w}_{\bomega}$ are learned from $\mathcal{T}$ through regression. Once trained, the surrogate predicts observable expectation values for unseen inputs using only classical computation, amortizing the quantum measurement cost across subsequent predictions. Under suitable structural assumptions, classical surrogates admit provable guarantees on sample complexity and prediction accuracy, potentially reducing the computational cost of repeated quantum evaluations~\cite{du2025efficient,liao2026demonstration,wang2026learnable}.

Beyond these structured machine-learning approaches, recent studies have explored deep learning and large language models to develop more flexible surrogates for both qubit-based and continuous-variable quantum systems~\cite{zhao2025rethink,gao2026classifying}.

\medskip
\noindent\textbf{Nonlinear properties of quantum states.}
Beyond linear properties, many physically relevant quantities depend nonlinearly on the quantum state \(\rho\). A representative class consists of degree-\(k\) polynomial functions of \(\rho\), such as
\begin{equation}
    f_O(\rho^k)=\Tr\left[ O \rho^{k}\right], \nonumber
\end{equation}
where $O$ is a fixed observable and $k$ is a positive integer. Such nonlinear functions arise naturally in the characterization of mixed states, with important examples including purity~\cite{elben2019statistical} and R\'enyi entropies~\cite{brydges2019probing}.

A key approach to estimating nonlinear properties is to reformulate them as linear-property estimation problems on multiple copies of the quantum state. Specifically, a homogeneous polynomial of degree $k$ in the matrix elements of $\rho$ can be represented as a linear functional of the replicated state $\rho^{\otimes k}$. This gives $
f_O(\rho^k)=\Tr\left[\mathcal{A}_k(O)\rho^{\otimes k}\right]$,
where $\mathcal{A}_k(O)$ is an operator acting on $k$ copies of the original Hilbert space. For example, the purity of a quantum state is defined as $\Tr(\rho^2)$, which can be expressed as  $\Tr(\rho^2)=\Tr(\mathbb{S}\rho^{\otimes 2})$, where $O=\mathbb{I}$ and $\mathcal{A}_2(O)$ is the swap operator $\mathbb{S}$.
By writing $f_O(\rho^k)$ as a linear property in terms of $\rho^{\otimes k}$, it could be characterized using classical shadow methods as
\begin{equation}
    \hat{f}_O(\rho^k)=\frac{(T-k)!}{T!}\sum_{\substack{t_1\ne t_2\cdots\ne t_k\in [T]}}
\Tr\left[\mathcal{A}(O)\hat\rho_{t_1}\otimes\cdots\otimes\hat\rho_{t_k}\right]. \nonumber
\end{equation}
where $\{\hat{\rho}_{t_i}\}_{i=1}^T$ refers to $T$ independent snapshots of $\rho$, and $\mathbb{E}[\hat{f}_O(\rho^k)]=f_O(\rho^k)$.  Beyond this basic approach, subsequent developments have introduced more sophisticated measurement protocols to improve the statistical efficiency of estimating $f_O(\rho^k)$ and reduce the measurement resources required for higher-order quantities~\cite{liu2026replica}.

\medskip
\noindent\textbf{Hamiltonian Learning.}
Hamiltonian learning aims to identify the interactions governing a quantum system from experimental observations, such as measurements of its time evolution~\cite{wang2017experimental}. Consider an $N$-qubit Hamiltonian $H=\sum_{l=1}^{L}h_l P_l$, where $P_l\in\{\mathbb{I},X,Y,Z\}^{\otimes N}\setminus\{\mathbb{I}^{\otimes N}\}$ are non-identity Pauli strings and $\bm{h}=(h_1,\ldots,h_L)\in\mathbb{R}^L$ denotes the unknown interaction coefficients. When the candidate Pauli terms $\{P_l\}_{l=1}^L$ are known, Hamiltonian learning reduces to estimating $\bm{h}$ from measurement data.

A common approach to Hamiltonian learning involves two stages. First, collecting time-dependent measurements for a set of initial-state and observable pairs $\{(\rho_j,O_j)\}_{j=1}^M$. Second, inferring the Hamiltonian coefficients $\bm{h}$ from these measurements. For each pair, the initial state evolves as $\rho_j(t)=e^{-\imath Ht}\rho_j e^{\imath Ht}$, and measurements of $O_j$ yield estimates of $f_{O_j}(\rho_j(t))=\Tr[O_j\rho_j(t)]$. The key insight is that the initial time derivative of $f_{O_j}(\rho_j(t))$ depends linearly on the unknown coefficients $\bm{h}$, enabling Hamiltonian learning through linear regression. Mathematically, we have
\begin{equation}
\bm{d}_j:=\left.\frac{\mathrm{d} f_{O_j}(\rho_j(t))}{dt}\right|_{t=0}
=\imath\cdot  \Tr(\rho_j[H,O_j])
=\sum_{l=1}^{L}\bm{K}_{jl}\bm{h}_l, \nonumber
\end{equation}
where $\bm{K}_{jl}=\imath \cdot \Tr(\rho_j[P_l,O_j])$ defines the entries of a matrix $\bm{K}\in\mathbb{R}^{M\times L}$ determined by the known input states, observables, and candidate Pauli strings. This relation reduces Hamiltonian learning to a linear inverse problem~\cite{gu2024hamiltonian}. Specifically, by measuring $f_{O_j}(\rho_j(t))$ at sufficiently short evolution times and fitting its time dependence, one can estimate the initial derivatives $\hat{\boldsymbol d}=(\hat{\bm{d}}_1,\ldots,\hat{\bm{d}}_M)$, which approximately satisfy $\hat{\bm{d}}\approx \bm{d}=\bm{K}\bm{h}$.
The Hamiltonian coefficients can then be inferred by solving a regression problem, such as least-squares estimation, i.e.,
\begin{equation}
\hat{\bm h}=\arg\min_{\bm h\in\mathbb{R}^{L}}
\|\bm{K} \bm{h}-\hat{\bm d}\|_2^2. \nonumber
\end{equation}

\medskip
\noindent\textbf{Quantum error mitigation.}
The objective of quantum error mitigation (QEM) is to estimate ideal observable expectation values $f_O(\rho)$ in Eq.~\eqref{SI:A:eq-1} using measurement data collected from noisy quantum devices \cite{cai2023quantum}. In practice, hardware noise leads to a noisy implementation $\tilde{\mathcal U}$, yielding $f_O(\tilde{\rho})=\Tr[O\tilde{\mathcal U}(\rho_0)]$, which generally deviates from $f_O(\rho)$. QEM seeks to reduce this discrepancy through suitably designed noisy experiments and classical post-processing.

A representative QEM approach is zero-noise extrapolation (ZNE)~\cite{giurgica2020digital}, which estimates $f_O(\rho)$ by measuring a quantum circuit at multiple noise levels and extrapolating to the zero-noise limit. Let $\lambda\geq 0$ denote the effective noise strength and $f_O(\tilde{\rho}_{\lambda})=\Tr[O\tilde{\rho}_{\lambda}]$, where $\tilde{\rho}_{\lambda}=\tilde{\mathcal U}_{\lambda}(\rho_0)$ and $\tilde{\mathcal U}_0=\mathcal U$ corresponds to the ideal circuit. ZNE first collects estimates $\{\hat f_O(\tilde{\rho}_{\lambda_j})\}_{j=1}^{M}$ at different noise levels, typically generated using techniques such as gate folding or pulse stretching. It then extrapolates these estimates toward $\lambda=0$. In particular, Richardson extrapolation constructs the estimator as
\begin{equation}
\hat f_O^{\mathrm{ZNE}}(\tilde{\rho}_{0})
=\sum_{j=1}^{M}\bm{c}_j \cdot \hat f_O(\tilde{\rho}_{\lambda_j}), \nonumber
\end{equation}
where the extrapolation coefficients $\{\bm{c}_j\}_{j=1}^{M}$ satisfy $
\sum_{j=1}^{M}\bm{c}_j=1$ and
$\sum_{j=1}^{M}\bm{c}_j\lambda_j^k=0$ for $k\in [M-1]$.
These constraints cancel the leading $M-1$ orders of noise contributions, reducing the extrapolation bias at the potential cost of amplified statistical errors. Beyond ZNE, various other QEM approaches have been developed, including probabilistic error cancellation~\cite{temme2017mitigation}, symmetry verification~\cite{Bonet2018}, and virtual distillation~\cite{Huggins2021}, each employing different strategies to suppress or compensate for the effects of quantum noise.

\medskip
\noindent\textbf{Quantum process characterization.}
This task aims to infer the properties of an unknown quantum process from measurements of its input-output behavior. Mathematically, a quantum process is described by a completely positive and trace-preserving (CPTP) map $\mathcal{E}$, which transforms an input quantum state $\rho$ into an output state $\mathcal{E}(\rho)$. Depending on the characterization objective, one may seek to reconstruct the entire quantum channel or predict selected properties of its output states.

\smallskip
\noindent\emph{Quantum process tomography.}
Quantum process tomography aims to reconstruct an unknown quantum channel $\mathcal{E}$ from measurements of its outputs for a set of known input states. A convenient representation of the channel is its Choi matrix, defined as
\begin{equation}
J_{\mathcal E}
=\sum_{a,b=1}^{d}|a\rangle\langle b|\otimes\mathcal E(|a\rangle\langle b|), \nonumber
\end{equation}
where $d$ denotes the dimension of the input Hilbert space. For a CPTP map, the Choi matrix satisfies the physical constraints $J_{\mathcal E}\succeq \bm{0}$ and $\Tr_{\mathrm{out}}(J_{\mathcal E})=\mathbb{I}_d$. These properties allow quantum process tomography to be formulated as a constrained estimation problem. Specifically, for a known input state $\rho_s$ and an output measurement effect $M_b$, the probability of obtaining measurement outcome $b$ is given by $
\mathbb{P}(b|s)
=\Tr[M_b\mathcal E(\rho_s)]
=\Tr[(\rho_s^{\top}\otimes M_b)J_{\mathcal E}].$ By preparing different input states and performing measurements on their corresponding outputs, one obtains a collection of measurement statistics that constrain the unknown Choi matrix. The channel can then be reconstructed by fitting a physical Choi matrix to the observed data, for example through maximum-likelihood estimation~\cite{torlai2023process}.

\smallskip
\noindent\emph{Prediction of output properties.}
Different from full process tomography, output-property prediction focuses on estimating selected observable expectation values for previously unseen input states. In particular, for an unknown quantum channel $\mathcal E$ and a target observable $O$, the prediction task is to learn the function
\begin{equation}
f_{\mathcal E,O}(\rho)
=\Tr[O\mathcal E(\rho)]
=\Tr[\mathcal E^\dagger(O)\rho], \nonumber
\end{equation}
where $\mathcal E^\dagger$ denotes the adjoint channel. This relation indicates that predicting the expectation value of $O$ at the output is equivalent to estimating the expectation value of the effective observable $\mathcal E^\dagger(O)$ at the input. Therefore, instead of reconstructing the entire channel, one can construct a classical predictive model by learning an approximation of this effective observable.

A representative approach collects training data consisting of known input states and estimates of the corresponding output expectation values~\cite{huang2023processes}. The effective observable can then be approximated using a selected operator basis $\{P_j\}$ as $\hat O_{\mathrm{eff}}=\sum_j \bm{w}_j P_j$, where the coefficients $\{\bm{w}_j\}$ are learned from the training data. For a new input state $\rho$, the resulting predictor takes the form
\begin{equation}
\hat f_{\mathcal E,O}(\rho)
=\Tr(\hat O_{\mathrm{eff}}\rho)
=\sum_j \bm{w}_j\Tr(P_j\rho). \nonumber
\end{equation}
Once the coefficients are learned, the predictor can estimate the target observable expectation value without additional access to the unknown channel, provided that the required input-state features are available.

\subsection{AI agents and harnesses}\label{append:sec:agents}

Large language models (LLMs) have evolved from conversational assistants into autonomous agents capable of executing complex tasks, ranging from software engineering to scientific research. Here, we introduce three foundations of AI scientists relevant to our work, including LLM agents for software engineering, scientific-reproduction agents, and automated harness design and agent self-improvement.

\medskip
\noindent\textbf{LLM agents for software engineering.}
Unlike conventional LLMs that generate responses to individual prompts, LLM agents can execute complex tasks through iterative reasoning and tool interactions. Software engineering has become a prominent application and evaluation setting for such agents. Specifically, an LLM agent operates within a software framework, known as a \emph{harness}, that manages interactions between the LLM and its environment. At each step, the harness provides the model with task instructions and execution history, receives a proposed action, executes it through available tools, and returns the resulting feedback. This process continues until the task is completed or a predefined resource limit is reached, with inference costs determined by the tokens consumed across model calls.

A representative benchmark for LLM agents in software engineering is SWE-bench~\cite{jimenez2024swebench}, which evaluates agents on resolving real-world software issues, with correctness assessed through repository tests. Building on this setting, SWE-agent~\cite{yang2024sweagent} demonstrated that specialized tools for code navigation, search, and editing can substantially improve agent performance without changing the underlying LLM. These findings underpin the importance of harness design alongside model capabilities. 

Subsequent research has explored different design philosophies, ranging from minimal harnesses built around a single shell tool to adaptive agents that construct their own tools during execution~\cite{minisweagent2025,xia2025livesweagent}. In practice, harnesses can be developed by model providers for their own models, such as DeepSeek Harness (DSH), or by third parties to support different models through application programming interfaces (APIs).

\smallskip
\noindent\textbf{Scientific-reproduction agents.}
Beyond software engineering, LLM agents are increasingly being developed as AI scientists to automate scientific research. One important application is scientific reproduction, which aims to reconstruct computational procedures and reproduce results reported in published papers, thereby facilitating the verification and reuse of scientific findings. Previous research has advanced along two complementary directions, namely developing benchmarks to evaluate reproduction capabilities and designing agents to automate reproduction workflows.

Existing benchmarks differ in their task objectives and the amount of prior information provided to agents. Typically, benchmark development involves defining reproduction objectives, constructing tasks and reference materials, and establishing evaluation criteria to assess the performance of different agents. For example, PaperBench~\cite{starace2025paperbench} evaluates the reproduction of machine-learning research from published papers, supplemented by author-provided clarifications, using expert-developed grading rubrics. PRBench~\cite{qiu2026prbench} extends paper-based reproduction to physics, providing expert-curated task instructions that specify target results and implementation requirements. In contrast, CORE-Bench~\cite{siegel2024corebench} evaluates agents on executing and interpreting released research code and data, while AutoExperiment~\cite{kim2026reproduction} examines their ability to reconstruct missing components of existing implementations. Other benchmarks investigate computational reproduction in materials science and the social sciences, using domain-specific research materials and evaluation protocols~\cite{huang2026automat,alizadeh2026reproduction}.

Complementing these benchmarks, research on reproduction agents focuses on developing computational workflows for recovering published results. Depending on the available research materials and task specifications, agents may be required to implement methods from scientific papers, execute existing research code, or reconstruct incomplete implementations. A common design is to coordinate multiple stages of scientific reproduction, including interpreting research materials, planning computational procedures, implementing or adapting code, executing experiments, and validating the resulting outputs. For example, DeepCode~\cite{li2025deepcode} proposes an agentic framework for generating executable code from research papers and technical specifications, whereas InternAgent~\cite{internagent2025} develops an agent-based workflow intended to support scientific understanding, implementation, and experimental execution. Related efforts, such as Paper2Agent~\cite{miao2026paper2agent}, transform scientific papers and associated research artifacts into interactive agents that expose reusable computational capabilities.

\smallskip
\noindent\textbf{Automated harness design and agent self-improvement.}
The performance of LLM agents depends not only on the underlying language model but also on how the agent is organized and how it learns from previous executions. This has motivated research on two complementary approaches to automatically improving agent capabilities: optimizing the harness that governs agent execution and accumulating reusable experience from completed tasks.

The first approach, automated harness design, aims to improve agent performance by optimizing its prompts, tools, and control flow. Existing methods employ LLM-based search to generate candidate harnesses, evaluate their performance on benchmark tasks, and iteratively refine their designs~\cite{hu2024adas,wang2025hgm,lee2026metaharness}. The resulting harness can then be deployed to execute subsequent tasks. This approach primarily improves the agent's execution framework, with the optimized design typically remaining fixed during deployment.

The second approach focuses on improving agents through experience accumulated during task execution. Instead of modifying model parameters or harness code, experience-based methods extract useful information from previous runs, such as reflections on failures, successful problem-solving strategies, reusable workflows, and optimized instructions~\cite{shinn2023reflexion,zhao2024expel,wang2024awm,agrawal2025gepa}. This information is stored as external memory and retrieved to guide subsequent executions, enabling agents to adapt their behavior without retraining the underlying model. In this work, we adopt this experience-based perspective of self-improvement, emphasizing the accumulation and reuse of validated knowledge across scientific reproduction tasks.

\subsection{Literature review}\label{SI:subsec:literature_review}

AI scientists are increasingly being explored to automate scientific reasoning, computational implementation, and validation. Three research directions are particularly relevant to our work: autonomous quantum algorithm design, autonomous investigation of quantum systems, and quantum research reproduction. These directions pursue distinct scientific objectives: designing improved computational methods, discovering or explaining physical phenomena, and independently verifying published findings, respectively. Below, we review representative advances in each direction and highlight their differences from our study.

\smallskip
\noindent\textbf{Autonomous quantum computing design.}  Autonomous quantum computing design aims to discover and optimize quantum algorithms, circuit architectures, and error-correcting codes under computational and hardware constraints. A central challenge is the vast design space of quantum circuits, programs, and error-correcting codes. Existing approaches leverage AI models to guide candidate generation with such domain-specific reasoning, while computational evaluation provides feedback for iterative refinement.

For algorithm construction, LLMs have been trained to generate programs representing families of quantum circuits~\cite{arlt2026metadesign}, while iterative search has been applied to quantum machine learning architectures~\cite{knipfer2026aiagents,sakka2026llmvqc}, quantum state preparation~\cite{cao2025llmevolution}, and molecular ground-state energy estimation~\cite{finger2026algorithmdiscovery}. Quantum resource optimization instead improves existing computational procedures by improving Hamiltonian simulation accuracy~\cite{zhang2025quantum} and reducing gate and qubit counts in quantum arithmetic subroutines~\cite{gu2026discovering}.
In quantum error correction, AI-guided search has been used to discover quantum error-correcting codes~\cite{yan2026omniqec,cruzbenito2026bbcodes,liu2026sce}.

\smallskip
\noindent\textbf{Autonomous investigation of quantum systems.} 
Autonomous investigation of quantum systems aims to advance scientific understanding through two complementary tasks: predicting physical behaviour from specified models and inferring explanations for observed phenomena. For the former, AI agents integrate established numerical methods with computational validation to investigate quantum phase transitions~\cite{miao2025physmaster} and dynamical evolution~\cite{li2026autonomous}. For the latter, agents propose candidate physical models and computationally assess their consistency with observations. This capability has been explored through controlled rediscovery of known physical models~\cite{naegele2026sciexplorer} and applied to interpreting experimental optical measurements in quantum materials~\cite{zhou2026aitheorist}, where human scientists guided the investigation and collected measurements, while AI agents developed and tested candidate explanations.

\smallskip
\noindent\textbf{Benchmark for quantum science.} As AI scientists become increasingly capable of conducting computational research, systematic benchmarks are needed to assess their scientific reasoning, implementation, and validation capabilities. Existing benchmarks evaluate AI agents on tailored research questions derived from scientific papers~\cite{miao2026prlbench}, simulated quantum engineering tasks~\cite{guo2026verifiedautonomy}, and the reproduction of computational results reported in physics~\cite{qiu2026prbench} and materials science~\cite{huang2026automat}. Automated scientific checks have also been introduced to assess computational correctness and provide feedback for code refinement~\cite{deng2026physvec}. Quantum programming benchmarks further assess whether LLMs can translate task descriptions into functionally correct programs across different software packages~\cite{vishwakarma2024qiskithumaneval,guo2025quanbench,yang2025qcircuitbench,slim2026quanbenchplus}.
These efforts establish structured evaluation environments for measuring the capabilities of AI scientists.

\medskip 

Our work differs from the three prior research directions by advancing autonomous paper-to-result reproduction in computational quantum science. 
Although PRBench~\cite{qiu2026prbench} benchmarks end-to-end paper reproduction across broader areas of physics, its reliance on expert-curated task instructions specifying implementation requirements and expected outputs limits its ability to assess whether AI scientists can independently reconstruct research procedures from original papers.
Our objective is to develop an AI scientist capable of reproducing published results directly from scientific papers and target figures with minimal human intervention. Importantly, our investigation reveals that ambiguities and even errors in published research can substantially hinder autonomous reproduction. Beyond benchmarking, the proposed \RA systematically investigates the causes of reproduction failures and distils validated corrections into reusable knowledge, enabling continual improvement across research tasks.

\section{Implementation details of \textsf{OmniQ}}\label{SI:secB-omniQ}

This section provides additional implementation details of \omniQ and an analysis of the challenges encountered during scientific reproduction. In SI~\ref{append:sec:curation-workflow}, we first describe the workflow for selecting and organizing research papers. Next, in SI~\ref{append:sec:cons-dataset}, we present the construction of the computational quantum science dataset for evaluating paper-to-result reproduction. Finally, in SI~\ref{append:sec:paper-levels}, we examine representative reproduction issues identified during our experiments.

\subsection{Workflow of \omniQ for selecting and organizing research papers}\label{append:sec:curation-workflow}

\begin{figure}[!htbp]
\centering
\includegraphics[width=\textwidth]{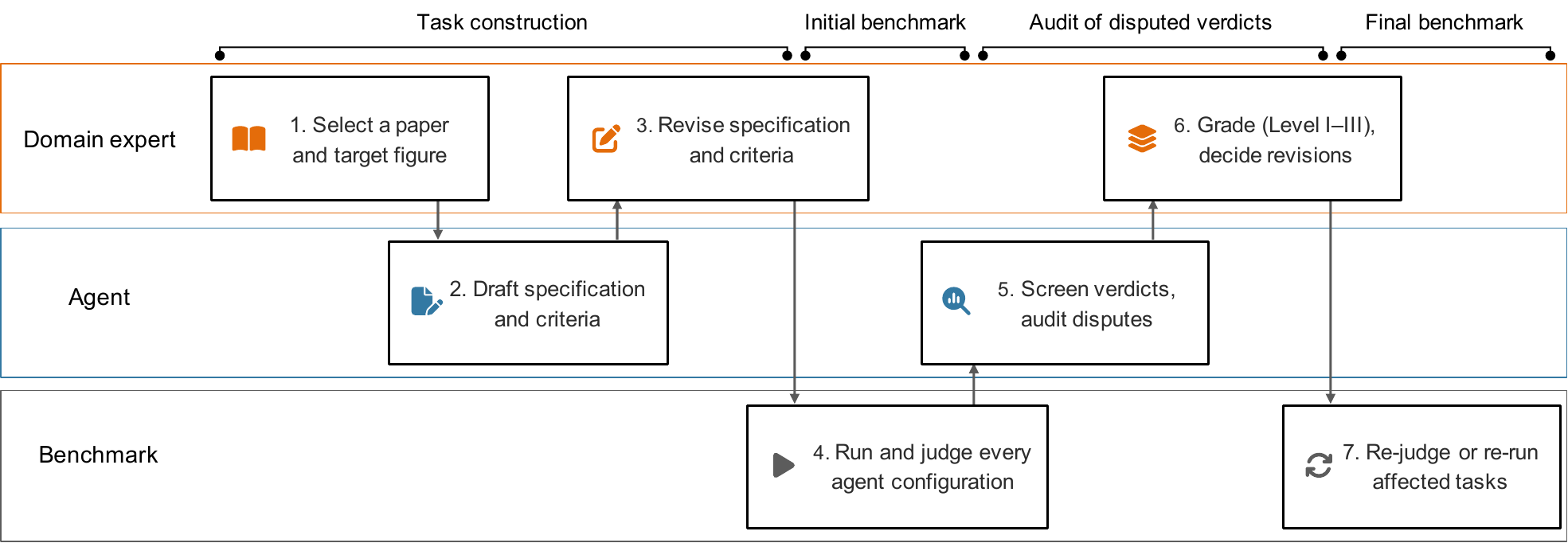}
\caption{\small\textbf{Workflow of \omniQ.} The workflow consists of seven steps involving domain experts, agents, and benchmark evaluation, distinguished by different colors. Arrows indicate the interactions and progression between steps.}
\label{fig:curation-workflow}
\end{figure}

To ensure the reliability and consistency of the reproduction benchmark, \omniQ employs a seven-step workflow that integrates multi-agent cross-validation with expert verification, as illustrated in Fig.~\ref{fig:curation-workflow}. The workflow begins with task construction in Steps 1--3, where a domain expert selects a research paper and its target figure, an LLM agent drafts the task specification and evaluation criteria, and the expert subsequently reviews and refines both. Next, Step 4 involves initial benchmarking, in which multiple agent configurations independently execute each reproduction task, with each configuration evaluated once. In Step 5, their evaluation results are cross-checked by an auditing agent to identify potentially disputed judgments and assess their consistency with the original paper, task specification, and evaluation criteria. Based on this audit, domain experts review the flagged papers in Step 6, classify confirmed reproduction issues into three levels (i.e., Levels I--III, as defined in the main text) according to their severity, and determine necessary revisions to the task specification or evaluation criteria. Finally, in Step 7, the affected tasks are re-evaluated or re-executed using the revised specifications and criteria, yielding a more reliable benchmark for assessing scientific reproduction capabilities.

Beyond the overall workflow, three key mechanisms underpin the construction and reliability of the \omniQ benchmark, i.e., task specification and evaluation criteria, automated judging, and cross-agent auditing with expert verification. We detail these mechanisms below.

\smallskip
\noindent\textbf{Task specification and evaluation criteria.}
Steps 1--3 of \omniQ establish each reproduction task through agent-assisted construction and expert review. As illustrated in Fig.~1, each task comprises two Markdown documents with distinct purposes: a preliminary document that guides the AI scientist and an evaluation criteria document reserved for the automated judge. This separation allows the reproduction objectives to be specified explicitly while keeping the expected results inaccessible to the AI scientist.

The preliminary document specifies the target figure and provides both general guidelines and paper-specific instructions. The general guidelines, shared across all tasks, require the AI scientist to follow the methods and settings reported in the original paper, prohibit fabricated or externally obtained results, and require all approximations to be documented. The paper-specific instructions define the target problem, experimental scope, and relevant conventions, while leaving the reconstruction of computational procedures and numerical results to the AI scientist.

The evaluation criteria document defines the requirements for assessing reproduction correctness. These include the structural properties of the target figure, quantitative checkpoints with reference values and tolerances, and verification of the experimental parameters and essential computational procedures. Both documents are initially drafted by an LLM agent and subsequently reviewed by domain experts to ensure consistency with the original paper. In particular, expert review addresses potentially misleading criteria, such as purely visual assessments, numerical tolerances that are overly restrictive relative to computational variability, or reference values incorrectly extracted from published figures. Importantly, the evaluation criteria are made available to the automated judge only after the reproduction is completed, ensuring that the AI scientist cannot access the reference values during execution.

\smallskip
\noindent\textbf{Automated judging.} To assess scientific reproduction beyond visual similarity, Step 4 of \omniQ employs an automated judge to verify both the reproduced results and their underlying computational procedures. Implemented using Claude Code with Opus~4.8, the judge examines the complete execution workspace, including source code, execution logs, numerical data, and experimental reports, against the predefined evaluation criteria. This evidence-based assessment examines not only whether the reproduced figure matches the published results but also whether those results are supported by the prescribed computational methods and experimental settings.

The judge classifies identified issues into two categories according to their impact on scientific correctness. In particular, critical issues include quantitative results outside the specified tolerances, deviations from essential methodological steps, inconsistent experimental parameters, and numerical values that cannot be traced to the executed computations. Every quantitative checkpoint is treated as a mandatory requirement, and documenting an approximation does not exempt the reproduction from satisfying the corresponding criteria. In contrast, non-critical issues are limited to visual presentation and do not affect scientific validity.

A reproduction is considered successful only when no critical issues are identified. For tasks involving multiple target figures, each figure is evaluated independently, and the task is considered successfully reproduced only if all target figures satisfy their respective criteria. This protocol prioritizes the fidelity of scientific results and computational procedures over superficial agreement with published figures.

\smallskip
\noindent\textbf{Cross-agent verification and expert review.}
To further establish the reliability of scientific reproduction assessments, Steps 5--7 of \omniQ introduce a cross-agent verification mechanism to systematically examine potential inconsistencies in the original papers, task specifications, and evaluation criteria. Specifically, this mechanism leverages evaluation results from multiple independent agent configurations to identify potential issues in the benchmark itself.

Across the 60 selected research papers, a verification agent examines 3,376 judging outcomes collected from multiple agent configurations and target figures, focusing on systematic failures where different configurations, including the strongest-performing ones, fail to satisfy the same evaluation criteria. This process identifies 19 papers for further investigation. For each flagged paper, the verification agent systematically compares the original publication, task specification, evaluation criteria, and computational results across configurations. Where possible, it also independently checks key numerical values using analytical calculations, authors' released code, or data extracted from published figures.

Domain experts subsequently review these findings, classify identified issues into three levels according to their severity, and determine whether revisions to the task specifications or evaluation criteria are warranted. Expert review confirms issues in 18 of the 19 flagged papers. Depending on their underlying causes, revisions may align task specifications with settings recovered from the authors' released code or data, restrict computational requirements to feasible resource budgets, or adjust evaluation checkpoints that depend on unstated or idealized conventions. Note that tasks whose evaluation outcomes are unaffected by the identified issues are retained without substantive changes. Finally, affected tasks are re-evaluated when the criteria are revised or re-executed when the task specifications are modified.

\subsection{Construction of computational quantum science dataset}\label{append:sec:cons-dataset}

To systematically evaluate the capabilities of AI scientists in computational quantum science, we construct a dataset comprising 60 research papers and their corresponding figure-level reproduction tasks. The selected papers span three major research areas with relatively balanced representation: classical simulation of quantum systems (20 papers), quantum algorithms (19 papers), and quantum property characterization (21 papers). This selection captures diverse computational paradigms and methodological challenges in computational quantum science. We prioritize influential studies published in leading journals while requiring that their numerical results be reproducible within a computational budget of approximately one day per task. This constraint largely excludes computationally intensive deep learning-based approaches. An overview of the dataset is provided in Table~\ref{tab:quantum-papers}.

For each paper, we select representative figures that capture its central scientific findings or methodological contributions. Reproducing these figures requires translating scientific descriptions into executable workflows, implementing the underlying methods, and validating the numerical results within \omniQ. By requiring AI scientists to reconstruct computational results from the underlying scientific principles, this design provides a systematic assessment of their scientific reasoning and implementation capabilities. In the following, we describe the dataset composition and representative reproduction tasks across the three research areas.

\medskip
\noindent \textbf{Classical simulation of quantum systems.} To evaluate the agent's ability to reproduce research-level results in classical quantum simulation, we select papers based on two complementary criteria. The first prioritizes state-of-the-art numerical methods that introduce novel algorithmic ideas, require technically demanding implementations, or play an important role in assessing claims of quantum utility. The second emphasizes physics-informed methods, where successful reproduction requires understanding the underlying quantum systems and how their physical properties, such as symmetries, locality, and entanglement structure, are exploited to enable efficient numerical simulation. These criteria enable us to systematically assess the AI scientist's ability to understand and faithfully reproduce sophisticated classical simulation methods.

The selected papers provide balanced coverage of representative classical quantum simulation methods discussed in SI~\ref{SI:subsec:classical_simulator}, spanning tensor networks, neural quantum states, Clifford simulation and Pauli propagation, free-fermion simulation, quantum Monte Carlo, and open-system simulation. More specifically, tensor-network methods cover diverse representations and computational tasks, ranging from quantum many-body dynamics and open-system evolution to Feynman-diagram calculations~\cite{tindall2024eagle,li2026matrix,tkcp-p5br,mckeever2021stable}. Neural quantum states explore learning-based representations for simulating quantum systems and characterizing their physical properties~\cite{kufel2025approximately,PhysRevLett.128.090501,mauron2025predicting}. Clifford simulation and Pauli propagation encompass complementary strategies for reducing the computational cost of quantum-circuit and many-body simulations~\cite{Bravyi_2019,rall2019simulation,angrisani2025classically,beguvsic2024fast,qian2025clifford}. Free-fermion methods leverage fermionic structure to investigate measurement-induced dynamics and non-stabilizerness~\cite{poboiko2023theory,collura2026non}. Quantum Monte Carlo covers stochastic methods for entanglement characterization and noisy quantum-circuit simulation~\cite{shen2026realtime,wang2024bipartite}, while open-system simulation addresses dissipative dynamics, finite-temperature effects, and nonequilibrium steady states~\cite{sander2025large,tamascelli2019efficient,gangat2017steady,pocklington2025efficient}.

\medskip
\noindent \textbf{Quantum algorithms.}
Our selection in this category covers both advances in quantum algorithm design and methods for reducing their quantum resource requirements. Together, these studies help assess the practical utility of quantum computing on both near-term and fault-tolerant devices. In this context, AI scientists capable of independently reproducing these results could accelerate the discovery of novel quantum algorithms and advance their practical applications.

The selected papers cover the five categories of quantum algorithms introduced in SI~\ref{SI:subsec:quantum_algorithm}.
Concretely, non-variational quantum simulation includes product formulas for Hamiltonian evolution, along with adiabatic evolution, imaginary-time evolution and spectral filtering for ground-state preparation~\cite{bosse2025efficient,bagherimehrab2026faster,lu2026digital,motta2020determining,karacan2025enhancing}.
Quantum phase estimation papers study the circuit depth and query cost required to estimate phases and amplitudes to a given precision~\cite{ni2023low,giurgica2022low,ramoa2025bayesian}.
For variational quantum simulation, the papers cover ground-state preparation, combinatorial optimization and quantum dynamics, including adaptive ansatz construction, tensor-network initialization and circuit knitting~\cite{grimsley2019adaptvqe,rudolph2023synergistic,farhi2022quantum,gentinetta2024overhead}.
Work on quantum machine learning examines the trainability and predictive performance of quantum neural networks and kernel methods, including comparisons with classical models and the effects of data and measurement resources~\cite{larocca2023theory,cerezo2021cost,holmes2021barren,schreiber2023surrogates,huang2021power,thanasilp2024exponential}.
The quantum-enhanced arkov chain Monte Carlo study compares Markov chain Monte Carlo convergence under quantum and classical proposal methods~\cite{layden2023quantum}.

\medskip
\noindent\textbf{Property characterization of quantum systems.} The selection of papers on quantum property characterization is guided by two complementary criteria. The first emphasizes measurement-efficient protocols for characterizing individual quantum systems, enabling the estimation of important physical properties with favorable scaling in measurement cost under suitable assumptions. The second focuses on learning unknown quantum systems and processes, including methods that extract predictive models from experimental data and generalize across different states, dynamics, or system configurations. These capabilities are essential for advancing quantum hardware development, evaluating quantum algorithms, and deepening our understanding of quantum physics. Accordingly, reproducing these results provides a meaningful test of the AI scientist's ability to implement and validate computational methods with direct relevance to quantum research.   

The selected papers provide balanced coverage of five representative research topics discussed in SI~\ref{SI:subsec:property_characterization}. In particular, linear-property estimation spans classical-shadow protocols, measurement optimization, and learning-based prediction of quantum observables, covering both theoretical foundations and practical estimation strategies~\cite{huang2020predicting,bertoni2024shallow,gresch2025shadowgrouping,veltheim2025multisets,tran2023analog,huang2021information,du2025efficient,lewis2024ground,huerta2022inference}. Nonlinear-property estimation focuses on measurement-efficient characterization of entanglement and related quantities using randomized measurements, replica-based estimators, and permutation moments~\cite{liu2026replica,rath2021importance,liu2022permutation,vermersch2024entropies}. Hamiltonian learning addresses the inference of unknown Hamiltonian parameters from quantum dynamics or thermal-state information~\cite{gu2024hamiltonian,hu2025ansatzfree}. Quantum error mitigation investigates computational strategies for recovering ideal observables from noisy quantum measurements~\cite{liao2024machine,jnane2024mitigated}. Quantum process characterization encompasses the estimation of gate-set properties, identification of conservation laws, and prediction of quantum-process observables~\cite{helsen2023gateset,zhan2024conservation,huang2023processes}.

\bigskip
\begingroup
\setlength{\LTcapwidth}{\textwidth}
\setlength{\tabcolsep}{2pt}
\renewcommand{\arraystretch}{1}
\setlength{\LTleft}{0pt}
\setlength{\LTright}{0pt}
\setlength{\LTpre}{0pt}
\setlength{\LTpost}{0pt}

\newcommand{\PaperIDWidth}{0.055\textwidth}
\newcommand{\PaperSubWidth}{0.135\textwidth}
\newcommand{\PaperTitleWidth}{0.455\textwidth}
\newcommand{\PaperJournalWidth}{0.155\textwidth}
\newcommand{\PaperFigWidth}{0.065\textwidth}
\newcommand{\PaperCodeWidth}{0.075\textwidth}

\newcommand{\PaperRow}[7]{%
  \parbox[c]{\PaperSubWidth}{\scriptsize\raggedright\strut #2\strut} &
  \parbox[c]{\PaperIDWidth}{\scriptsize\raggedright\strut \##1\strut} &
  \parbox[c]{\PaperTitleWidth}{\scriptsize\raggedright\strut #3~\cite{#4}\strut} &
  \parbox[c]{\PaperJournalWidth}{\scriptsize\raggedright\strut\textit{#5}\strut} &
  \parbox[c]{\PaperFigWidth}{\scriptsize\raggedright\strut #6\strut} &
  \parbox[c]{\PaperCodeWidth}{\scriptsize\raggedright\strut #7\strut} \\[2pt]
}

% Column headings: original table font size
\newcommand{\PaperColumnHeader}{%
  \parbox[t]{\PaperSubWidth}{\scriptsize\raggedright\strut\textbf{Subcategory}\strut} &
  \parbox[t]{\PaperIDWidth}{\scriptsize\raggedright\strut\textbf{Paper ID}\strut} &
  \parbox[t]{\PaperTitleWidth}{\scriptsize\raggedright\strut\textbf{Paper title}\strut} &
  \parbox[t]{\PaperJournalWidth}{\scriptsize\raggedright\strut\textbf{Journal}\strut} &
  \parbox[t]{\PaperFigWidth}{\scriptsize\raggedright\strut\textbf{Fig.}\strut} &
  \parbox[t]{\PaperCodeWidth}{\scriptsize\raggedright\strut\textbf{Code}\strut} \\[2pt]
  \midrule
}

 \newcommand{\AreaHeader}[1]{%
  \specialrule{0.9pt}{0pt}{0pt}%
  \rowcolor{black!8}%
  \multicolumn{6}{@{}c@{}}{%
    \scriptsize\bfseries
    \rule{0pt}{11pt}#1\rule[-3pt]{0pt}{4pt}%
  }\\
  \specialrule{0.65pt}{0pt}{0pt}%
  \PaperColumnHeader
}

\begin{longtable}{@{}llllll@{}}
\caption{\textbf{Collection of representative papers in computational quantum science.}
Journal names are abbreviated for conciseness. Public code availability is classified as
Y (Yes), a publicly available implementation; P (Partial), an implementation covering only a subset of the reported numerical targets or computational workflow; and N (No), no public implementation identified, including code available only upon request. For papers with publicly accessible implementations, the code availability entries provide direct hyperlinks to the corresponding repositories or software archives.}
\label{tab:quantum-papers}\\

\endfirsthead
\endhead
\endfoot
\bottomrule
\endlastfoot

\AreaHeader{Classical simulation of quantum systems}

% Tensor network

\PaperRow{2}{Tensor network}
{Efficient tensor network simulation of IBM's Eagle kicked Ising experiment}
{tindall2024eagle}{PRX Quantum}{3}{N}

\PaperRow{7}{}
{Matrix product states and first quantization}
{li2026matrix}{PRL}{4}{N}

\PaperRow{8}{}
{Low-rank quantics tensor train representations of Feynman diagrams for multiorbital electron-phonon models}
{tkcp-p5br}{PRL}{1}{N}

\PaperRow{14}{}
{Stable iPEPO tensor-network algorithm for dynamics of two-dimensional open quantum lattice models}
{mckeever2021stable}{PRX}{5}{N}

\midrule

% Neural quantum state

\PaperRow{16}{Neural quantum state}
{Predicting topological entanglement entropy in a Rydberg analog simulator}
{mauron2025predicting}{Nat. Phys.}{6}
{\href{https://doi.org/10.5281/zenodo.14697667}{Y}}

\PaperRow{17}{}
{Approximately-symmetric neural networks for quantum spin liquids}
{kufel2025approximately}{PRL}{3}
{\href{https://github.com/dom-kufel/Approximate-Symmetries-TC}{Y}}

\PaperRow{18}{}
{Autoregressive Transformer Neural Network for Simulating Open Quantum Systems via a Probabilistic Formulation}
{PhysRevLett.128.090501}{PRL}{3}
{\href{https://github.com/Rodin1000/POVM-Quantum-Simulation}{Y}}

\midrule

% Pauli propagation

\PaperRow{1}{Pauli propagation}
{Simulation of quantum circuits by low-rank stabilizer decompositions}
{Bravyi_2019}{Quantum}{4}{N}

\PaperRow{4}{}
{Simulation of Qubit Quantum Circuits via Pauli Propagation}
{rall2019simulation}{PRA}{6}
{\href{https://github.com/sebastianptz/Bachelor-Project-Pauli-Propagation}{P}}

\PaperRow{5}{}
{Classically estimating observables of noiseless quantum circuits}
{angrisani2025classically}{PRL}{2}
{\href{https://github.com/SparqleSim/PauliPropagation.jl}{P}}

\PaperRow{6}{}
{Fast and converged classical simulations of evidence for the utility of quantum computing before fault tolerance}
{beguvsic2024fast}{Sci. Adv.}{5A}{N}

\PaperRow{11}{}
{Clifford Circuits Augmented Time-Dependent Variational Principle}
{qian2025clifford}{PRL}{3}{N}

\midrule

% Free fermion simulation

\PaperRow{3}{Free fermion simulation}
{Theory of free fermions under random projective measurements}
{poboiko2023theory}{PRX}{3}{N}

\PaperRow{20}{}
{The non-stabilizerness of fermionic Gaussian states}
{collura2026non}{Quantum}{4}{N}

\midrule

% Quantum Monte Carlo

\PaperRow{9}{Quantum Monte Carlo}
{Real-time Sign-Problem-Suppressed Quantum Monte Carlo Algorithm For Noisy Quantum Circuit Simulations}
{shen2026realtime}{PRL}{2}
{\href{https://github.com/USCqserver/OpenQMC-simulator}{Y}}

\PaperRow{10}{}
{Bipartite reweight-annealing algorithm of quantum Monte Carlo to extract large-scale data of entanglement entropy and its derivative}
{wang2024bipartite}{Nat. Commun.}{3}{N}

\midrule

% Open quantum systems

\PaperRow{12}{Simulation for open quantum system}
{Large-scale stochastic simulation of open quantum systems}
{sander2025large}{Nat. Commun.}{7}
{\href{https://github.com/munich-quantum-toolkit/yaqs}{Y}}

\PaperRow{13}{}
{Efficient simulation of finite-temperature open quantum systems}
{tamascelli2019efficient}{PRL}{2}
{\href{https://github.com/tfmlaX/Chaincoeffs}{P}}

\PaperRow{15}{}
{Steady States of Infinite-Size Dissipative Quantum Chains via Imaginary Time Evolution}
{gangat2017steady}{PRL}{1}{N}

\PaperRow{19}{}
{Efficient simulation of non-trivial dissipative spin chains via stochastic unraveling}
{pocklington2025efficient}{PRX Quantum}{6}{N}

\AreaHeader{Quantum algorithms}

% Non-variational quantum simulation

\PaperRow{23}{Non-variational quantum simulation}
{Efficient and practical Hamiltonian simulation from time-dependent product formulas}
{bosse2025efficient}{Nat. Commun.}{1}{N}

\PaperRow{31}{}
{Faster Algorithmic Quantum and Classical Simulations by Corrected Product Formulas}
{bagherimehrab2026faster}{PRX Quantum}{2, 3}
{\href{https://github.com/mohsenbm/CPFs}{Y}}

\PaperRow{34}{}
{Digital adiabatic evolution is universally accurate}
{lu2026digital}{Nat. Commun.}{3}
{\href{https://github.com/QuantyyLu/AQC\_Simulation}{Y}}

\PaperRow{30}{}
{Determining eigenstates and thermal states on a quantum computer using quantum imaginary time evolution}
{motta2020determining}{Nat. Phys.}{3}
{\href{https://github.com/mariomotta/QITE}{Y}}

\PaperRow{33}{}
{Enhancing Scalability of Quantum Eigenvalue Transformation of Unitary Matrices for Ground State Preparation through Adaptive Finer Filtering}
{karacan2025enhancing}{Quantum}{6}
{\href{https://github.com/erenaykrcn/AFF}{P}}

\midrule

% Quantum phase estimation

\PaperRow{32}{Quantum phase estimation}
{On low-depth algorithms for quantum phase estimation}
{ni2023low}{Quantum}{5, 6}{N}

\PaperRow{35}{}
{Low depth algorithms for quantum amplitude estimation}
{giurgica2022low}{Quantum}{5, 6}{N}

\PaperRow{36}{}
{Bayesian Quantum Amplitude Estimation}
{ramoa2025bayesian}{Quantum}{4, 5, 6}
{\href{https://github.com/alexandra-frca/BAE}{Y}}

\midrule

% Variational quantum simulation

\PaperRow{28}{Variational quantum simulation}
{An adaptive variational algorithm for exact molecular simulations on a quantum computer}
{grimsley2019adaptvqe}{Nat. Commun.}{2}
{\href{https://github.com/mayhallgroup/adapt-vqe}{P}}

\PaperRow{29}{}
{Synergistic pretraining of parametrized quantum circuits via tensor networks}
{rudolph2023synergistic}{Nat. Commun.}{2}
{\href{https://github.com/MSRudolph/Synergy-PQC-TN}{P}}

\PaperRow{22}{}
{The Quantum Approximate Optimization Algorithm and the Sherrington-Kirkpatrick Model at Infinite Size}
{farhi2022quantum}{Quantum}{1}
{\href{https://github.com/leologist/QAOA-SK}{P}}

\PaperRow{24}{}
{Overhead-constrained circuit knitting for variational quantum dynamics}
{gentinetta2024overhead}{Quantum}{4}
{\href{https://github.com/gentinettagian/circuit-cutting-pvqd}{Y}}

\midrule

% Quantum machine learning

\PaperRow{25}{Quantum machine learning}
{Theory of overparametrization in quantum neural networks}
{larocca2023theory}{Nat. Comput. Sci.}{3, 5}
{\href{https://github.com/DiegoGM91/Theory-of-overparametrization}{Y}}

\PaperRow{38}{}
{Cost Function Dependent Barren Plateaus in Shallow Parametrized Quantum Circuits}
{cerezo2021cost}{Nat. Commun.}{6}{N}

\PaperRow{27}{}
{Barren plateaus preclude learning scramblers}
{holmes2021barren}{PRL}{2}{N}

\PaperRow{26}{}
{Classical surrogates for quantum learning models}
{schreiber2023surrogates}{PRL}{4}{N}

\PaperRow{21}{}
{Power of data in quantum machine learning}
{huang2021power}{Nat. Commun.}{4}
{\href{https://www.tensorflow.org/quantum/tutorials/quantum\_data}{P}}

\PaperRow{60}{}
{Exponential concentration in quantum kernel methods}
{thanasilp2024exponential}{Nat. Commun.}{6}
{\href{https://github.com/Supanut-Thanasilp/Exponential-concentration-in-quantum-kernel-methods}{P}}

\midrule

% Quantum-enhanced sampling

\PaperRow{37}{Quantum-enhanced Markov chain Monte Carlo}
{Quantum-enhanced Markov chain Monte Carlo}
{layden2023quantum}{Nature}{2}
{\href{https://doi.org/10.5281/zenodo.7799889}{Y}}

\AreaHeader{Quantum property characterization}

% Linear state properties

\PaperRow{39}{Linear state properties}
{Information-theoretic bounds on quantum advantage in machine learning}
{huang2021information}{PRL}{3}{N}

\PaperRow{40}{}
{Predicting Many Properties of a Quantum System from Very Few Measurements}
{huang2020predicting}{Nat. Phys.}{3}
{\href{https://github.com/hsinyuan-huang/predicting-quantum-properties}{Y}}

\PaperRow{41}{}
{Inference-Based Quantum Sensing}
{huerta2022inference}{PRL}{2, 3}{N}

\PaperRow{44}{}
{Shallow shadows: Expectation estimation using low-depth random Clifford circuits}
{bertoni2024shallow}{PRL}{3}
{\href{https://github.com/christianbertoni/shallow-shadows}{Y}}

\PaperRow{47}{}
{Efficient Learning for Linear Properties of Bounded-Gate Quantum Circuits}
{du2025efficient}{Nat. Commun.}{2, 3}
{\href{https://github.com/yuxuan-du/Efficient_Predicting_Bounded_Gate_QC}{P}}

\PaperRow{48}{}
{Improved machine learning algorithm for predicting ground state properties}
{lewis2024ground}{Nat. Commun.}{2}
{\href{https://github.com/lllewis234/improved-ml-algorithm}{Y}}

\PaperRow{49}{}
{Guaranteed efficient energy estimation of quantum many-body Hamiltonians using ShadowGrouping}
{gresch2025shadowgrouping}{Nat. Commun.}{3}
{\href{https://gitlab.com/GreschAI/shadowgrouping}{Y}}

\PaperRow{50}{}
{Optimizing quantum measurements by partitioning multisets of observables}
{veltheim2025multisets}{PRL}{3}{N}

\PaperRow{55}{}
{Measuring Arbitrary Physical Properties in Analog Quantum Simulation}
{tran2023analog}{PRX}{6}{N}

\midrule

% Nonlinear state properties

\PaperRow{51}{Nonlinear state properties}
{Importance sampling of randomized measurements for probing entanglement}
{rath2021importance}{PRL}{2}
{\href{https://github.com/bvermersch/RandomMeas}{Y}}

\PaperRow{52}{}
{Auxiliary-Free Replica Shadows: Efficient Estimation of Multiple Nonlinear Quantities}
{liu2026replica}{PRL}{2}{N}

\PaperRow{53}{}
{Detecting entanglement in quantum many-body systems via permutation moments}
{liu2022permutation}{PRL}{3}{N}

\PaperRow{54}{}
{Many-body entropies and entanglement from polynomially-many local measurements}
{vermersch2024entropies}{PRX}{4}{N}

\midrule

% Hamiltonian learning

\PaperRow{43}{Hamiltonian learning}
{Practical Hamiltonian learning with unitary dynamics and Gibbs states}
{gu2024hamiltonian}{Nat. Commun.}{3, 4, 5}
{\href{https://github.com/andigu/hamiltonian-learning}{P}}

\PaperRow{56}{}
{Ansatz-free Hamiltonian learning with Heisenberg-limited scaling}
{hu2025ansatzfree}{PRX Quantum}{2}{N}

\midrule

% Quantum error mitigation

\PaperRow{46}{Quantum error mitigation}
{Machine learning for practical quantum error mitigation}
{liao2024machine}{Nat. Mach. Intell.}{3}
{\href{https://github.com/qiskit-community/ml-qem/tree/research}{Y}}

\PaperRow{59}{}
{Quantum Error Mitigated Classical Shadows}
{jnane2024mitigated}{PRX Quantum}{3}{N}

\midrule

% Quantum process characterization

\PaperRow{42}{Quantum processes characterization}
{Quantum circuit complexity and unsupervised machine learning of topological order}
{che2026topological}{Nat. Commun.}{2, 3}
{\href{https://doi.org/10.6084/m9.figshare.29828336}{Y}}

\PaperRow{45}{}
{Shadow estimation of gate-set properties from random sequences}
{helsen2023gateset}{Nat. Commun.}{2}{N}

\PaperRow{57}{}
{Learning conservation laws in unknown quantum dynamics}
{zhan2024conservation}{PRX Quantum}{2}{N}

\PaperRow{58}{}
{Learning to predict arbitrary quantum processes}
{huang2023processes}{PRX Quantum}{2, 3}
{\href{https://github.com/hsinyuan-huang/learning-quantum-process}{Y}}

\end{longtable}
\endgroup

\subsection{Analysis of paper-to-result reproduction issues}\label{append:sec:paper-levels}
In this subsection, we provide a detailed analysis of the reproduction issues identified through AI-assisted cross-agent verification and independently confirmed by domain experts. Specifically, our examination of the initial reproduction results revealed issues affecting 18 of the 60 selected papers, including inconsistencies between published methodological descriptions and released implementations, incomplete experimental specifications, and discrepancies between reported numerical results and theoretical constraints. These issues span all three areas of computational quantum science considered in this study and are classified into three levels according to their severity and resolvability, as introduced in the main text, i.e., 
\begin{itemize}
\item \textbf{Level I (16 papers):} Recoverable methodological details or discrepancies in task specifications and evaluation criteria that can be resolved using the available scientific record or expert interpretation.
\item \textbf{Level II (1 paper):} Localized inconsistencies in the published study that prevent faithful reproduction of the reported results but can be resolved through expert inspection of the available evidence.
\item \textbf{Level III (1 paper):} Fundamental inconsistencies between the reported results, methodological descriptions, and released code that remain unresolved even after expert inspection, preventing faithful reproduction from the available scientific record.
\end{itemize}
A summary of the major issues of these 18 research papers is provided in Table~\ref{tab:paper-levels}.

In the following, we examine representative studies at each level to illustrate the corresponding failure modes. For Level I, we select three representative papers, one from each of the three research areas, while for Levels II and III, we examine the sole paper in each category. For each paper, we first introduce the scientific background and then follow the steps of the curation workflow described in SI.~\ref{append:sec:curation-workflow} to illustrate how AI scientists and domain experts collaborate to identify, verify, and resolve potential reproduction issues.

\begin{table}[!h]
\caption{\small\textbf{Papers with an issue confirmed in the curation audit.} Paper numbers are those of Table~\ref{tab:quantum-papers}, and levels follow the three-level scale of SI~\ref{append:sec:paper-levels}. The last column shows how we handled each issue. Re-judge means that the criteria were revised and the affected runs re-judged. Re-run means that the task specification was corrected and the affected tasks re-run. Keep means that no verdict depended on the issue, so that neither a re-judgement nor a re-run was needed. Removed means that the paper was excluded from the benchmark.}
\label{tab:paper-levels}
\footnotesize
\setlength{\tabcolsep}{4pt}
\begin{tabular}{@{}llll@{}}
\toprule
\parbox[c]{0.06\textwidth}{\raggedright\strut \textbf{Level}\strut} & \parbox[c]{0.08\textwidth}{\raggedright\strut \textbf{Paper}\strut} & \parbox[c]{0.70\textwidth}{\raggedright\strut \textbf{Issue}\strut} & \parbox[c]{0.11\textwidth}{\raggedright\strut \textbf{Handling}\strut} \\
\midrule
I & \#39~\cite{huang2021information} & \parbox[c]{0.70\textwidth}{\raggedright\strut The reported results reflect an idealized setting rather than the actual execution of the algorithm described in the paper\strut} & Re-judge \\ \addlinespace[3pt]
 & \#40~\cite{huang2020predicting} & \parbox[c]{0.70\textwidth}{\raggedright\strut Underspecified experimental details in the paper and overly restrictive figure-based evaluation criteria\strut} & Re-judge \\ \addlinespace[3pt]
 & \#26~\cite{schreiber2023surrogates} & \parbox[c]{0.70\textwidth}{\raggedright\strut The initial criteria require the classical model to have at least ten times lower error than the quantum models, a requirement not explicitly stated in the paper, and incorrectly marks quantum-model results with errors below the prescribed range as failures. \strut} & Re-judge \\ \addlinespace[3pt]
 & \#9~\cite{shen2026realtime} & \parbox[c]{0.70\textwidth}{\raggedright\strut The excited-state convention is reversed relative to the standard qubit convention\strut} & Re-judge \\ \addlinespace[3pt]
 & \#14~\cite{mckeever2021stable} & \parbox[c]{0.70\textwidth}{\raggedright\strut Hamiltonian bond-counting convention differs from the standard interpretation \strut} & Re-judge \\ \addlinespace[3pt]
 & \#51~\cite{rath2021importance} & \parbox[c]{0.70\textwidth}{\raggedright\strut Underspecified experimental details in the paper and overly restrictive figure-based evaluation criteria\strut} & Re-judge \\ \addlinespace[3pt]
 & \#52~\cite{liu2026replica} & \parbox[c]{0.70\textwidth}{\raggedright\strut Unstated experimental setting about the number of measurements\strut} & Re-judge \\ \addlinespace[3pt]
 & \#28~\cite{grimsley2019adaptvqe} & \parbox[c]{0.70\textwidth}{\raggedright\strut The initial criteria set an incorrect energy target for BeH$_2$. \strut} & Re-judge \\ \addlinespace[3pt]
 & \#38~\cite{cerezo2021cost} & \parbox[c]{0.70\textwidth}{\raggedright\strut The paper leaves details of the optimization procedure unspecified, but the initial criteria require the resulting measurement counts to fall within fixed ranges and mark results that correctly reproduce the growth trend as failures. \strut} & Re-judge \\ \addlinespace[3pt]
 & \#34~\cite{lu2026digital} & \parbox[c]{0.70\textwidth}{\raggedright\strut The paper specifies the truncation order as $K=\lfloor 4\log T\rfloor$ without defining the logarithm base, while the released code uses $K=4\lfloor\log_{10}T\rfloor$, giving different truncation orders. \strut} & Re-judge \\ \addlinespace[3pt]
 & \#21~\cite{huang2021power} & \parbox[c]{0.70\textwidth}{\raggedright\strut The paper omits the input rescaling needed to reproduce the reported quantum kernel classification performance and gives conflicting rules for generating class labels. \strut} & Re-run \\ \addlinespace[3pt]
 & \#10~\cite{wang2024bipartite} & \parbox[c]{0.70\textwidth}{\raggedright\strut Incorrect lattice dimensions in the preliminary task specification\strut} & Re-run \\ \addlinespace[3pt]
 & \#33~\cite{karacan2025enhancing} & \parbox[c]{0.70\textwidth}{\raggedright\strut The paper confuses the ground-state overlap with its square, and the initial task specification recommends a random state that does not reliably reproduce the reported speedup. \strut} & Re-run \\ \addlinespace[3pt]
 & \#46~\cite{liao2024machine} & \parbox[c]{0.70\textwidth}{\raggedright\strut Inconsistency of hyperparameter settings between the paper's claim and the released code\strut} & Re-run \\ \addlinespace[3pt]
 & \#17~\cite{kufel2025approximately} & \parbox[c]{0.70\textwidth}{\raggedright\strut Incorrect entropy units in the preliminary task specification\strut} & Keep \\ \addlinespace[3pt]
 & \#50~\cite{veltheim2025multisets} & \parbox[c]{0.70\textwidth}{\raggedright\strut Unspecified detailed about the random number generator\strut} & Keep \\
\midrule
II & \#42~\cite{che2026topological} & \parbox[c]{0.70\textwidth}{\raggedright\strut Analysis admits several readings, and the published values need two unstated steps\strut} & Re-run \\
\midrule
III & \#60~\cite{thanasilp2024exponential} & \parbox[c]{0.70\textwidth}{\raggedright\strut One figure plots the variance of the distance term $C$ in the kernel $\kappa^{\mathrm{PQ}}=e^{-\gamma C}$ but incorrectly labels it as kernel variance. The released code and documented settings are insufficient to regenerate the underlying data. \strut} & Removed \\
\bottomrule
\end{tabular}
\end{table}

\medskip
\subsubsection{Level I (Paper~\#46): Hyperparameter inconsistency in quantum error mitigation}\label{append:sec:paper46}
Quantum error mitigation (QEM) plays an important role in improving the accuracy of quantum computations on noisy quantum devices, yet achieving accurate mitigation with limited experimental overhead remains challenging~\cite{cai2023quantum}. Paper~\#46~\cite{liao2024machine}, published in \textit{Nature Machine Intelligence} in 2024, investigated machine-learning-based QEM (ML-QEM) to improve mitigation accuracy and computational efficiency, with particular emphasis on generalization beyond the training regime. Here, we examine the reproduction of its random-forest (RF) results, where an inconsistency between the hyperparameter settings described in the paper and those used in the released implementation led to substantial reproduction discrepancies.

\begin{figure}[!htbp]
\centering
\includegraphics[width=0.98\textwidth]{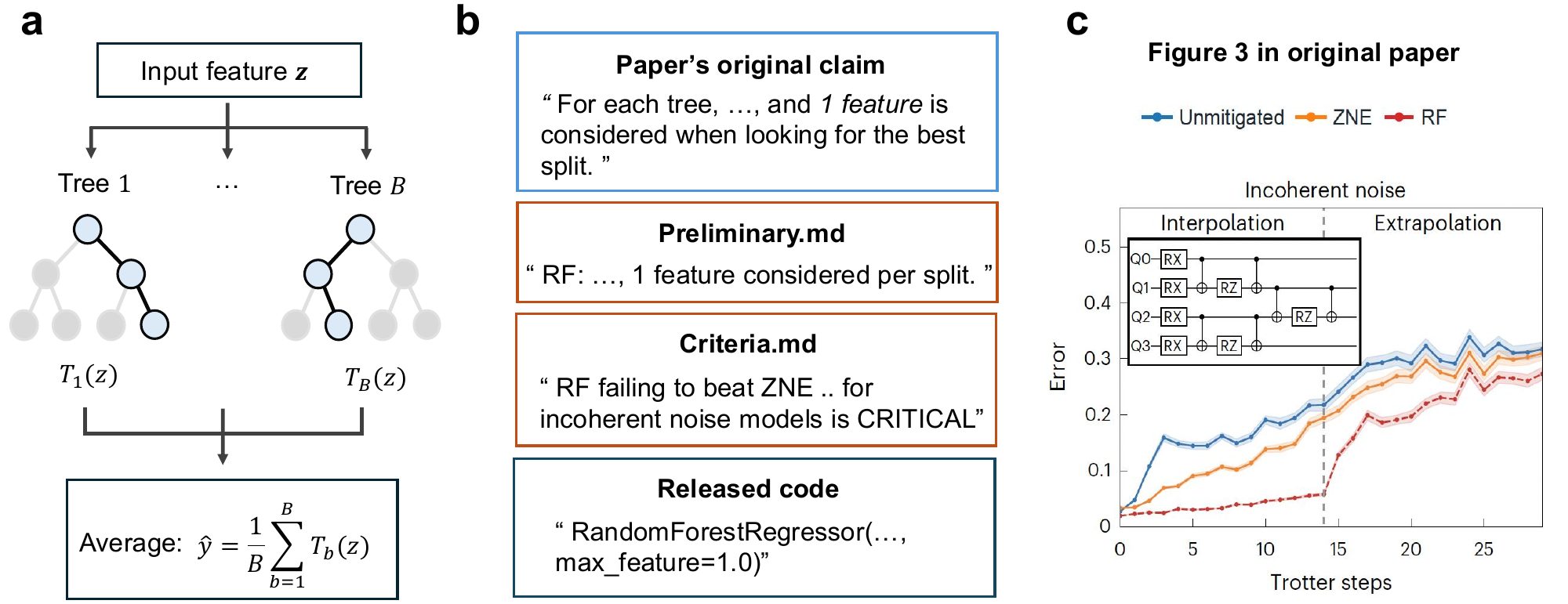}
\caption{\textbf{Overview of the reproduction task in Paper~\#46 for ML-based quantum error mitigation.} \textbf{a.} Schematic of random-forest regression for quantum error mitigation, where the predictions of multiple decision trees are averaged to estimate ideal observable expectation values from noisy measurement features. At each split of a decision tree, the hyperparameter \texttt{max\_features} controls the number of candidate features considered when selecting the optimal split. \textbf{b.} Description of the \texttt{max\_features} hyperparameter settings in the original paper, initial task specification (\texttt{Preliminary.md}), and released implementation, together with the evaluation criteria (\texttt{Criteria.md}) for successful reproduction. \textbf{c.} The upper-left panel of original Figure~3 of Paper~\#46~\cite{liao2024machine}, showing four-qubit Trotterized Ising dynamics with incoherent noise and no readout error. The red, orange, and blue curves represent the errors obtained using random-forest (RF) mitigation, zero-noise extrapolation (ZNE), and unmitigated measurements, respectively. The vertical dashed line at Trotter step 14 separates interpolation from extrapolation regimes.}
\label{fig:si-qem-evidence}
\end{figure}

\medskip
\noindent\textbf{Algorithmic background.}
Recall from SI.~\ref{SI:subsec:property_characterization} that QEM aims to recover the ideal expectation value $f_O(\rho)=\Tr(OU\rho_0U^\dagger)$ from noisy measurements, which yield $f_O(\tilde{\rho})=\Tr[O\widetilde{\mathcal U}(\rho_0)]$ with $\widetilde{\mathcal U}$ representing the corresponding noisy quantum channel. ML-QEM approaches this task by learning a classical predictive model from training data $\mathcal{T}=\{(\bm{z}^{(i)},y^{(i)})\}_{i=1}^{n}$, where $\bm{z}^{(i)}\in\mathbb{R}^d$ encodes noisy measurement outcomes and circuit-related features, while $y^{(i)}$ denotes the corresponding ideal expectation value. As shown in Fig.~\ref{fig:si-qem-evidence}\textbf{a}, given the dataset $\mathcal{T}$, RF constructs an ensemble of $B$ decision trees and averages their predictions to obtain the mitigated expectation value,
\[
\hat f_O^{\mathrm{RF}}(\bm z)=\frac{1}{B}\sum_{b=1}^{B}T_b(\bm z),
\]
where $T_b$ denotes the prediction of the $b$-th decision tree. Each tree recursively partitions the feature space by selecting a feature and a splitting threshold that minimize the prediction error. At each split, only a subset of candidate features $\mathcal{S}\subseteq[d]$ is considered. The hyperparameter \texttt{max\_features} controls the number of these candidate features $|\mathcal{S}|$, thereby influencing the construction of the decision trees and their predictive performance.

\medskip
\noindent \textbf{Task construction with \omniQ.} The domain expert selected Fig.~3 of the original paper as the reproduction target. Fig.~\ref{fig:si-qem-evidence}\textbf{c} shows its upper-left panel, which compares the mitigation errors of random forest (RF), digital zero-noise extrapolation (ZNE), and unmitigated measurements for four-qubit Trotterized Ising dynamics under incoherent noise, with the number of Trotter steps spanning both interpolation and extrapolation regimes. The target reproduction figure was selected for two reasons. First, it requires reproducing distinct computational procedures, including noisy-circuit simulation, RF model training, and noise-scaled circuit simulations for ZNE. Second, it directly examines the paper's central claim that ML-QEM can achieve superior mitigation accuracy and generalize to circuits deeper than those used for training. 

Based on the methodology described in the paper, an agent drafted the initial task specification (\texttt{Preliminary.md}) and evaluation criteria (\texttt{Criteria.md}), as illustrated in Fig.~\ref{fig:si-qem-evidence}\textbf{b}. The former specified the RF hyperparameter \texttt{max\_features=1}, consistent with the paper's statement that only one feature is considered when selecting the best split. The latter specified quantitative evaluation criteria requiring RF to outperform ZNE in both interpolation and extrapolation regimes. However, some experimental details, such as the initial quantum state $\rho_0$, were not fully specified in the original paper. To make the task executable, \texttt{Preliminary.md} required a product state but allowed the reproducing agent to choose the specific state and document it in \texttt{report.md}. Since the mitigation errors depend on the choice of initial state, the evaluation criteria in \texttt{Criteria.md} focused on reproducing the relative performance of the mitigation methods instead of requiring strict numerical agreement with the published error values.

\begin{figure}[h!]
\centering
\includegraphics[width=0.98\textwidth]{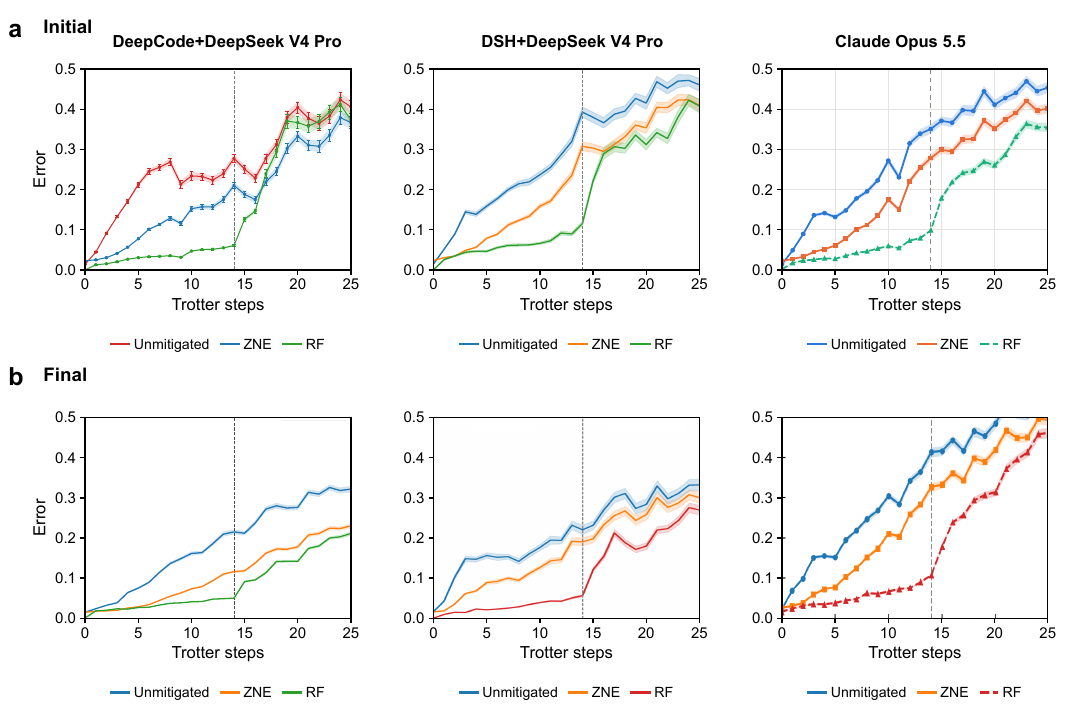}
\caption{\textbf{Reproduction results for Paper~\#46 on ML-based quantum error mitigation.} \textbf{a}. Initial benchmark results, with one candidate feature per RF split prescribed in the preliminary (\texttt{max\_features=1}). \textbf{b}. Final benchmark results after expert revision of the preliminary to specify all features (\texttt{max\_features=1.0}). Columns show DeepCode+DeepSeek V4 Pro, DSH+DeepSeek V4 Pro, and Claude Opus 5.5, from left to right. Each plot compares RF, ZNE, and unmitigated measurements under incoherent noise without readout error, corresponding to the original Figure 3 of Paper~\#46 (as shown in Fig.~\ref{fig:si-qem-evidence}\textbf{c}). Vertical dashed lines at Trotter step 14 separate interpolation from extrapolation regimes. The curve colors representing different methods in each panel may vary across agents, as each agent independently selects its plotting colors. The corresponding methods are identified in the legend of each panel.}
\label{fig:si-qem-agents}
\end{figure}

\medskip
\noindent\textbf{Initial benchmark.}
Fig.~\ref{fig:si-qem-agents}\textbf{a} presents the initial reproduction results of three representative AI scientists selected from the 23 agents evaluated on this task, which are DeepCode with DeepSeek V4 Pro, DSH with DeepSeek V4 Pro, and Claude Code with Opus 5.5. Both DeepCode and DSH followed the single-feature instruction by setting \texttt{max\_features=1}. Their reproduced results deviated from the published findings in the extrapolation regime. That is, DeepCode largely lost the reported RF advantage over ZNE at larger circuit depths, as its mitigation error approaches the unmitigated baseline, while the RF and ZNE errors obtained by DSH became nearly indistinguishable at the largest displayed depths. 

Different from DeepCode with DeepSeek V4 Pro and DSH with DeepSeek V4 Pro, Claude Code with Opus 5.5 independently recognized the distinction between the integer \texttt{1} and floating-point \texttt{1.0} settings, tested the prescribed single-feature configuration separately, and used \texttt{max\_features=1.0} in its main reproduction. Consequently, its results verified the RF advantage in the extrapolation regime. The initial judge marked this run as successful because it classified the documented deviation from the prescribed setting as \textit{non-critical}. These contrasting outcomes revealed a potential inconsistency in the task specification and motivated a closer examination of the published methodology and released implementation.

\medskip
\noindent\textbf{Cross-agent verification of reproduction inconsistencies.} To investigate the discrepancy, the verification AI scientist cross-checked the original paper, task specification, evaluation criteria, reproduction records, and authors' released RF implementation. As illustrated in Fig.~\ref{fig:si-qem-evidence}\textbf{b}, this examination revealed an inconsistency in the RF hyperparameter \texttt{max\_features}. Specifically, the paper and initial task specification required a single candidate feature at each split, whereas the released implementation used the scikit-learn default, which considers all available features. In scikit-learn, the integer setting \texttt{max\_features=1} corresponds to $|\mathcal{S}|=1$, while the floating-point setting \texttt{max\_features=1.0} corresponds to $|\mathcal{S}|=d$. The released RF notebook instantiated the regressor without explicitly specifying \texttt{max\_features}, thereby adopting the all-feature configuration.

Domain experts independently reviewed this evidence and confirmed that the task specification should follow the released implementation by explicitly requiring \texttt{max\_features=1.0}. They further determined that the discrepancy did not invalidate the paper's central scientific claim, since the underlying RF method remained unchanged and the reported extrapolation advantage could be recovered using the authors' implementation. The issue was therefore classified as Level I, as the available source code provided sufficient evidence for a reliable correction.

\medskip
\noindent\textbf{Final benchmark.}
Following expert verification, the task specification was revised to require \texttt{max\_features=1.0}, and the affected reproduction tasks were rerun and evaluated under the corrected specification. Fig.~\ref{fig:si-qem-agents}\textbf{b} presents the resulting reproduction outcomes. All three agents recovered the qualitative ordering reported in the original paper, with RF outperforming ZNE throughout the displayed extrapolation regime. In particular, DeepCode recovered the RF advantage that it had failed to reproduce in its initial run, DSH showed a clearer separation between RF and ZNE at larger circuit depths, and Claude Code with Opus 5.5 maintained the performance ordering observed in its initial reproduction. These results demonstrate that correcting the task specification enables recovery of the reported qualitative mitigation advantage of ML methods. 

Beyond recovering the qualitative advantage of RF, the DeepCode rerun with DeepSeek V4 Pro also produced mitigation errors whose magnitudes more closely matched those reported in the original figure. The domain experts further examined the \texttt{report.md} and found that it employed a different initial quantum state from that used in its initial run, despite the initial-state specification in \texttt{Preliminary.md} remaining unchanged. This observation indicates that the improved numerical agreement cannot be attributed solely to the correction of \texttt{max\_features}. More broadly, it highlights how different implementations of underspecified experimental settings can affect numerical reproducibility, even when the central scientific claim is successfully recovered.

\subsubsection{Level I (Paper~\#14): Unclear Hamiltonian specification in dynamics simulation}\label{append:sec:paper14}
Understanding the dynamics of interacting open quantum systems in the thermodynamic limit is crucial for exploring nonequilibrium phases and dissipative critical phenomena~\cite{diehl2008quantum,sieberer2013dynamical,fitzpatrick2017observation}, yet accurately simulating such dynamics remains challenging, particularly in higher dimensions. Paper~\#14~\cite{mckeever2021stable}, published in \textit{Physical Review X} in 2021, developed an infinite projected entangled pair operator (iPEPO) algorithm that incorporates optimal bond truncation techniques, enabling accurate simulations of correlated open-system dynamics in regimes beyond the reach of existing methods. Here, we examine the reproduction of its Figure~5 benchmark, where an unclear specification of the system Hamiltonian led to substantial discrepancies between the reproduced and reported results.

\begin{figure}[!htbp]
\centering
\includegraphics[width=0.98\textwidth]{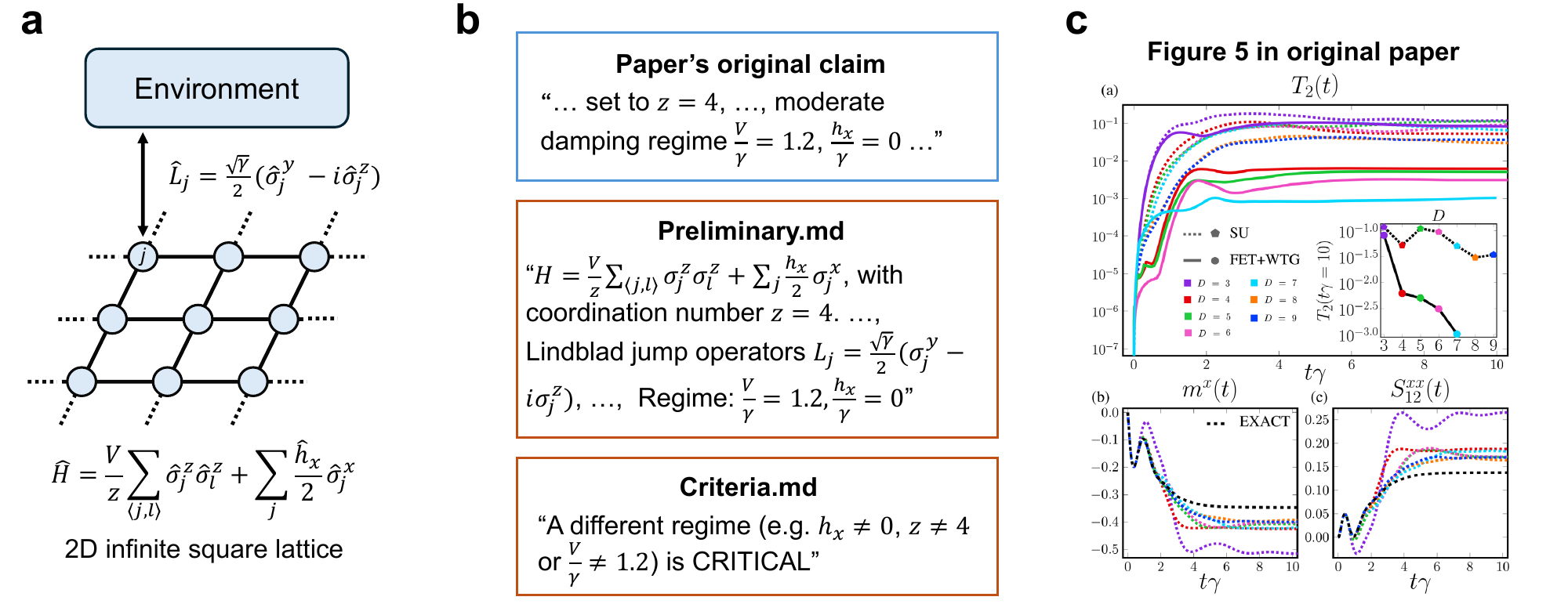}
\caption{\textbf{Overview of the reproduction task in Paper~\#14 for iPEPO 2D-dynamics simulation.}
\textbf{a.} Schematic of the dissipative transverse-field Ising model on an infinite two-dimensional square lattice, with nearest-neighbor Ising interactions $V/z$, a transverse magnetic field $h_x$, and local Lindblad jump operators $L_j$ describing dissipation.
\textbf{b.} Parameter settings reported in the original paper and specified in the initial reproduction instructions (\texttt{Preliminary.md}), together with the evaluation criteria (\texttt{Criteria.md}). 
\textbf{c.} Original Figure~5 of Paper~\#14, comparing the simple update (SU) and full-environment truncation with weighted trace gauge (FET+WTG) algorithms at different iPEPO bond dimensions $D$. The upper panel shows the trace distance $T_2(t)$ between the simulated and exact nearest-neighbor reduced density matrices, with the inset showing its dependence on $D$ at $t\gamma=10$. The lower panels show the time evolution of the local magnetization $m^x(t)$ and nearest-neighbor correlation $S_{12}^{xx}(t)$, respectively, with black dashed curves indicating the exact solutions.
}
\label{fig:si-dynamics-evidence}
\end{figure}

\medskip
\noindent\textbf{Physical model background.}
As reviewed in SI.~\ref{SI:subsec:classical_simulator}, the evolution of a Markovian open quantum system is governed by the Lindblad master equation $\frac{\partial \rho_t}{\partial t}= \mathcal{L}(\rho_t)= -\imath[H,\rho_t]+ \sum_j (L_j\rho_t L_j^\dagger -\frac12\{L_j^\dagger L_j,\rho_t\})$, where $H$ governs the coherent dynamics and $L_j$ are jump operators describing the coupling to the environment. As Fig.~\ref{fig:si-dynamics-evidence}\textbf{a} shows, Paper~\#14 considers a dissipative transverse-field Ising model on an infinite two-dimensional square lattice. Its Hamiltonian and local jump operators are given by
\begin{equation}
        H = \frac{V}{z}\sum_{\langle j,l\rangle}
        \sigma_j^z\sigma_l^z
        +\frac{h_x}{2}\sum_j\sigma_j^x,\quad \textit{with}~
        L_j = \frac{\sqrt{\gamma}}{2}
        (\sigma_j^y-\imath\sigma_j^z). \nonumber
\end{equation}
Here, $\sigma_j^{x,y,z}$ are Pauli operators, $\langle j,l\rangle$ denotes nearest-neighbor pairs, and $z=4$ is the lattice coordination number. The parameters $V$, $h_x$, and $\gamma$ control the interaction strength, transverse field, and dissipation rate, respectively. In particular, the regime $h_x=0$ admits an exact solution for local observables, providing a benchmark for evaluating the accuracy of the iPEPO simulations.

\medskip
\noindent\textbf{Task construction with \omniQ.}
The domain expert selected Fig.~5 of the original paper as the reproduction target. As shown in Fig.~\ref{fig:si-dynamics-evidence}\textbf{c}, this reproduction task requires the agent to correctly implement iPEPO evolution using different update strategies, including full-environment truncation with weighted trace gauge (FET+WTG) and simple update (SU), and benchmark their predictions against exact solutions for the trace distance $\langle T_2(t)\rangle$, magnetization $\langle m^x(t)\rangle$, and nearest-neighbor spin correlations $\langle S^{xx}_{12}(t)\rangle$. This task captures the paper's core technical contributions and enables a direct and rigorous assessment of reproduction accuracy against exact solutions.

When submitted this task to \omniQ, the agent generates the initial task specification (\texttt{Preliminary.md}) and evaluation criteria (\texttt{Criteria.md}), as shown in Fig.~\ref{fig:si-dynamics-evidence}\textbf{b}. In particular, following the description in the original paper, the agent specified the model parameters as $z=4$, $V/\gamma=1.2$, and $h_x=0$ in \texttt{Preliminary.md} to ensure that the reproduction was conducted in the same moderate-damping regime. The evaluation criteria in \texttt{Criteria.md} also required these parameter settings to be matched exactly for the reproduction to be considered successful.

\begin{figure}[!htbp]
\centering
\includegraphics[width=0.98\textwidth]{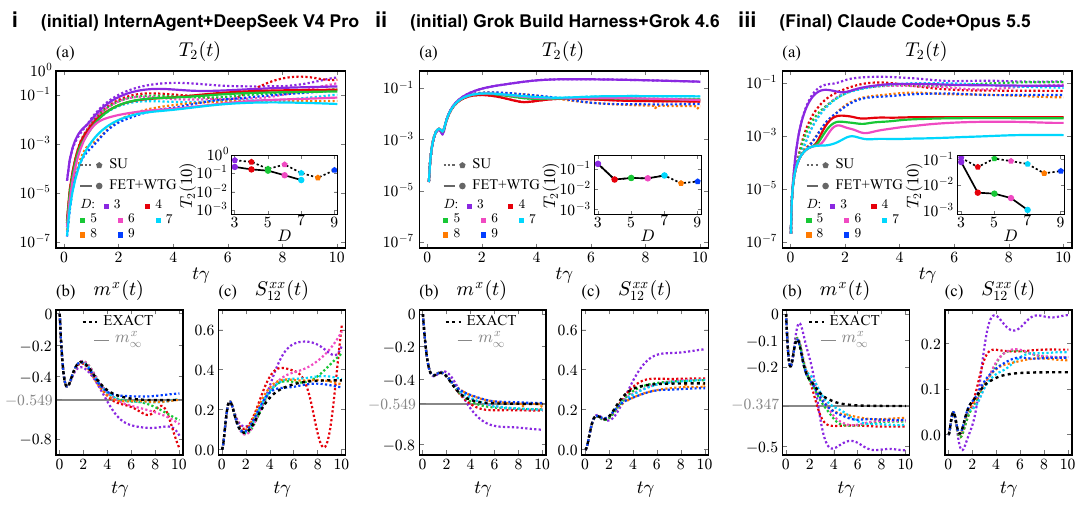}
\caption{
\textbf{Initial and corrected reproduction results for Paper~\#14 on iPEPO 2D-dynamics simulation.}
\textbf{i.} Initial reproduction results of InterAgent obtained by following the reported Hamiltonian parameters and counting each undirected nearest-neighbor bond once. \textbf{ii.} Initial reproduction results of Grok Build Harness obtained by following the reported Hamiltonian parameters and counting each undirected nearest-neighbor bond once. \textbf{iii.} Final reproduction results of Claude Code after expert revision to count each nearest-neighbor bond twice. In each group, panel (a) shows the trace distance $T_2(t)$ between the iPEPO and exact reduced density matrices, comparing SU (dotted lines) and FET+WTG (solid lines) for bond dimensions $D=3,\ldots,9$. The inset shows $T_2(10)$ as a function of $D$. Panels (b) and (c) show the time evolution of the local magnetization $m^x(t)$ and nearest-neighbor correlation $S_{12}^{xx}(t)$, respectively, with black dotted curves indicating the exact trajectories. The gray horizontal lines in panel (b) indicate the analytically determined steady-state magnetizations, approximately $-0.549$ and $-0.347$ for the initial and corrected reproductions, respectively. After correcting the bond-counting convention, the exact trajectories and the performance of the iPEPO algorithms become consistent with the results reported in the original paper.
}
\label{fig:si-dynamics-agents}
\end{figure}

\medskip
\noindent\textbf{Initial benchmark.}
Fig.~\ref{fig:si-dynamics-agents}\textbf{i} and Fig.~\ref{fig:si-dynamics-agents}\textbf{ii} present the initial reproduction results of InternAgent with DeepSeek V4 Pro and Grok Build Harness with Grok 4.6 as representative examples among the 23 agents evaluated on this task, where all reproduced curves deviate from the published results at late times ($\gamma t=10$). Notably, both InternAgent and Grok Build Harness obtained nearly identical values of approximately $-0.549$ for the exact magnetization $\langle m^x(t)\rangle$, as shown in panels \textbf{i}(b) and \textbf{ii}(b). Both agents independently obtained nearly identical exact trajectories that differed substantially from the published results. This agreement suggests that their exact simulations were correctly implemented, motivating the domain expert to investigate whether ambiguities in the definition of the model Hamiltonian could explain the discrepancy.

\medskip
\noindent\textbf{Cross-agent verification of reproduction inconsistencies.}
The domain expert independently derived the exact dynamics of the magnetization $m^x(t)$ and compared the resulting trajectory with the reproduced and original results to investigate the discrepancy. Following the Heisenberg Lindblad evolution $\partial_t\langle O\rangle =\langle\mathcal L^\dagger(O)\rangle$, the domain expert found that the evolution of magnetization $m^x(t)=\langle\sigma_j^x\rangle$ can be fully described in the subspace span by  $\{\langle\sigma_j^y\sigma_a^z\rangle, \langle\sigma_j^x\sigma_a^z\sigma_b^z\rangle, \langle\sigma_j^y\sigma_a^z\sigma_b^z\sigma_c^z\rangle, \langle\sigma_j^x\sigma_a^z\sigma_b^z\sigma_c^z\sigma_d^z\rangle\}$ where $a,b,c,d$ are denoted as the four distinct nearest neighbors of lattice site $j$. In particular, for the initial product state $\rho_0 = |\!\!\uparrow\rangle\langle\uparrow\!\!|^{\otimes N}$ along z-axis and counting each undirected nearest-neighbor bond once, their dynamics obeys
\begin{equation*}
\begin{aligned}
\partial_t\langle\sigma_j^x\rangle
={}&-\gamma(1+\langle\sigma_j^x\rangle)
-\frac{8V}{z}\langle\sigma_j^y\sigma_a^z\rangle,
\\
\partial_t\langle\sigma_j^y\sigma_a^z\rangle
={}&-\gamma\langle\sigma_j^y\sigma_a^z\rangle+\frac{2V}{z}
\bigl(\langle\sigma_j^x\rangle
+3\langle\sigma_j^x\sigma_a^z\sigma_b^z\rangle\bigr),
\\
\partial_t\langle\sigma_j^x\sigma_a^z\sigma_b^z\rangle
={}&-2\gamma
\langle\sigma_j^x\sigma_a^z\sigma_b^z\rangle
-\gamma e^{-\gamma t}-\frac{4V}{z}
\bigl(\langle\sigma_j^y\sigma_a^z\rangle
+\langle\sigma_j^y\sigma_a^z\sigma_b^z\sigma_c^z\rangle\bigr),
\\
\partial_t\langle\sigma_j^y\sigma_a^z\sigma_b^z\sigma_c^z\rangle
={}&-2\gamma
\langle\sigma_j^y\sigma_a^z\sigma_b^z\sigma_c^z\rangle\frac{2V}{z}
\bigl(3\langle\sigma_j^x\sigma_a^z\sigma_b^z\rangle
+\langle\sigma_j^x\sigma_a^z\sigma_b^z
\sigma_c^z\sigma_d^z\rangle\bigr),
\\
\partial_t\langle\sigma_j^x\sigma_a^z\sigma_b^z
\sigma_c^z\sigma_d^z\rangle
={}&-3\gamma
\langle\sigma_j^x\sigma_a^z\sigma_b^z
\sigma_c^z\sigma_d^z\rangle-\frac{8V}{z}
\langle\sigma_j^y\sigma_a^z\sigma_b^z
\sigma_c^z\rangle
-\gamma e^{-2\gamma t}.
\end{aligned}
\label{eq:audit-exact-dynamics}
\end{equation*}
Starting at all zero expectation values when $t=0$, $m^x(t)$ achieves the steady-state limit when $t\to\infty$ and all time derivatives equal to zero, which
yields the exact analytical expression
\begin{equation*}
m^x(\infty)=
-\frac{
3+62\left(\frac{V}{z\gamma}\right)^2
+96\left(\frac{V}{z\gamma}\right)^4
}{
3+110\left(\frac{V}{z\gamma}\right)^2
+512\left(\frac{V}{z\gamma}\right)^4
}.
\label{eq:audit-exact-steady-state}
\end{equation*}

Following the convention above, this formula gives $m^x(\infty)=-0.548923$ when the parameters are $V/\gamma=1.2$ and $z=4$ as reported in Paper~\#14. In contrast, counting each bond twice with effectively $V/\gamma =2.4$ yields $m^x(\infty)=-0.346579$. As shown in Fig.~\ref{fig:si-dynamics-evidence}\textbf{c}(b), Fig.~\ref{fig:si-dynamics-agents}\textbf{i}(b) and Fig.~\ref{fig:si-dynamics-agents}\textbf{ii}(b), the original and initially reproduced trajectories approach approximately $-0.347$ and $-0.549$, respectively. These values agree with the analytical predictions under the double- and single-counting conventions, respectively. This suggests that \textit{the original simulations effectively counted each nearest-neighbor bond twice, revealing a factor-of-two inconsistency between the reported Hamiltonian under its conventional single-counting interpretation and the numerical results}. This issue can therefore be corrected by specifying the Hamiltonian parameters from an exact calculation and is classified as Level~I.

\medskip
\noindent\textbf{Final benchmark.}
To resolve this ambiguity, the domain expert revised the task specification to explicitly require each nearest-neighbor bond to be counted twice in the Hamiltonian. As shown in Fig.~\ref{fig:si-dynamics-agents}\textbf{iii}-(b), the corrected exact trajectories implemented by Claude Code with Opus 5.5 agree with the independent analytical predictions, and the iPEPO simulations from all three reproduction runs closely match the results reported in the original paper. This confirms that the initial reproduction failures were mainly caused by ambiguity in the Hamiltonian bond-counting convention rather than errors in the implementation of the iPEPO algorithms.

\subsubsection{Level I (Paper~\#38): Inconsistent measurement-cost criteria in the training of variational quantum algorithms}
\label{append:sec:paper38}

Variational quantum algorithms (VQAs) can suffer from barren plateaus, where gradient variance decreases exponentially with the number of qubits~\cite{mcclean2018barren}. Paper~\#38~\cite{cerezo2021cost}, published in \textit{Nature Communications} in 2021, studied when barren plateaus arise for global and local cost functions at different circuit depths. The authors also simulated quantum autoencoder training~\cite{romero2017quantum}, claiming that increasing circuit depth slows training and raises the number of measurement shots needed to reach a given cost threshold under a local cost function. Here, we leverage \omniQ to reproduce this experiment, revealing an inconsistency between the acceptance criteria based on absolute shot counts and the optimization choices permitted by the task specification.

\begin{figure}[!htbp]
\centering
\includegraphics[width=0.98\textwidth]{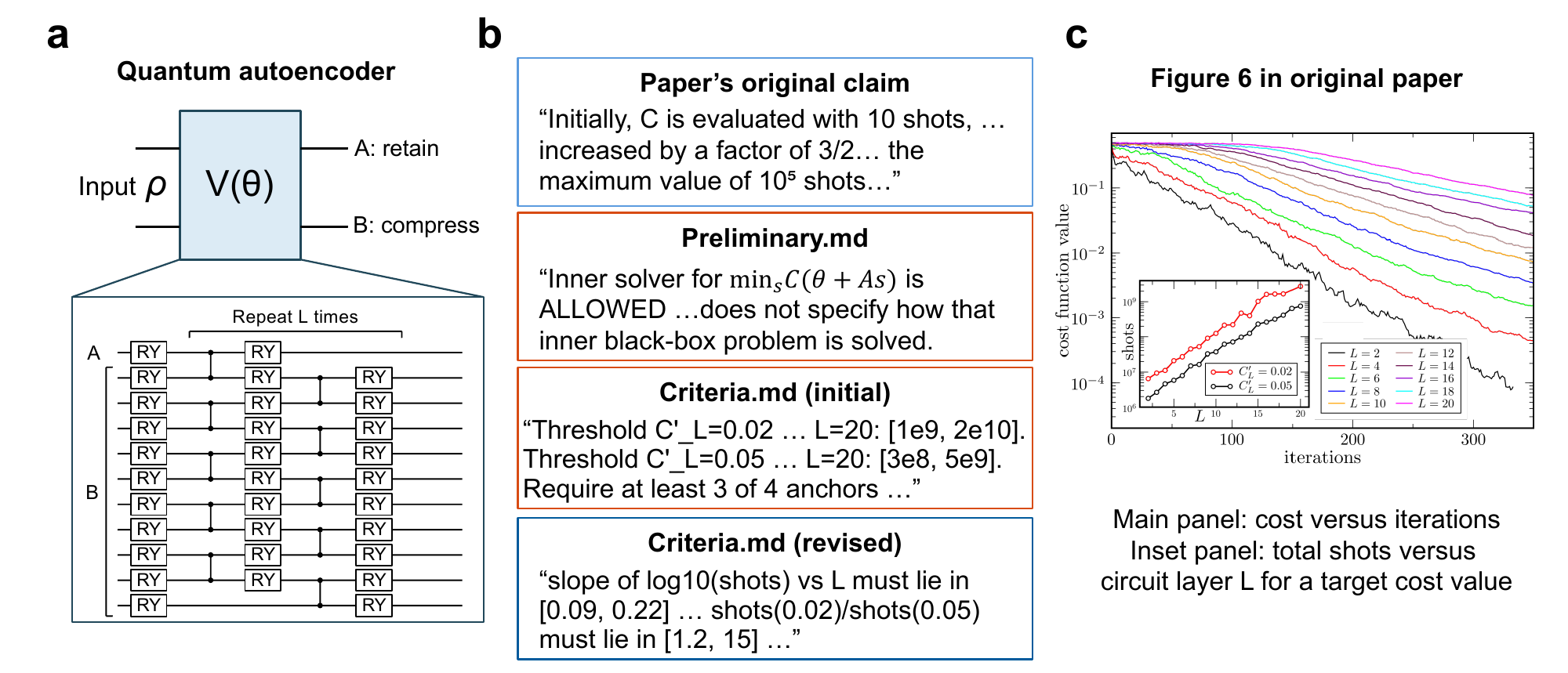}
\caption{\textbf{Overview of the reproduction task in Paper~\#38 for quantum autoencoder training.}
\textbf{a.} Schematic of the quantum autoencoder, which compresses the input quantum information into subsystem $A$ while bringing subsystem $B$ close to the reference state $|0\rangle^{\otimes n_B}$. The expanded circuit shows the ansatz used in the numerical simulation, with initial RY rotations followed by $L$ layers of alternating nearest-neighbor controlled-$Z$ gates and RY rotations.
\textbf{b.} Optimization settings described in the original paper and task specification (\texttt{Preliminary.md}), together with the initial and revised evaluation criteria (\texttt{Criteria.md}).
\textbf{c.} Original Figure~6 of Paper~\#38~\cite{cerezo2021cost}. The main panel shows the local cost during training for different circuit depths. The inset shows the cumulative number of shots needed to reach target costs 0.02 and 0.05 as a function of circuit depth $L$.}
\label{fig:si-qae-evidence}
\end{figure}

\medskip
\noindent\textbf{Algorithmic background.}
A quantum autoencoder uses a variational circuit $V(\bm\theta)$ to compress an ensemble of quantum states $\{(p_\mu,|\psi_\mu\rangle)\}_\mu$, where $p_\mu$ is the probability of state $|\psi_\mu\rangle$ and $\sum_\mu p_\mu=1$. As shown in Fig.~\ref{fig:si-qae-evidence}\textbf{a}, the circuit retains the encoded information in subsystem $A$ while bringing subsystem $B$ close to a fixed reference state so that $B$ can be discarded. Training minimizes a cost that measures how closely $B$ approaches this reference.

In Paper~\#38, the reference state is $|0\rangle^{\otimes n_B}$, and the local cost is
\begin{equation}
C'_{L}(\bm\theta)
=1-\frac{1}{n_B}\sum_{j=1}^{n_B}
\Tr\!\left[\left(|0\rangle\langle0|_j\otimes\mathbb{I}_{\bar j}\right)\Tr_A[V(\bm\theta)\rho V(\bm\theta)^\dagger] \right], 
\label{eq:si-qae-local-cost}
\end{equation}
where $\rho=\sum_\mu p_\mu|\psi_\mu\rangle\langle\psi_\mu|$ is the ensemble-average input state, and $\mathbb{I}_{\bar j}$ is the identity on all qubits in $B$ except $j$. The simulations use two input states, $|\psi_1\rangle=|0\rangle_A|0\rangle_B^{\otimes n_B}$ and $|\psi_2\rangle=|1\rangle_A|110\cdots0\rangle_B$, with probabilities $p_1=2/3$ and $p_2=1/3$. The authors fix $n_A=1$ and $n_B=10$ and vary the number of circuit layers over $L=2,4,\ldots,20$, using a brickwork ansatz of $R_y$ rotations and nearest-neighbor controlled-$Z$ gates as shown in Fig.~\ref{fig:si-qae-evidence}\textbf{a}. The circuit parameters $\bmt$ are optimized using gradient-free random-subspace search. Mathematically, at iteration $t$, a random isometry $A(t)\in\mathbb R^{P\times d}$ defines a $d=10$-dimensional search subspace of the $P$ circuit parameters. An inner optimizer then solves the following $d$-dimensional optimization problem
\begin{equation}
\bm{s}^* = \arg\min_{\bm s\in\mathbb R^d} C'_{L}\Big(\bm\theta(t)+A(t)\bm s\Big). 
\label{eq:si-qae-subspace-search}
\end{equation}
The parameters are then updated as $\bm\theta(t+1)=\bm\theta(t)+A(t)\bm{s}^*$. The number of shots used for each evaluation of the cost function starts at $10$ and increases by a factor of $3/2$ whenever the optimization reaches a plateau, up to a maximum of $10^5$.

\medskip
\noindent\textbf{Task construction with \omniQ.}
The domain expert selected Fig.~6 of Paper~\#38 as the reproduction target to examine how circuit depth affects training convergence and measurement cost. As shown in Fig.~\ref{fig:si-qae-evidence}\textbf{c}, the main panel of the original figure presents cost trajectories during optimization for different circuit depths, while the inset plots the cumulative number of measurement shots required to reach a target cost of $0.02$ or $0.05$ as a function of the circuit depth $L$. The reproduction task aims to recover the slower convergence and approximately exponential growth in measurement cost as circuit depth $L$ increases. Reproducing the inset also requires accounting for all shots used during the inner optimizations in Eq.~(\ref{eq:si-qae-subspace-search}).

With \omniQ, an agent prepared \texttt{Preliminary.md} based on the published input ensemble, circuit architecture, cost function, random-subspace search, and shot schedule, and drafted the corresponding \texttt{Criteria.md}, as summarized in Fig.~\ref{fig:si-qae-evidence}\textbf{b}. \textit{Since the original paper did not specify the solver for the inner optimization problem in Eq.~(\ref{eq:si-qae-subspace-search}), the task allowed reproducing agents to choose the solver, convergence tolerances, and evaluation budget, provided that these choices were fixed and documented.} The plateau-detection rule and shot-allocation convention were also left to the agents, but the shots used in every sampled cost evaluation had to be included in the cumulative count.

The initial \texttt{Criteria.md} defined numerical checkpoints for the training curves and their expected ordering across circuit depths. For the inset, each shot-count curve was required to fall within at least three of four absolute acceptance windows at $L\approx2,10,15,20$, for example $[10^9,2\times10^{10}]$ for the target cost $0.02$ at $L=20$. The criteria further required $S_{0.02}(L)\geq S_{0.05}(L)$ and at least a hundredfold increase in each series from $L=2$ to $L=20$.

\begin{figure}[!htbp]
\centering
\includegraphics[width=0.98\textwidth]{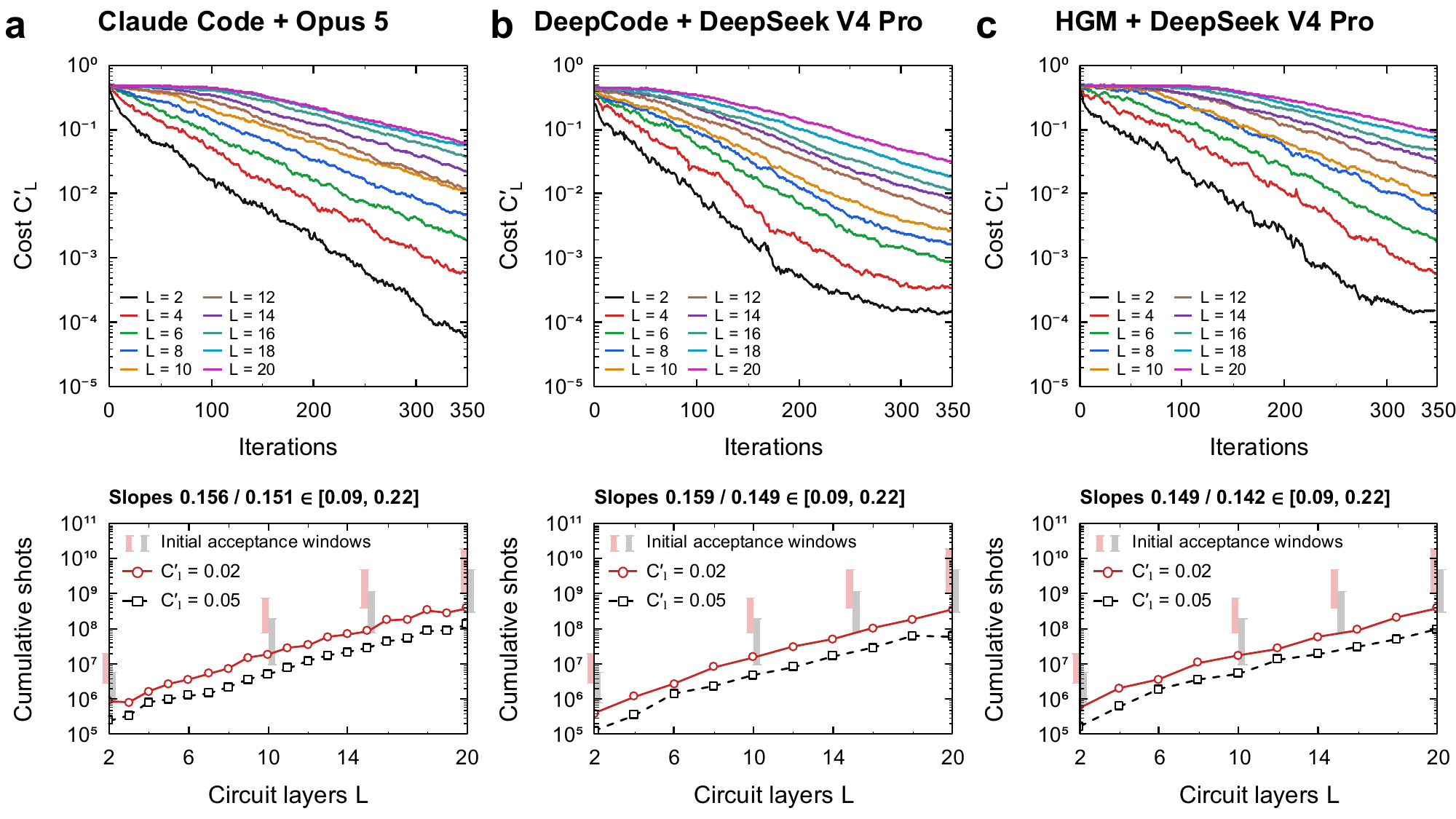}
\caption{\textbf{Reproduction results for Paper~\#38 on quantum autoencoder training.}
Panels \textbf{a-c} show results from Claude Code with Opus 5, DeepCode with DeepSeek V4 Pro, and HGM with DeepSeek V4 Pro, respectively, corresponding to the original Figure~6 in Fig.~\ref{fig:si-qae-evidence}\textbf{c}. The upper panels show the local cost during training for $L=2,4,\ldots,20$. The lower panels show cumulative shots needed to reach target costs 0.02 (red open circles, solid lines) and 0.05 (black open squares, dashed lines). Pale red and gray bars mark the initial acceptance windows for targets 0.02 and 0.05, respectively, at $L\approx2,10,15,20$. The annotations give the slopes of $\log_{10}S_\tau$ versus $L$ and the revised acceptance interval $[0.09,0.22]$.}
\label{fig:si-qae-agents}
\end{figure}

\medskip
\noindent\textbf{Initial benchmark.}
The three representative runs reproduced the reported depth-dependent training behavior but failed the absolute shot-count criteria. As shown in Figs.~\ref{fig:si-qae-agents}\textbf{a--c}, Claude Code with Opus 5, DeepCode with DeepSeek V4 Pro, and HGM with DeepSeek V4 Pro all obtained decreasing local costs, with convergence slowing as circuit depth increased. All three satisfied the requirements for threshold ordering and growth in cumulative shots with depth. However, their shot counts fell below the acceptance windows specified in the initial \texttt{Criteria.md} at all four depth checkpoints, as indicated by the pale bars in the lower panels. At $L=20$, all three runs required fewer than $5\times10^8$ cumulative shots to reach $C'_L\leq0.02$, compared with the required range $[10^9,2\times10^{10}]$. The initial judge classified these discrepancies as \textit{critical}, resulting in three failed reproduction assessments. This pattern suggested that the absolute shot-count requirements might be incompatible with the optimization choices permitted by the task specification.

\medskip
\noindent\textbf{Cross-agent verification of reproduction inconsistencies.}
The verification AI scientist examined whether the absolute shot-count requirements were justified by the published method and task specification. The paper's Methods section described the random-subspace search and shot schedule but did not specify the inner solver, convergence tolerances, or evaluation budget. As shown in Fig.~\ref{fig:si-qae-evidence}\textbf{b}, the task permitted reproducing agents to select and document these settings. The reproduction records showed that all three agents used the Nelder--Mead method for the inner optimization and included the shots from every sampled cost evaluation in the cumulative count. Claude Code set a limit of 400 cost evaluations per random-subspace optimization step, compared with 200 for DeepCode and HGM. Both limits were permitted by the task specification. Since each cost evaluation consumes measurement shots, these choices can affect the cumulative shot count even under the same shot schedule.

Expert review confirmed that the published method did not uniquely determine the absolute shot counts imposed by the initial criteria. As such, the expert approved replacing the absolute acceptance windows with checks on shot-count growth with circuit depth and the relative measurement costs at the two thresholds. The inconsistency was classified as Level I because it could be identified and resolved using the original paper and reproduction records.

\medskip
\noindent\textbf{Final benchmark.} 
Reassessment of the existing runs under the revised evaluation criteria changed all three outcomes from failure to success. The revised \texttt{Criteria.md} in Fig.~\ref{fig:si-qae-evidence}\textbf{b} requires the least-squares slope of $\log_{10}S_\tau(L)$ against $L$ to fall within $[0.09,0.22]$ for each target over $2\leq L\leq20$. It also requires the ratio $S_{0.02}(L)/S_{0.05}(L)$ to lie within $[1.2,15]$. A broader absolute shot-count range of $[10^7,10^{11}]$ at $L=20$ was included as a \textit{non-critical} check, while the threshold ordering and hundredfold-growth requirement remained unchanged. For targets $0.02$ and $0.05$, respectively, the fitted slopes in Figs.~\ref{fig:si-qae-agents}\textbf{a--c} were $(0.156,0.151)$ for Claude Code, $(0.159,0.149)$ for DeepCode, and $(0.149,0.142)$ for HGM. All six slopes fall within the revised acceptance interval.

\subsubsection{Level II (Paper~\#42): Underspecified implementation details in topological phase characterization}\label{append:sec:paper42}
Paper~\#42~\cite{che2026topological}, published in \textit{Nature Communications} in 2026, investigates the unsupervised identification of topological order, an important task in quantum many-body physics. The paper proposes two quantum kernels based on reduced-state fidelities and entanglement entropies to distinguish topologically ordered states from topologically trivial states.  In particular, they examined the robustness of these kernels under random two-qubit unitary circuits of increasing depth. However, the initial reproduction experiments failed to recover the reported clustering performance despite following the methodological description in the paper. Subsequent AI-assisted auditing identified several consequential differences between the published description and the released implementation, which required domain-expert intervention to resolve.

\begin{figure}
\centering
\includegraphics[width=0.98\textwidth]{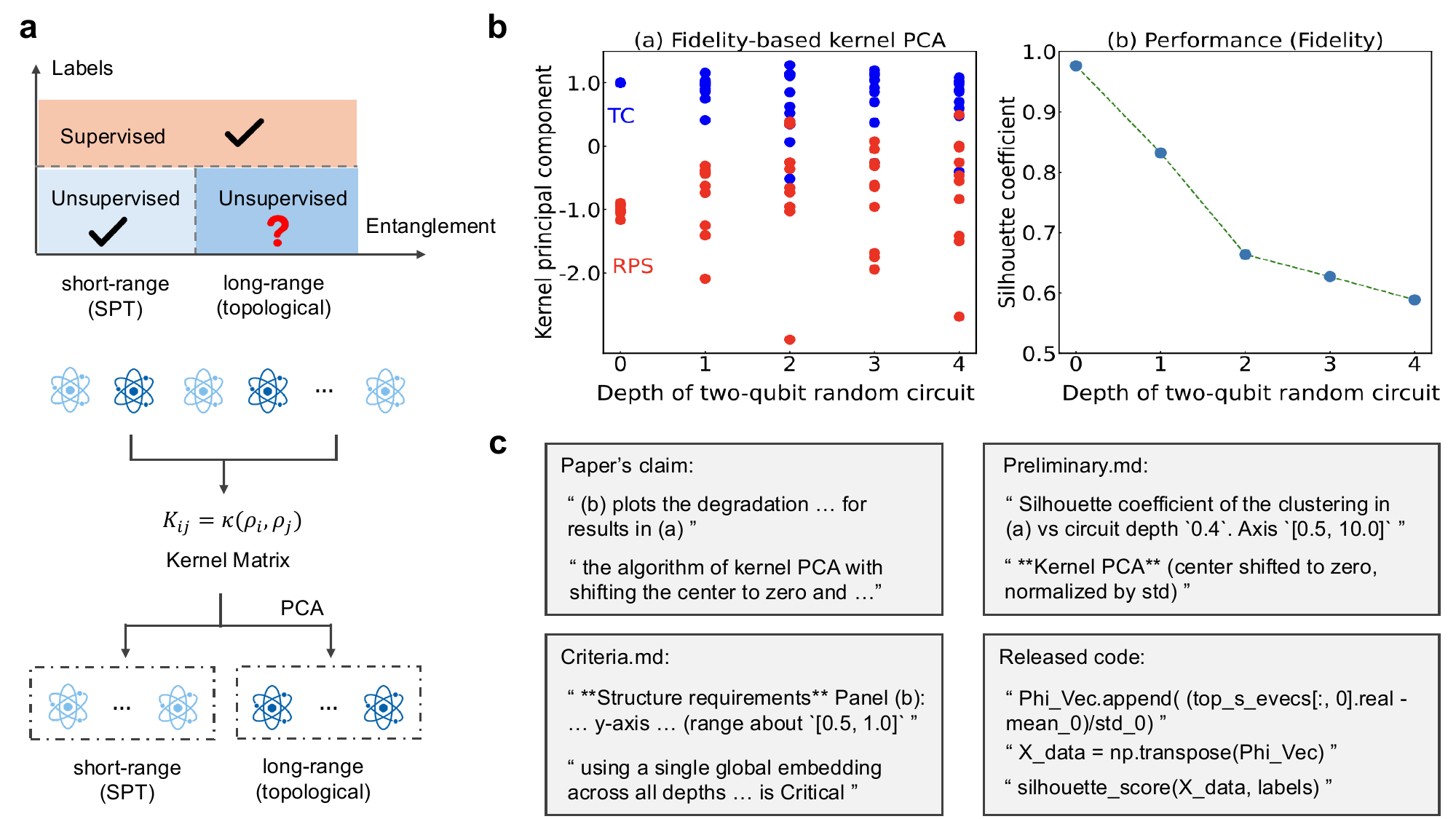}
\caption{\textbf{Overview of the reproduction task in Paper~\#42 for topological phase characterization.} \textbf{a.} Schematic of quantum-kernel-based unsupervised learning for topological phase characterization. Pairwise similarities between quantum states are used to construct a kernel matrix, followed by kernel principal component analysis (kernel PCA) to obtain low-dimensional representations for clustering states belonging to distinct quantum phases.  \textbf{b.} Figure~3(a)\&(b) from Paper~\#42~\cite{che2026topological}. The left panel illustrates the separation of Kitaev's toric code (TC) ground states and random product states (RPS) using the fidelity-based quantum kernel under random unitary perturbations, while the right panel quantifies the degradation in clustering performance with increasing circuit depth using the silhouette coefficient. \textbf{c.} Comparison of the methodological descriptions in the original paper, the task specification (\texttt{Preliminary.md}), the evaluation criteria (\texttt{Criteria.md}), and the authors' released implementation.}
\label{fig:si-tpc-evidence}
\end{figure}

\medskip
\noindent\textbf{Algorithmic background.}
Topological phases are characterized by nonlocal properties, such as long-range entanglement, that cannot generally be distinguished using conventional local order parameters. Unsupervised topological phase characterization aims to identify different quantum phases by clustering quantum states without using their phase labels during learning. As shown in Fig.~\ref{fig:si-tpc-evidence}\textbf{a}, given a dataset of $N$ quantum states $\{\rho_i\}_{i=1}^N$, 
kernel-based methods construct a similarity matrix $K$ with entries $K_{ij}=\kappa(\rho_i,\rho_j)$, where $\kappa$ denotes a quantum kernel. Kernel principal component analysis (PCA) then maps these states into a low-dimensional representation that can reveal their underlying clustering structure. To evaluate robustness against random unitary perturbations, let $\rho_i(d)$ denote the $i$-th state after applying a random circuit of depth $d$, and let $K(d)$ be the corresponding kernel matrix. Under a standard centered kernel-PCA procedure, the kernel matrix is first centered as $K_c(d)=HK(d)H$, where $H=\mathbb{I}-\boldsymbol{1}\boldsymbol{1}^{\mathsf T}/N$. The leading eigenvector $\bm v_1(d)$ of $K_c(d)$ is then standardized to obtain a low-dimensional representation $\bm z(d)\in\mathbb{R}^N$. Its clustering performance can be evaluated using the silhouette coefficient~\cite{rousseeuw1987silhouettes}, i.e.,
\begin{equation}\label{eq:sil_coefficient}
    s(d)=\operatorname{Silhouette} \left(\bm{z}(d),\bm{y}(d) \right),
\end{equation}
where $\bm{y}(d)=[\bm{y}_1(d),\cdots,\bm{y}_N(d)]$ contains the phase labels at depth $d$ used for evaluation. A larger silhouette coefficient indicates more compact and better-separated clusters.

\medskip

\noindent \textbf{Task construction with \omniQ.} Figure~3 of Paper~\#42 is selected as the reproduction target to evaluate the AI scientist's ability to reproduce the reported clustering performance and its robustness against random circuit perturbations.  Two panels from this figure are shown in Fig.~\ref{fig:si-tpc-evidence}\textbf{b}. That is, the left panel, i.e., Fig.~3(a) of Paper~\#42, illustrates the ability of the fidelity-based quantum kernel to distinguish the ground states of Kitaev's toric code (TC) from random product states (RPS) under random unitary perturbations. The right panel, i.e., Fig.~3(b) of Paper~\#42, quantifies the degradation of clustering performance using the silhouette coefficient, which remains above $0.55$ at depth $d=4$ in the published results. These panels support the paper's claim that entanglement-based kernels exhibit greater robustness to random unitary perturbations than fidelity-based kernels.

Following the methodological description in the paper, as illustrated in Fig.~\ref{fig:si-tpc-evidence}\textbf{c}, the task specification in \texttt{Preliminary.md} requires standard kernel PCA and a dataset of $N=20$ quantum states. Although the paper did not explicitly specify whether clustering evaluations at different circuit depths were performed independently or jointly, Fig.~3(a) of Paper~\#42 presents separate one-dimensional representations for each depth, while Fig.~3(b) is described as quantifying their clustering performance. Accordingly, the initial task specification adopted a depth-wise evaluation protocol, consistent with the interpretation introduced above. As illustrated in Fig.~\ref{fig:si-tpc-evidence}\textbf{c}, the evaluation criteria also required the reproduced silhouette plot to retain the original vertical-axis range of $[0.5,1]$. Although no strict numerical tolerance was imposed on the silhouette coefficients at every depth, this plotting requirement implicitly assumed that the reproduced values would remain within the displayed range.

\begin{figure}
\centering
\includegraphics[width=0.98\textwidth]{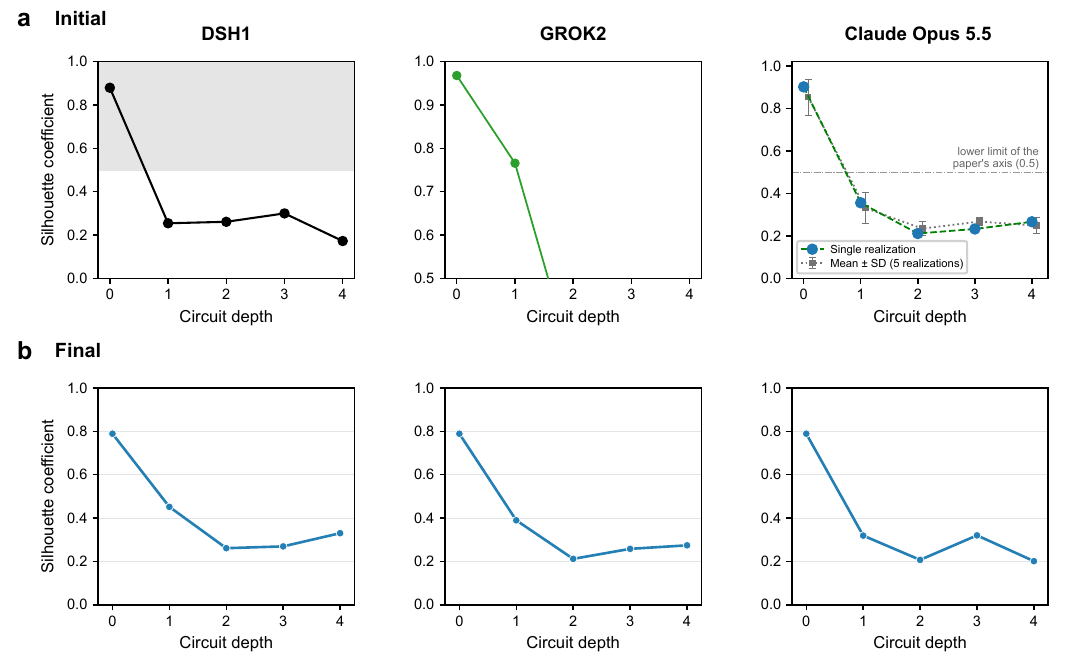}
\caption{\textbf{Reproduction results for Paper~\#42 on unsupervised learning for topological phase characterization.} \textbf{a}. Initial benchmark results under the depth-wise protocol and the original $[0.5,1]$ axis requirement. \textbf{b}. Final benchmark results under the revised specification, which retains independent analysis at each depth and expands the permitted $y$-axis range to $[0,1]$. Columns show DSH with DeepSeek V4 Pro, Grok2, and Claude Code with Opus 5.5, from left to right. Each main curve shows the fidelity-kernel silhouette coefficient computed from 10 TC and 10 RPS samples at each circuit depth, corresponding to the target quantity in original Figure~3(b). Initial panels retain the agents' plotting styles and $y$-axis ranges: $[0,1]$, $[0.5,1]$, and $[0,1.02]$, respectively. Consequently, Grok2's values at depths 2--4 fall outside its displayed window. Grey shading in the DSH panel and the dash-dotted line in the Opus panel mark the original paper's display window and its lower limit, respectively. The Opus initial panel additionally shows the mean and population standard deviation over five independent realizations, and the main curves show single realizations.}
\label{fig:si-tpc-agents}
\end{figure}

\medskip
\noindent\textbf{Initial benchmark.}
As shown in Fig.~\ref{fig:si-tpc-agents}\textbf{a}, we present the initial reproduction results of three representative AI scientists: DSH with DeepSeek V4 Pro, a Grok-based agent, and Claude Code with Opus 5.5. All three followed the prescribed depth-wise protocol in \texttt{Preliminary.md}, independently constructing the fidelity-kernel representation and calculating the silhouette coefficient at each circuit depth. However, their results exhibited a substantially stronger degradation in clustering performance than reported in the original paper. At depth $d=4$, the reproduced silhouette coefficients were approximately $0.173$, $0.283$, and $0.267$, respectively, all falling outside the displayed $y$-axis range of $[0.5,1]$ in the original figure.

The agents adopted in \omniQ responded differently to this discrepancy. The Grok-based agent keeped the prescribed vertical-axis range of $[0.5,1]$, causing the silhouette values at depths 2--4 to fall outside the visible plotting window. In contrast, DSH and Claude Code with Opus 5.5 expanded the axis range to display their computed values. These different treatments also led to inconsistent evaluation outcomes: clipped or missing curves were reported as critical issues in some runs, whereas axis extensions were accepted as non-critical deviations in others. Notably, the initial judge accepted the Claude Code reproduction, despite its lower silhouette values and modified axis range. These outcomes exposed a discrepancy between faithfully executing the specified computational procedure and reproducing the appearance of the published figure, motivating further examination of the original implementation and evaluation criteria.

\medskip
\noindent\textbf{Cross-agent verification of reproduction inconsistencies.}
To identify the source of these discrepancies, the auditing AI scientist in \omniQ examined how the authors' released implementation evaluated clustering performance across different circuit depths. As illustrated in Fig.~\ref{fig:si-tpc-evidence}\textbf{c}, the released implementation combines kernel-PCA representations from all depths up to $d$, contrary to the requirement in \texttt{Preliminary.md} that each depth be evaluated independently. Specifically, instead of computing the silhouette coefficient from the depth-specific representation $\bm z(d)$ as in Eq.~\eqref{eq:sil_coefficient}, the released implementation constructs an accumulated feature matrix $X=[\bm{z}(0),\cdots,\bm{z}(d)]$ to calculate
\begin{equation}
    s(d)=\operatorname{Silhouette} \left(X,\bm{y} \right), \nonumber 
\end{equation}
where $\bm{y}=[\bm{y}(1),\cdots,\bm{y}(d)]$ denotes the cluster labels obtained by applying K-means to the accumulated feature matrix $X$. This procedure preserves information from earlier depths when evaluating clustering performance at later depths, potentially obscuring the degradation caused by increasingly deep random circuits.

The AI scientist further identified a difference in the kernel-PCA implementation. Specifically, the released code of Paper~\#42 diagonalizes the uncentered kernel matrix and subsequently standardizes its leading eigenvector, instead of centering the kernel before eigendecomposition. Considering that these operations are not equivalent, this additional difference can also affect the resulting clustering representations and silhouette coefficients. Together, these implementation choices explain \textit{why following the initial task specification does not necessarily reproduce the published numerical results}.

Following the AI-assisted identification of these discrepancies, domain experts determined that the independent-depth protocol should be retained to preserve the stated correspondence between Fig.~3(a) and Fig.~3(b) of Paper~\#42. Accordingly, the revised \texttt{Preliminary.md} explicitly requires centered kernel PCA and independent silhouette evaluation for the 20 quantum states at each depth. Meanwhile, the evaluation criteria in \texttt{Criteria.md} were revised to allow a $y$-axis range of $[0,1]$ and assess the relative robustness of the two kernels without requiring the published fidelity-kernel silhouette magnitudes.  

Accordingly, this issue was classified as Level II because the discrepancy between the published description and released implementation affected the experimental methodology and could not be resolved through straightforward corrections, requiring domain experts to determine the appropriate reproduction protocol.

\medskip
\noindent\textbf{Final benchmark.}
With the clarified depth-wise evaluation protocol and revised criteria, the final benchmark assesses whether AI scientists can recover the reported clustering behavior under a consistent experimental setting via \omniQ. 

As shown in Fig.~\ref{fig:si-tpc-agents}\textbf{b}, all three AI scientists reproduced the degradation of fidelity-kernel clustering performance with increasing circuit depth. Specifically, the silhouette coefficients decreased from approximately $0.789$ at depth $d=0$ to $0.330$, $0.274$, and $0.201$ at $d=4$ for DSH with DeepSeek V4 Pro, the Grok-based agent, and Claude Code with Opus 5.5, respectively. Meanwhile, the corresponding entanglement-kernel results remained above $0.98$ at nonzero depths, recovering the reported qualitative advantage of the entanglement-based kernel under random circuit perturbations. All three agents passed the revised evaluation for Fig.~3 of Paper~\#42, although their fidelity-kernel silhouette coefficients remained substantially below the values reported in the original paper. These results demonstrate that the expert-revised benchmark enables faithful reproduction of the central scientific conclusion under a consistent depth-wise evaluation protocol, rather than exact recovery of the published numerical results.

\subsubsection{Level III (Paper~\#60): Unresolved inconsistencies in the reproduction of quantum kernel concentration}
\label{append:sec:paper60}

Quantum kernel methods measure similarities between classical inputs using quantum states prepared by a data-encoding circuit. Paper~\#60~\cite{thanasilp2024exponential}, published in \textit{Nature Communications} in 2024, identified several mechanisms that cause kernel values to concentrate exponentially with the number of qubits, increasing the measurement cost of resolving their differences. The authors also investigated how encoding expressivity affects kernel concentration through numerical simulations with different qubit counts and circuit depths. Here, we examine the reproduction of this experiment, which identified an inconsistency in the plotted statistic and a separate discrepancy in the underlying data that remained unresolved.

\begin{figure}[!htbp]
\centering
\includegraphics[width=0.98\textwidth]{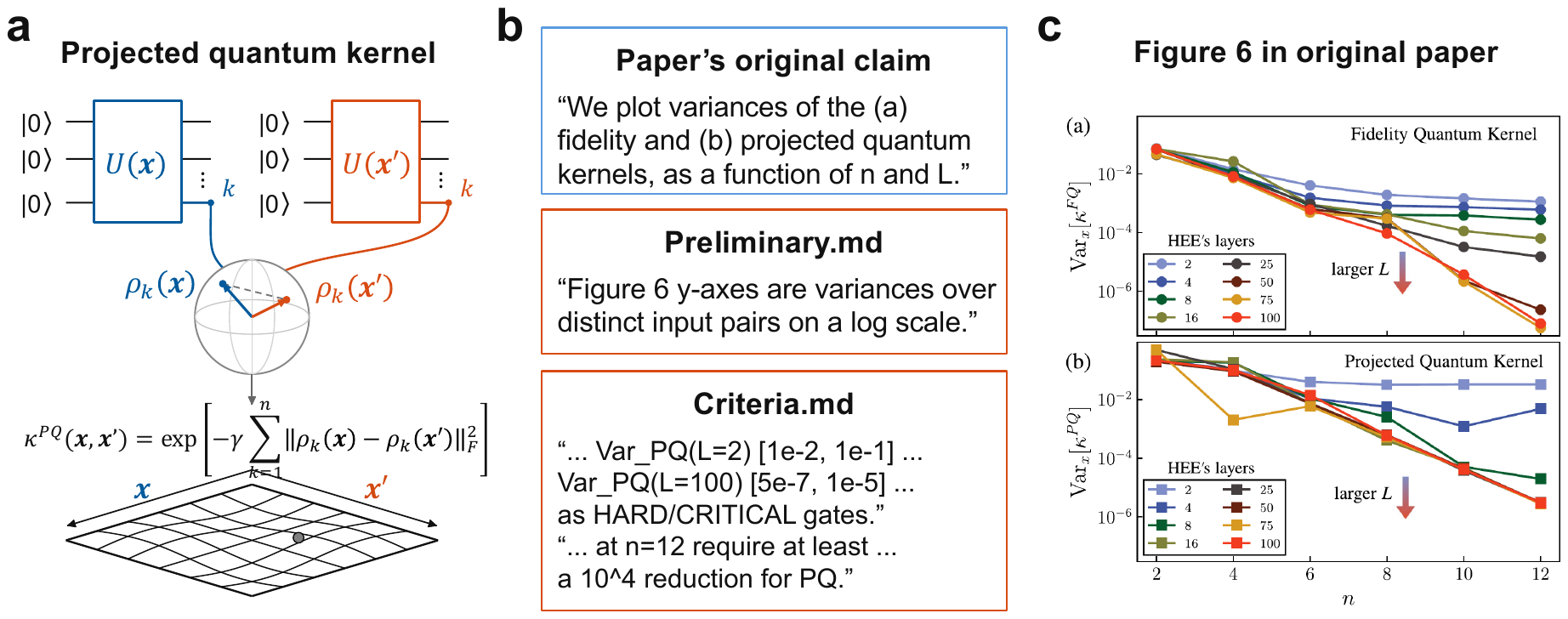}
\caption{\textbf{Reproduction task for Paper~\#60 on quantum kernel concentration.}
\textbf{a.} Schematic of the projected quantum kernel. Two encoding circuits prepare $n$-qubit states for inputs $\bm{x}$ and $\bm{x}'$. The corresponding quantum projected kernel follows from Eq.~\eqref{eq:si-kernel-definition}. The lower surface illustrates kernel values over input pairs.
\textbf{b.} Excerpts from the figure caption in the original paper (\texttt{Paper's original claim}), task specification (\texttt{Preliminary.md}), and evaluation criteria (\texttt{Criteria.md}). The magnitude windows apply to $n=12$ qubits.
\textbf{c.} Original Figure~6 of Paper~\#60~\cite{thanasilp2024exponential}, showing the reported variance curves for the fidelity and projected quantum kernels versus qubit count. The experiment uses 40 MNIST images of digits 0 and 1, reduced by PCA to $n=2,4,6,8,10,12$ features, with $L=2,4,8,16,25,50,75,100$ hardware-efficient encoding layers.}
\label{fig:si-kernel-evidence}
\end{figure}

\medskip
\noindent\textbf{Algorithmic background.} 
For a classical input $\bm{x}$, the encoding circuit $U(\bm{x})$ prepares $\rho(\bm{x})=U(\bm{x})\rho_0 U(\bm{x})^\dagger$ from an $n$-qubit initial state $\rho_0$. Given a dataset $\{\bm{x}^{(i)}\}$, the kernel Gram matrix has entries $K_{ij}=\kappa(\bm{x}^{(i)},\bm{x}^{(j)})$. The fidelity quantum kernel measures the overlap between the full encoded states, $\kappa^{{FQ}}(\bm{x}^{(i)},\bm{x}^{(j)})={\rm Tr}[\rho(\bm{x}^{(i)})\rho(\bm{x}^{(j)})]$. The projected quantum kernel compares the single-qubit reduced states $\rho_k(\bm{x})=\operatorname{Tr}_{\ne k}\rho(\bm{x})$, as illustrated in Fig.~\ref{fig:si-kernel-evidence}\textbf{a}:
\begin{equation}
C_{ij}=\sum_{k=1}^{n}\bigl\|\rho_k(\bm{x}^{(i)})-\rho_k(\bm{x}^{(j)})\bigr\|_F^2,
\qquad
\kappa^{{PQ}}_{ij}=\exp(-\gamma C_{ij}).
\label{eq:si-kernel-definition}
\end{equation}
Here, $\gamma>0$ is a constant and $\|\cdot\|_F$ denotes the Frobenius norm. The encoding circuit used in the Paper~\#60 consists of $L$ hardware-efficient encoding (HEE) layers~\cite{thanasilp2023subtleties}, each applying data-dependent $RX$ rotations followed by nearest-neighbor entangling gates.
For the $B=40$ inputs used in the experiment, let $\mathcal P=\{(i,j):1\leq i<j\leq B\}$ denote the $M=|\mathcal P|=780$ unordered pairs of distinct inputs. The empirical variance for either $Z=C$ or $Z=\kappa^{{PQ}}$ is defined as
\begin{equation}
\operatorname{Var}_{\mathcal P}[Z]
=\frac{1}{M}\sum_{(i,j)\in\mathcal P}(Z_{ij}-\bar Z)^2,
\qquad
\bar Z=\frac{1}{M}\sum_{(i,j)\in\mathcal P}Z_{ij},
\label{eq:si-kernel-variance}
\end{equation}
where the figure axes abbreviate the variance as $\operatorname{Var}[Z]$.

\medskip
\noindent\textbf{Task construction with \omniQ.}
The reproduction task targets the complete Figure~6 of Paper~\#60, as shown in Fig.~\ref{fig:si-kernel-evidence}\textbf{c}. Reproducing it requires classical simulation of the fidelity and projected quantum kernels across qubit counts $n=2, 4, 6, 8, 10, 12$ and circuit depths $L=2, 4, 8, 16, 25, 50, 75, 100$, followed by variance calculations over distinct input pairs. The task tests both the implementation of the encoding and reduced-state calculations and the reported dependence of kernel concentration on qubit count and circuit depth.

The task specification required agents to generate the kernel values independently and document the preprocessing, input selection, random seeds, entangling gates, and kernel parameter. 
For settings not specified in the paper, it recommended selecting 20 images of each digit (0 and 1) as the input subset, using a CNOT ring as the entangling pattern, and setting the kernel parameter to $\gamma=1$. The \texttt{Criteria.md}, summarized in Fig.~\ref{fig:si-kernel-evidence}\textbf{b}, specified absolute acceptance windows for the quantity labeled as projected-kernel variance at $n=12$: $[10^{-2},10^{-1}]$ at $L=2$ and $[5\times10^{-7},10^{-5}]$ at $L=100$. It also required a reduction of at least $10^4$ between these depths. The absolute windows applied only under the recommended settings. When agents documented alternative choices for parameters unspecified in the paper, these windows were treated as advisory, but the depth-reduction requirement remained critical.

\begin{figure}[!htbp]
\centering
\includegraphics[width=0.98\textwidth]{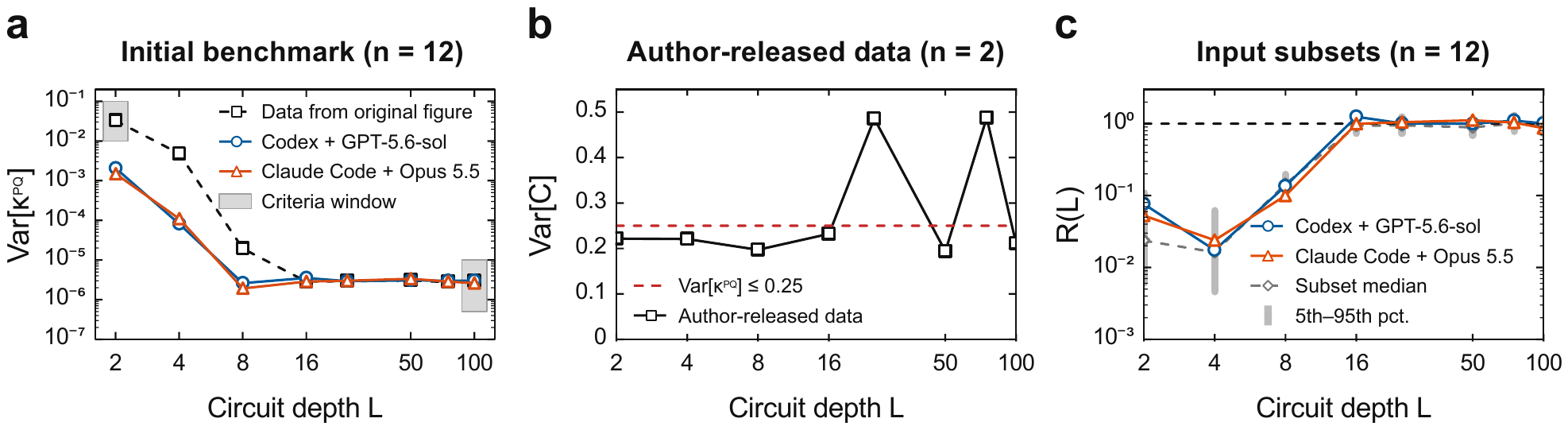}
\caption{\textbf{Initial benchmark and numerical checks of the projected-kernel result in Paper~\#60.}
\textbf{a.} Projected-kernel variance at $n=12$ across circuit depths for Codex with GPT-5.6-sol (blue circles), Claude Code with Opus~5.5 (orange triangles), and the values reported in original Figure~6b (black squares). Gray bands indicate the acceptance windows at $L=2$ and $L=100$.
\textbf{b.} Variance computed from the authors' released data for Figure~6b of Paper~\#60 at $n=2$. The red dashed line marks the upper bound $\operatorname{Var}[\kappa^{{PQ}}]\leq1/4$ for projected-kernel variance, as given in Eq.~\eqref{eq:si-kernel-bound}.
\textbf{c.} Normalized distance variances at $n=12$ from the two agent runs and 3,000 balanced subsets of 20 images per digit. Each normalized value, denoted by $R(L)$ on the axis, is the ratio of the corresponding distance variance to the reference variance computed from the authors' released data at the same depth. The subsets were drawn from an audit pool of 1,500 images per digit. Gray diamonds indicate the subset medians, and gray bars show the empirical 5th--95th percentile ranges. The dashed line marks $R=1$.}
\label{fig:si-kernel-checks}
\end{figure}

\medskip
\noindent\textbf{Initial benchmark.}
At $n=12$ and $L=2$, Codex with GPT-5.6-sol and Claude Code with Opus~5.5 each obtained projected-kernel variances below $3\times10^{-3}$, as shown in Fig.~\ref{fig:si-kernel-checks}\textbf{a}. Both values were more than an order of magnitude smaller than the value reported in the original paper and outside the acceptance window $[10^{-2},10^{-1}]$ specified in \texttt{Criteria.md}. At $L=100$, both results fell within the acceptance window. However, the variance decreased by less than a factor of $10^3$ between these depths in each run, falling short of the required $10^4$ reduction. Both runs therefore failed the initial numerical criteria. We then examined the statistic plotted in the original figure and the data used to generate it.

\medskip
\noindent\textbf{Cross-agent verification of reproduction inconsistencies.}
Equation~(\ref{eq:si-kernel-definition}) confines every projected-kernel value to $[0,1]$. Writing $\kappa=\kappa^{{PQ}}$ and taking averages over $\mathcal{P}$, we have
\begin{equation}
\operatorname{Var}_{\mathcal{P}}[\kappa]
={\mathbb{E}}_{\mathcal{P}} \left[ \kappa^2 \right] - \left( {\mathbb{E}}_{\mathcal{P}} \left[ \kappa \right] \right)^2 \leq {\mathbb{E}}_{\mathcal{P}} \left[ \kappa \right] - \left( {\mathbb{E}}_{\mathcal{P}} \left[ \kappa \right] \right)^2 = {\mathbb{E}}_{\mathcal{P}} \left[ \kappa \right] \left( 1- {\mathbb{E}}_{\mathcal{P}} \left[ \kappa \right] \right) \leq\frac{1}{4}.
\label{eq:si-kernel-bound}
\end{equation}
This bound holds for any finite collection of input pairs, regardless of the distribution $\mathcal{P}$. At $n=2$, however, the variance computed from the authors' released data for the original Figure~6b is approximately $0.5$ at both $L=25$ and $75$, as shown in Fig.~\ref{fig:si-kernel-checks}\textbf{b}. Both values exceed the upper bound of $1/4$ and should not represent the projected-kernel variance labeled in the original figure.
The released scripts explain this inconsistency. The function \texttt{proj\_kernel} in the authors' code returns the distance sum $C_{ij}$, while the plotting routine applies \texttt{np.var} directly to these distances and labels the result as projected-kernel variance.

We compared the distance variances $\operatorname{Var}[C]$ obtained by Codex with GPT-5.6-sol and Claude Code with Opus~5.5 with those calculated from the data released by the authors at $n=12$, as shown in Fig.~\ref{fig:si-kernel-checks}\textbf{c}. The ratio $R(L)$ is defined as the variance from each run divided by the reference variance at the same depth, with values close to $1.0$ indicating agreement with the reference results. For both agents, $R(L)<10^{-1}$ at $L=2$ and $4$, and approaches $1.0$ for $L\geq16$. To examine whether input selection could explain these differences, we calculated distance variances for 3,000 balanced subsets of 20 images per digit and normalized them by the reference variance at each depth. At both $L=2$ and $4$, the reference variance exceeds the 95th percentile of the sampled subsets. These results suggest that random input selection is unlikely to explain the shallow-depth discrepancies under the settings examined.

\medskip
\noindent\textbf{Final benchmark.}
Expert review classified the case as Level~III because \textit{the plotted-statistic mismatch could be explained from the released scripts, but no reproducible procedure was identified for generating the reference variances}.  Consequently, we exclude this task from the cleaned computational quantum science dataset used to evaluate the reproduction capabilities of different AI scientists.

\section{Implementation of \RA}\label{append:sec:ra-impl}
In this section, we provide the implementation details of \RA, whose overall workflow is illustrated in Fig.~1\textbf{b} in the main text. We begin with an overview of the framework of \RA in SI.~\ref{sec:overview}, followed by a detailed description of the implementation of its components in SI.~\ref{append:sec:planner}--\ref{append:sec:ra-distiller}.

\subsection{Overview}\label{sec:overview}
\RA consists of five functional components: a \textit{planner}, an \textit{executor}, a \textit{critic}, a \textit{judge agent}, and an \textit{experience distiller}. These components collaborate to reconstruct computational workflows from scientific papers, validate reproduction results, resolve implementation issues, and accumulate reusable experience. Specifically, the planner, executor, and critic form the core reproduction workflow, while the judge agent independently evaluates the completed reproduction and the experience distiller extracts generalizable lessons from verified feedback.

Given a scientific paper, a target figure, and the corresponding task specification, the planner decomposes the reproduction task into a set of sub-plans with explicit dependencies, represented by a \textit{dependency graph}. Each node corresponds to a computational sub-plan, while each edge specifies a dependency between sub-plans. The planner also establishes shared methodological interpretations and assigns completion requirements to individual sub-plans to maintain consistency throughout execution.

Following this plan, the executor and critic interact iteratively to implement and validate the sub-plans. The executor performs the required computations, while the critic independently checks the resulting implementations and outputs against the prescribed methodology and task requirements. Based on its review, the critic determines whether a sub-plan can be accepted, requires local repair, or needs a new computational approach. This iterative process continues until the required computations are completed or the available computational budget is exhausted.

After execution, the judge agent independently evaluates the assembled reproduction results against the prescribed evaluation criteria, which are not accessible to the executor or critic. The resulting feedback, together with execution records and verified failures, provides the evidence from which the experience distiller extracts reusable lessons. During development, these lessons are reviewed by domain experts before being incorporated into the experience layer, which guides subsequent reproduction tasks without modifying the underlying base LLM or orchestration code. Through this process, \RA combines within-task issue resolution with experience-based self-improvement across tasks.

\begin{figure}[!htbp]
    \centering
    \includegraphics[width=0.98\textwidth]{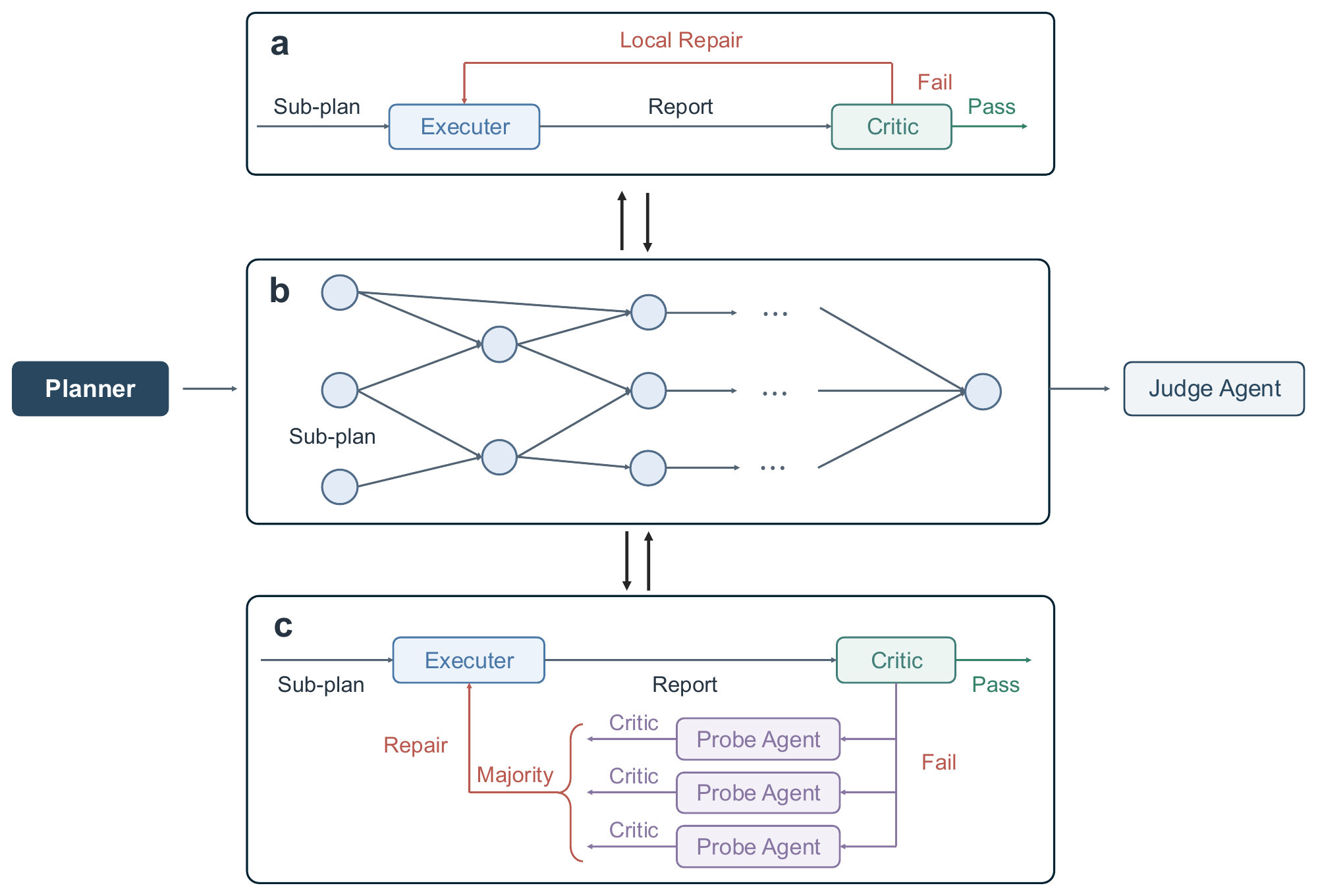}
    \caption{\small
    \textbf{Schematic of the planner, executor, and critic workflow.}
    The planner decomposes the reproduction task into interdependent sub-plans organized in a dependency graph, while the executor and critic collaborate to execute, review, and correct each sub-plan. \textbf{a.} Local repair mechanism, where the critic returns defective results to the executor for correction. \textbf{b.} Schematic of the dependency graph constructed by the planner to organize sub-plans and their execution dependencies.
    Each blue circle represents a sub-plan. \textbf{c.} Experiment-driven repair mechanism, where three independent probe agents investigate failures through small experiments, followed by a majority vote among three critics to select a new approach for execution.
    }
    \label{fig:si-schemetic-planener-ececutor-critic}
\end{figure}

\subsection{Planner}\label{append:sec:planner}

The planner is responsible for translating a paper-to-result reproduction task into a structured computational plan, which will be passed to the executor and critic for implementation and verification. Given the scientific paper, target figure, and task specification, it first identifies the computational requirements needed to reproduce the reported results and decomposes them into a set of sub-plans. Each requirement, such as a numerical value, parameter-dependent curve, or figure panel, is then assigned to exactly one sub-plan to ensure that all prescribed outputs are covered.

For each sub-plan, the planner sets a computational resource budget and defines completion criteria based on two sources. First, the planner establishes criteria based on explicit requirements in the task specification and derives additional criteria from the scientific claims to be reproduced when necessary. Moreover, these criteria can be further specified using additional information extracted from published figures to support self-validation without treating the plotted values as fitting targets.

The dependencies among these sub-plans are organized into a dependency graph, which determines their execution order and enables intermediate results to be shared across related computations. When the paper admits multiple methodological interpretations, the planner establishes a consistent interpretation shared by the relevant sub-plans, while preserving experimental settings explicitly fixed by the task specification. These shared interpretations may be revised when supported by new evidence, with the affected sub-plans subsequently revalidated. 

\subsection{Interactive mechanism of critic and executor}

The executor and critic form the core iterative mechanism of \RA for implementing, validating, and correcting computational sub-plans. As illustrated in Fig.~\ref{fig:si-schemetic-planener-ececutor-critic}, this mechanism operates on the dependency graph generated by the planner, where each node represents a computational sub-plan and each edge specifies a dependency between sub-plans. During execution, the critic evaluates the results of individual nodes and determines whether they can be accepted, require local repair, or need a new computational approach. These decisions guide the progression of the dependency graph, with additional diagnostic branches introduced when necessary. In this way, the reproduction process can be viewed as an adaptive execution graph that progressively resolves computational sub-plans through validation and issue resolution.

\medskip
\noindent\textbf{Sub-plan execution and validation.}
Following the dependency graph, the executor implements each sub-plan once its prerequisite nodes have been accepted. Each executor operates in an independent session and receives the corresponding task requirements, shared methodological interpretations, and outputs from preceding sub-plans. Before performing full-scale computations, the executor validates its implementation on the smallest exactly solvable instance and estimates the computational cost by timing a representative step at the largest required system size. These checks help identify implementation errors and assess computational feasibility before committing substantial resources. For computationally demanding tasks, long-running calculations are launched as background jobs tracked by the orchestrator, allowing the executor session to terminate without interrupting the computation. The orchestrator subsequently resumes the executor when the job completes or stalls, without requiring continuous calls to the base LLM. To ensure reproducibility, all reported numerical results must be traceable to the corresponding code and data saved on disk.

\medskip
\noindent\textbf{Critic review and decision routing.}
The critic independently reviews the computational results generated by each executor before accepting the corresponding sub-plan. Unlike the judge agent, which evaluates the completed reproduction against the benchmark criteria, the critic only accesses the scientific paper, task specification, and execution records. Its review focuses on three aspects: (i) \textit{result provenance}, verifying that reported numerical values are supported by the generated code and data; (ii) \textit{methodological correctness}, checking whether the implementation follows the prescribed algorithms and experimental settings; and (iii) \textit{task completion}, assessing whether all requirements of the sub-plan have been satisfied. These checks are primarily conducted through inspection, supplemented by small verification experiments when necessary. The critic can also inspect ongoing computations and identify potential failures before they finish.

Based on these checks, the critic assigns each sub-plan to one of three outcomes: (i) \textit{Pass}, when all requirements are satisfied and no verified defects remain, allowing dependent sub-plans to proceed; (ii) \textit{Local repair}, when an identified defect can be corrected within the remaining computational budget; and (iii) \textit{New approach}, when the defect cannot be reliably resolved through local correction or requires a different computational strategy. A passing node is marked as completed, making its validated outputs available to dependent nodes in the graph. In contrast, the latter two outcomes trigger different corrective branches, as described below.

\smallskip
\noindent\textit{Local repair.}
When a defect has a known correction, the critic returns the corresponding sub-plan to the executor with specific feedback identifying the error and the required changes. The executor then modifies the existing implementation and resubmits its results for another critic review. This procedure allows the same node to be corrected without changing its dependencies or restarting the entire computational workflow. The repair process continues until the sub-plan passes the review or further local correction is deemed ineffective. Importantly, results containing fabricated or untraceable numerical values are not eligible for ordinary repair, since their computational provenance cannot be established.

\smallskip
\noindent\textit{Experiment-driven diagnosis and new approach.}
When local repair is insufficient, either because the source of the defect remains unclear or because repeated corrections have failed, the critic initiates an experiment-driven diagnosis. This process introduces three temporary diagnostic branches associated with the failed node, each handled by an independent \textit{probe agent}. These agents examine the failed sub-plan and its residual, consisting of verified defects and reusable computational artifacts. Each probe agent independently proposes a possible cause of the failure and tests its hypothesis through a small numerical experiment.

The resulting diagnoses are evaluated by a \textit{critic panel} consisting of three independent critic sessions. Before selecting a solution, the panel excludes proposals that violate fixed task requirements, substitute prescribed methodological components, or rely solely on improved agreement with the published figure without independent validation. The remaining diagnoses are ranked by majority vote, and the highest-ranked diagnosis forms the basis of a new computational approach, potentially incorporating compatible elements from other proposals. A fresh executor then implements this approach using the archived code and residual, with its results subjected to another round of critic review. These diagnostic branches support the resolution of the original failed node without changing the scientific requirements or dependency structure of the reproduction task.

\medskip
\noindent\textbf{Dependency-aware revalidation.} The three-way decision mechanism allows \RA to progressively complete the reproduction task specified by the dependency graph while correcting issues at individual nodes. Beyond these local decisions, changes to shared methodological interpretations may affect multiple previously accepted sub-plans. When such a change is supported by new evidence, the orchestrator reopens the sub-plans that used the shared information, and the critic reassesses their validity. Results that remain correct are retained, while affected computations undergo local repair or a new approach. This dependency-aware revalidation maintains consistency across the computational workflow without unnecessarily repeating completed work.

In exceptional cases where the prescribed requirements of a sub-plan are contradictory or cannot be satisfied, the critic may retire the affected node. If this prevents subsequent sub-plans from proceeding, the planner may revise the dependency graph to determine whether the remaining requirements can still be completed without violating the task specification. Throughout execution, all components operate under a shared computational budget, with part of the available time reserved for saving results and preparing the final report. Once all required sub-plans have passed the critic's review, their outputs are assembled into the final reproduction artifacts for independent evaluation by the judge agent. Together, these procedures enable \RA to combine independent validation, iterative issue resolution, and coordinated revisions within a reliable computational reproduction workflow.

\subsection{Judge agent}
\label{append:sec:ra-judge}

The Judge agent is used to evaluate the final reproduction results generated by \RA. Once the reproduction run is complete, it receives the task specification, evaluation criteria, and execution workspace, including the source code, logs, numerical data, figures, and final report. Only the Judge has access to the evaluation criteria, while the Planner, Executor, and Critic do not receive it during task execution.

For each target figure, the Judge first checks that all required panels and curves are present over the specified parameter range. It then traces the plotted values to the saved data and code to verify how they were calculated. The implementation is checked against the task specification, including the definitions of observables, experimental parameters, and required computational steps. Any permitted approximations or choices of convention are checked against the description in the report. The Judge then compares the saved numerical results with the prescribed checkpoints and assesses whether the curves reproduce the expected performance rankings and trends.

The Judge records any issues found in this assessment together with evidence from the code, data, or execution logs. Each issue is classified as \textit{critical} or \textit{non-critical} under the \omniQ protocol. Critical issues affect scientific correctness or leave a required part of the reproduction incomplete, such as an incorrect computational procedure, an unsupported numerical value, a missing required curve, or a result outside a mandatory tolerance. Non-critical issues concern presentation differences, such as colors or legend placement, that leave the scientific content unchanged. A figure passes when no critical issue is found. A task is counted as successful when all its target figures pass.

\begin{figure}
\centering
\includegraphics[width=\textwidth]{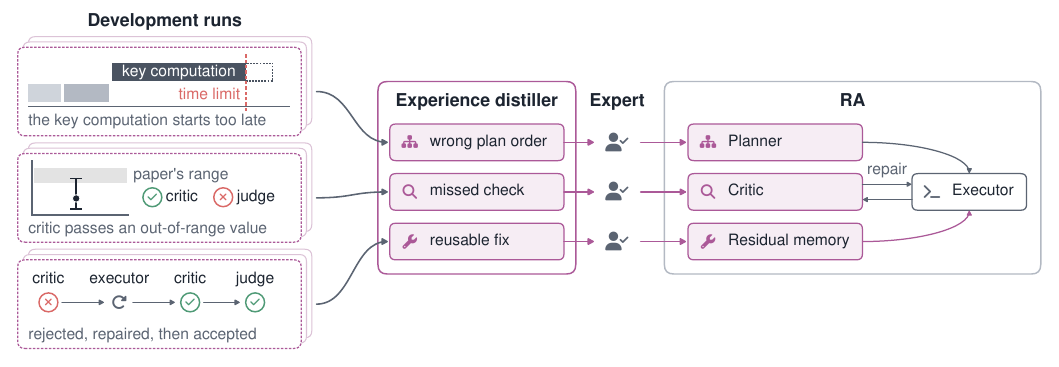}
\caption{\small\textbf{The scheme of the experience distiller.} The distiller identifies common patterns across development runs and transforms them into reusable experience, which is incorporated into the corresponding agent memories after human expert approval. Left, three representative patterns identified across multiple runs, with stacked copies indicating that similar patterns appear in other runs. Middle, the distiller evaluates each pattern and determines which agent role can benefit from it. Right, the workflow of \RA, where purple highlights components improved by accumulated experience. In particular, residual memory guides the executor in repairing failed steps. All runs are illustrated schematically.}
\label{fig:si-ra-distiller}
\end{figure}

\subsection{Experience distiller}
\label{append:sec:ra-distiller}
The experience distiller enables \RA to learn from previously encountered failures by converting verified issue resolutions into reusable guidance for subsequent reproduction tasks. It operates outside individual reproduction runs and is invoked after each round of development experiments. Importantly, the distiller updates only the experience layer, while the base LLM, agent roles, and orchestration code remain unchanged. To extract useful experience, it examines the judge agent's feedback and the complete execution records, including the computational plans, critic reviews, and implementations of individual sub-plans.

The distilled experience is organized into three types of memory: \textit{planner memory}, \textit{critic memory}, and \textit{residual memory}, as illustrated in Fig.~\ref{fig:si-ra-distiller}. The planner memory provides guidance for constructing reliable computational plans, such as scheduling validation on small, exactly solvable instances before full-scale computation and prioritizing tasks essential to the paper's central claims. The critic memory improves the identification of implementation issues by emphasizing checks on numerical provenance, methodological correctness, and consistency with the reported results. The residual memory stores generalizable failure signatures, each consisting of three elements: a description of the failure, a diagnostic check for identifying it, and a corresponding remedy. These checks are derived from how the critic or judge agent identified the issue, while the remedies are extracted from verified corrections that subsequently led to successful reproduction.

During task execution, relevant entries in the residual memory are retrieved to guide issue resolution. For example, when the critic requests a local repair, it identifies the entries corresponding to the verified defect, which are then provided to the executor to guide the correction. Similar guidance is supplied to a fresh executor when a new computational approach is required. This mechanism allows experience accumulated from previous failures to inform subsequent decisions without modifying the underlying execution framework.

To maintain the reliability and generalizability of the accumulated experience, the distiller excludes lessons specific to individual papers and merges those that overlap with existing entries. It can propose additions, modifications, or deletions across all three memories, but every update must be reviewed and approved by a human expert before being incorporated into the experience layer. In our implementation, the memories were developed exclusively from five development tasks, resulting in ten entries in the residual memory. The resulting experience layer was then fixed throughout evaluation on the remaining 54 unseen tasks, allowing us to assess whether the distilled experience generalizes beyond the tasks used for its development.

\section{An autonomous platform for quantum computational science}\label{append:sec:platform}

 To make computational quantum science more accessible and reusable, we develop an autonomous research platform that integrates \omniQ in SI~\ref{SI:secB-omniQ}, \RA in SI~\ref{append:sec:ra-impl}, and the Model Context Protocol (MCP). The platform enables researchers and AI agents to reproduce published results and build upon previously verified computational methods. Specifically, it provides two core functionalities:
\begin{enumerate}
\item \textit{Reproducing new research}. Automatically reproducing computational results from user-submitted scientific papers.
\item \textit{Reusing existing algorithms}. Discovering and applying previously reproduced algorithms through simple MCP queries.
\end{enumerate}
Below, we briefly describe how these two functionalities are implemented and accessed. More details are provided in \url{https://omniq-qid-lab.github.io/}.

\medskip
\noindent\textbf{Reproducing new research.}
Building on the \omniQ framework, the platform integrates task specification, AI-driven computation, independent verification, and result delivery into a unified workflow. In particular, a user begins by submitting a scientific paper and defining a research objective, such as reproducing a published figure or investigating the same method under new experimental parameters. An analysis agent translates this request into two documents: a task specification that guides the AI scientist and evaluation criteria that define the requirements for successful reproduction. The user reviews and confirms both documents before execution, while the evaluation criteria remain inaccessible to the AI scientist.

Once the task is confirmed, the user selects an AI scientist, such as \RA or a first-party coding agent, together with the underlying LLM and computational budget. The platform executes the task in an isolated computational environment and independently evaluates the resulting code, data, and figures against the predefined criteria. 

Upon completion, the user receives an evaluation report and can download the source code, generated data, reproduced figures, and a runnable container. This workflow enables researchers to inspect, reproduce, and extend computational results without reconstructing the experimental environment.

\medskip
\noindent\textbf{Reusing existing algorithms.}
Beyond reproducing individual papers, the platform organizes successfully reproduced algorithms into a shared collection of verified computational workflows. To establish their reliability, successful reproductions can undergo additional verification using a fixed judge configuration, with optional review by domain experts. 

The platform is designed to issue digitally signed certificates linking each verified workflow to its source paper, evaluation criteria, AI scientist configuration, and executable results. These certificates establish the reproducibility of the specified computational results without making claims about the scientific validity of the original paper.

Researchers and external AI agents can access these workflows through a web interface or MCP. In particular, MCP enables an AI agent to search for relevant algorithms using natural-language queries, retrieve their implementations and experimental settings, and incorporate them into new research tasks. As such, researchers can directly build upon existing, verified implementations, reducing redundant computational effort and facilitating subsequent scientific investigations.

\medskip
\noindent\textbf{Extensibility.}
The platform adopts a modular architecture that supports integration with different AI scientists, underlying LLMs, computational environments, and scientific evaluation protocols. 

Although initially developed for computational quantum science, the framework can be extended to other computational disciplines by adapting their task specifications, evaluation criteria, and computational resources. The 59 reproductions from this study provide the initial collection of workflows, which can expand as new papers are reproduced and verified. We have implemented a prototype web interface and MCP tool definitions, with deployment of the full public service planned.

\section{More experimental results}\label{sec:appendix-tables}

This section provides additional experimental results to further evaluate the capabilities and limitations of AI scientists in computational quantum science. In SI~\ref{append:sec:harness-classes}, we present additional results of different AI scientists on the established computational quantum science dataset. In SI~\ref{append:sec:ra-cases}, we examine representative case studies to understand the mechanism of \RA.

\subsection{Additional results of AI scientists on the established computational quantum science dataset}\label{append:sec:harness-classes}

To investigate how harness design influences the scientific reproduction capabilities of AI scientists, here we conduct a model-controlled comparison in which the underlying LLM is held fixed while the agent harness varies. In doing so, this comparison allows us to assess how alternative agent architectures and execution strategies affect reproduction performance on the cleaned computational quantum science dataset. 

The experimental settings are as follows. We use DeepSeek V4 Pro, a text-only model, to compare six third-party harnesses across three categories introduced in SI~A, i.e.,  general software harnesses with mini-swe-agent and live-swe-agent, scientific-reproduction harnesses with InternAgent and DeepCode, and evolved harnesses with HGM and Meta-Harness. We additionally include DeepSeek Harness (DSH), the model provider's first-party agent, as a reference. All third-party harnesses use their publicly released implementations and default configurations, except for the modifications described below. DSH operates with its full toolset, including shell commands, file operations, sub-agents, and background jobs.

The experimental settings are as follows. We use DeepSeek V4 Pro, a text-only model, to compare six third-party harnesses across three categories introduced in SI~\ref{append:sec:agents}: general software harnesses (i.e., mini-swe-agent~\cite{minisweagent2025} and live-swe-agent~\cite{xia2025livesweagent}), scientific-reproduction harnesses (i.e.,  InternAgent~\cite{internagent2025} and DeepCode~\cite{li2025deepcode}), and evolved harnesses (i.e., HGM~\cite{wang2025hgm} and Meta-Harness~\cite{lee2026metaharness}). We additionally include DeepSeek Harness (DSH), the model provider's first-party agent, as a reference. Specifically, DSH operates with its full toolset, including shell commands, file operations, sub-agents, and background jobs. Note that all third-party harnesses use their publicly released implementations and default configurations, except for the modifications described below. 

\begin{itemize}
    \item For general software harnesses, mini-swe-agent uses a single bash tool, with a limit of 250 steps and 1800\,s per command. live-swe-agent additionally supports tools constructed during execution, with 300 steps and 3600\,s per command.
    \item For scientific-reproduction harnesses, InternAgent operates in paper-reproduction mode with idea generation disabled. Its coding sessions use the OpenAI-compatible tool-calling interface, with initial limits of 70 steps and 45\,min per session and 65\,min per experiment run. For some tasks, the session limit is increased to 150 steps, with the time cap extended to the task-level limit. DeepCode follows its paper-to-code workflow, followed by execution of the generated code without its web interface. 
\item For evolved harnesses, HGM uses its released best agent with a single attempt instead of the default three and a 120\,s limit per bash call. Meta-Harness uses its bash, mark-complete, and image-read tools, with a 600\,s limit per command block.
\end{itemize}

\begin{table}[!htbp]
\caption{\small\textbf{Failure categories of AI scientists using DeepSeek V4 Pro.} 
The first row reports successful reproductions out of 59 tasks. Failure categories are non-exclusive, with each run counted once per applicable category; runs without reports are counted only under ``No report''. ``Reduced problem'' denotes reductions in problem size, sampling, evaluation points, or numerical precision, while ``Mismatch with the paper'' indicates discrepancies in reported values, rankings, or trends. The final two rows characterize the consistency of failed-run reports with their underlying results.}
\label{tab:tp-failures}
\footnotesize
\setlength{\tabcolsep}{4pt}
\begin{tabular}{@{}lccccccc@{}}
\toprule
\parbox[t]{0.24\textwidth}{\raggedright \strut} & DSH & \parbox[t]{0.08\textwidth}{\centering mini-swe-\\agent\strut} & \parbox[t]{0.08\textwidth}{\centering live-swe-\\agent\strut} & \parbox[t]{0.08\textwidth}{\centering Intern-\\Agent\strut} & \parbox[t]{0.08\textwidth}{\centering Deep-\\Code\strut} & HGM & \parbox[t]{0.08\textwidth}{\centering Meta-\\Harness\strut} \\[2pt]
\midrule
\parbox[t]{0.24\textwidth}{\raggedright Reproduced tasks\strut} & 29 & 28 & 18 & 16 & 21 & 24 & 11 \\
\parbox[t]{0.24\textwidth}{\raggedright Failed runs\strut} & 30 & 31 & 41 & 43 & 38 & 35 & 48 \\
\midrule
\parbox[t]{0.24\textwidth}{\raggedright No report\strut} & 6 & 0 & 1 & 13 & 0 & 1 & 1 \\
\parbox[t]{0.24\textwidth}{\raggedright Substitution or fabrication\strut} & 4 & 8 & 14 & 7 & 4 & 9 & 13 \\
\parbox[t]{0.24\textwidth}{\raggedright Reduced problem\strut} & 6 & 6 & 13 & 7 & 15 & 7 & 18 \\
\parbox[t]{0.24\textwidth}{\raggedright Implementation error\strut} & 9 & 13 & 22 & 13 & 12 & 10 & 11 \\
\parbox[t]{0.24\textwidth}{\raggedright Missing content\strut} & 2 & 3 & 7 & 4 & 10 & 6 & 6 \\
\parbox[t]{0.24\textwidth}{\raggedright Rendering error\strut} & 2 & 10 & 6 & 3 & 7 & 9 & 9 \\
\parbox[t]{0.24\textwidth}{\raggedright Mismatch with the paper\strut} & 24 & 25 & 37 & 30 & 34 & 28 & 42 \\
\midrule
\parbox[t]{0.24\textwidth}{\raggedright Report contradicts its data\strut} & 3 & 1 & 5 & 7 & 4 & 2 & 3 \\
\parbox[t]{0.24\textwidth}{\raggedright Report admits the gap\strut} & 16 & 18 & 21 & 12 & 21 & 23 & 31 \\
\bottomrule
\end{tabular}
\end{table}

\smallskip
Table~\ref{tab:tp-failures} summarizes the reproduction performance and failure patterns of different AI scientists on the cleaned computational quantum science dataset. To characterize these failures, we used Claude Opus~5.5 to classify 954 critical findings across 266 failed runs into predefined categories, without revealing harness identities. A repeated classification of 60 randomly sampled findings yielded 98.3\% agreement (59/60).

An interesting observation is that despite sharing the same underlying LLM, the seven harnesses exhibit substantial differences in reproduction performance. More concretely, DSH achieves the highest success rate, reproducing 29 of 59 tasks, closely followed by mini-swe-agent with 28 successful reproductions at approximately one-third of DSH's cost. In contrast, live-swe-agent reproduces only 18 tasks despite incurring roughly twice the cost of mini-swe-agent. Among the scientific-reproduction harnesses, DeepCode outperforms InternAgent, while HGM reproduces 24 tasks, exceeding both scientific-reproduction harnesses. Meta-Harness achieves the lowest performance, reproducing only 11 tasks.

To further characterize these differences, we analyze the critical issues identified in unsuccessful runs. As shown in Tab.~\ref{tab:tp-failures}, mismatches with the original paper are the most prevalent failure category across all harnesses, affecting $24-42$ runs per configuration. However, the failure patterns vary considerably. InternAgent fails to produce a report in 13 of its 43 unsuccessful runs, suggesting difficulties in completing the reproduction workflow. In contrast, live-swe-agent exhibits implementation errors in 22 of its 41 failed runs, while Meta-Harness reduces the original problem settings in 18 of its 48 failed runs. Notably, these failure categories are not mutually exclusive, as a single run may exhibit multiple issues. Collectively, these results highlight the importance of harness design in scientific reproduction and suggest that greater architectural complexity does not necessarily translate into higher reproduction performance.

\subsection{Case studies of \RA}\label{append:sec:ra-cases}

\smallskip

In this subsection, we examine how \RA identifies and corrects reproduction failures through three representative case studies from Paper~\#2, Paper~\#35, and Paper~\#29. On these tasks, DSH and most third-party harnesses evaluated with DeepSeek V4 Pro failed to reproduce the reported results, whereas \RA succeeded. Notably, none of the three papers had encountered reproduction issues in the curation audit in Table~\ref{tab:paper-levels}, allowing us to focus on failures in agent execution rather than known inconsistencies in the scientific record. In each case, \RA identified discrepancies between intermediate results and the methodological or quantitative requirements of the original paper, subsequently diagnosing and correcting the underlying errors without altering the prescribed task settings. These cases illustrate three complementary mechanisms of \RA, namely \textit{sub-plan decomposition}, \textit{critic-guided issue resolution}, and \textit{experiment-driven diagnosis}, which we examine separately below.

\begin{figure}
\centering
\includegraphics[width=\textwidth]{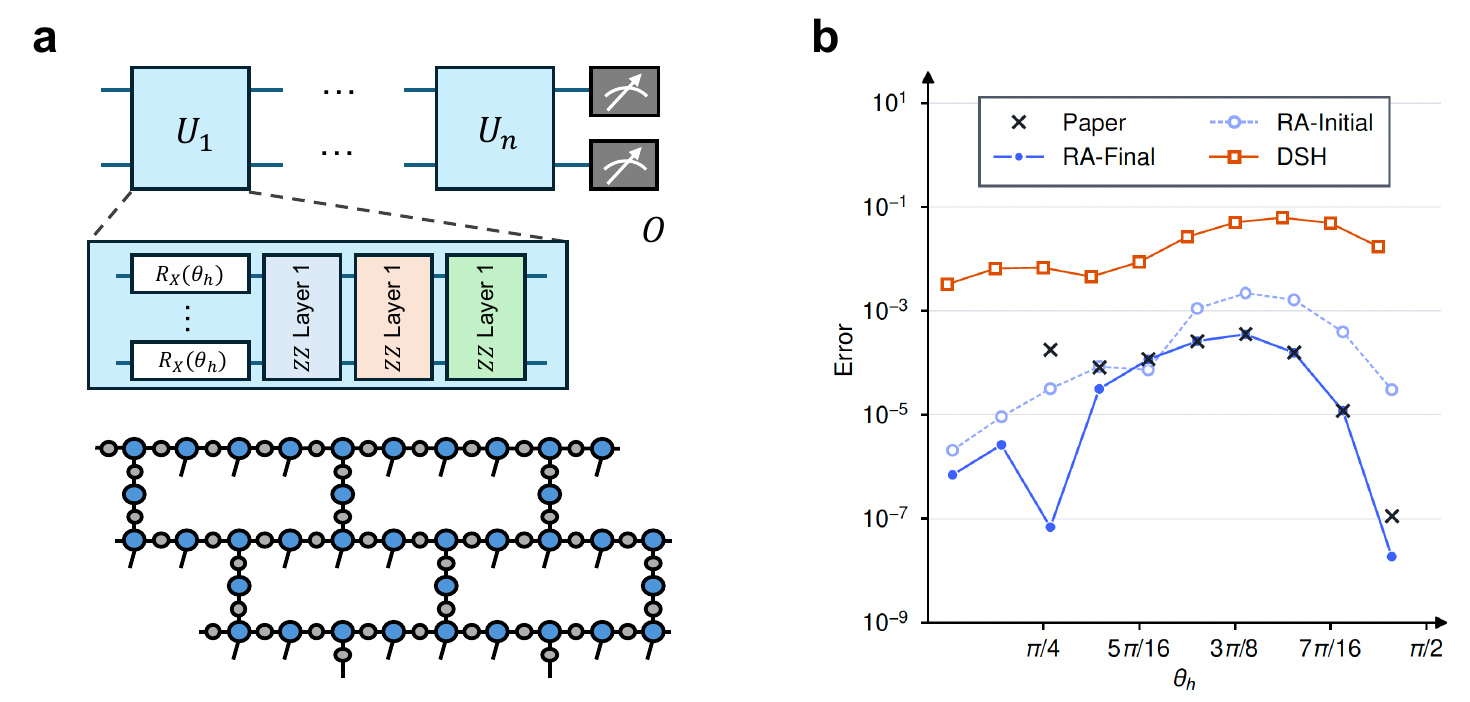}
\caption{\small\textbf{Simulation workflow and reproduction results for Paper~\#2.} \textbf{a.} Schematic of the 127-qubit kicked Ising circuit (upper panel) and the corresponding tensor-network simulation based on belief propagation (lower panels). \textbf{b.} Comparison of the reproduction results obtained by different AI scientists with the original results from Fig.~3(c) of Paper~\#2~\cite{tindall2024eagle}. The horizontal axis denotes the single-qubit rotation angle $\theta_h$, while the vertical axis shows the error between the tensor-network estimate and the exact light-cone benchmark on a logarithmic scale. The labels `RA-Initial' and `RA-Final' denote the results obtained by \RA before and after its internal corrections, respectively, while `DSH' denotes the baseline reproduction using DeepSeek Harness. The label `Paper' denotes the reference error values reported in the original paper.}
\label{fig:si-ra-paper2-results}
\end{figure}

\medskip
\noindent\textbf{Case study 1 (Paper~\#2): Sub-plan decomposition.} This case illustrates how \RA uses sub-plan decomposition to overcome a reproduction challenge that other AI scientists failed to resolve, by ensuring consistent methodological interpretations and independent validation across computational sub-plans.

\smallskip
\noindent\textit{Reproduction challenge and baseline failure.}
Paper~\#2~\cite{tindall2024eagle} investigated the classical simulation of IBM's 127-qubit kicked Ising experiment using tensor networks contracted through belief propagation, with simulation errors evaluated against an exact light-cone benchmark. As shown in Fig.~\ref{fig:si-ra-paper2-results}\textbf{a}, the circuit consists of repeated Trotter steps combining fixed Ising interactions with single-qubit $X$ rotations parameterized by an angle $\theta_h$, where $\theta_h=\pi/2$ corresponds to a Clifford circuit that provides an exact reference point. A valid comparison requires the tensor-network estimate $\langle O\rangle_{\mathrm{TN}}$ and the exact benchmark $\langle O\rangle_{\mathrm{exact}}$ to evaluate the same physical observable $O$. In particular, the weight-17 Pauli observable $O_{17}$ in Figure~3(c) of the original paper is constructed by propagating a single-site $Z$ operator backward through several Clifford steps, requiring the correct qubit and number of steps to be identified. However, DSH using the same base LLM failed to maintain this consistency. It initially simulated Figure~3(c) using a qubit selected during an early search, but subsequently discovered that this choice generated a weight-23 observable $O_{23}$. Although DSH then computed the exact benchmark for the correct observable, the corresponding simulation was not completed before the run ended. Consequently, its final figure incorrectly compared $\braket{O_{23}}_{\mathrm{TN}}$ with $\braket{O_{17}}_{\mathrm{exact}}$, resulting in substantially larger errors than those reported in the paper.

\smallskip
\noindent\textit{Sub-plan decomposition and reproduction results.}
\RA identified the correct target observable before execution and decomposed the task into five independently validated sub-plans, using shared methodological interpretations to maintain consistency across their computations. In particular, the exact benchmark was implemented independently of the simulation engine, while both evaluated the same observable. When the full-scale simulation revealed a separate error in the belief-propagation implementation, \RA coordinated corrections across the affected sub-plans while retaining the validated exact benchmark.

The reproduction results are presented in Fig.~\ref{fig:si-ra-paper2-results}\textbf{b}. DSH reported errors ranging from approximately $10^{-3}$ to $10^{-1}$, several orders of magnitude larger than the published reference values at many rotation angles. As discussed above, these discrepancies resulted from comparing different observables rather than directly reflecting the accuracy of the intended tensor-network simulation. By contrast, the initial \RA reproduction already achieved substantially smaller errors, although noticeable deviations from the published results remained at intermediate rotation angles, particularly around $\theta_h=3\pi/8$. Following the coordinated corrections, the final \RA results more closely reproduced both the reported error magnitudes and their dependence on $\theta_h$, including the peak near $\theta_h=3\pi/8$ and the sharp decrease towards the Clifford point $\theta_h=\pi/2$. Quantitatively, the final results agreed with the published error values to within approximately $2\%$ at 15 of the 23 evaluated points. These results illustrate how sub-plan decomposition and coordinated correction enabled \RA to recover the reported simulation behavior that DSH failed to reproduce.

\medskip
\noindent\textbf{Case study 2 (Paper~\#35). Critic-guided issue resolution.}
In this case, \RA corrected an underestimated oracle-call cost through critic-guided local repair: the critic identified an invalid circuit schedule, the executor corrected the implementation, and the critic reviewed the revised results before accepting the affected sub-plan.

\begin{figure}[!htbp]
    \centering
    \includegraphics[width=\textwidth]{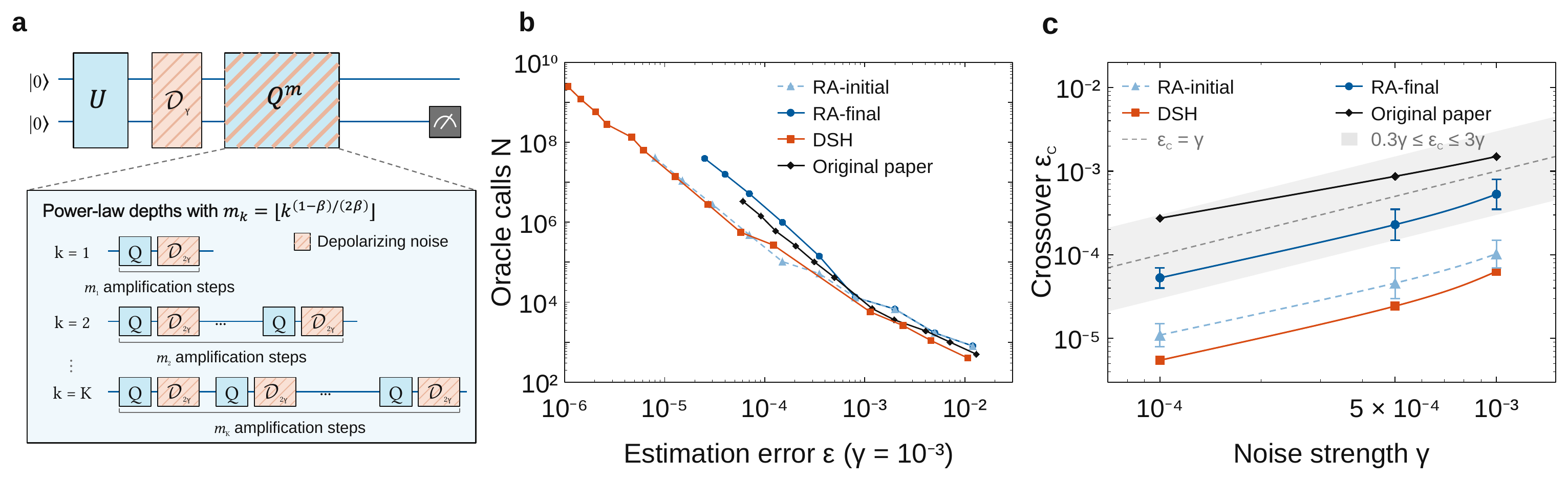}
    \caption{\textbf{Noisy amplitude estimation and critic-guided issue resolution for Paper~\#35.}
    \textbf{a}. Amplitude estimation with state preparation $U$ and $m_k=\lfloor k^{(1-\beta)/(2\beta)}\rfloor$ amplification steps at round $k$. Each $Q$ uses two oracle calls. In this equivalent circuit representation, noise is grouped into $\mathcal{D}_{\gamma}$ after the initial $U$ and $\mathcal{D}_{2\gamma}$ after each $Q$. Here $\mathcal{D}_t$ denotes global depolarizing noise with the factor $e^{-t}$.
    \textbf{b}. Reported oracle calls $N$ versus error $\epsilon$ at $\gamma=10^{-3}$ for the original paper, DSH and \RA before and after correction. 
     \textbf{c}. Crossover diagnostics versus noise strength. For both \RA stages, $\epsilon_C$ is the geometric midpoint between adjacent target errors where the chosen circuit family changes between power-law and exponential schedules. Vertical bars delimit these sampled intervals. The DSH values are crossover estimates extracted from its numerically evaluated theoretical cost curves. The dashed line marks $\epsilon_C=\gamma$, and the shaded region is the evaluation interval $[0.3\gamma,3\gamma]$.}
    \label{fig:si-ra-paper35}
\end{figure}

\smallskip
\noindent\textit{Background and baseline failure.}
Paper~\#35~\cite{giurgica2022low} develops low-depth algorithms for estimating the angle $\theta$ encoded by a state-preparation oracle $U$, i.e., $U|0\rangle^{\otimes n}
=\cos\theta\,|\psi_0\rangle|0\rangle
+\sin\theta\,|\psi_1\rangle|1\rangle$
, where $0\leq\theta\leq\pi/2$, and $\{|\psi_0\rangle, |\psi_1\rangle\}$ are normalized states of the remaining register. The task is to estimate $\theta$ within additive error $\epsilon$ using the procedure illustrated in Fig.~\ref{fig:si-ra-paper35}\textbf{a}. Specifically, each amplification step applies $Q=UR_0U^\dagger S_0$, where $R_0=2|0\rangle^{\otimes n}\langle0|^{\otimes n}-\mathbb{I}_{2^n}$ and $S_0=\mathbb{I}_{2^{n-1}}\otimes Z$. Each application of $Q$ uses two oracle calls of $U$.
In this amplitude estimation procedure, the number of amplitude amplification steps $m_k$ increases across measurement rounds $k=1,2,\cdots$ according to a power law $m_k= \lfloor k^{(1-\beta)/(2\beta)} \rfloor $ with a tunable exponent, which allows a tradeoff between circuit depth and total oracle calls.
The reproduction targets are the relationships between the number of oracle calls $N$ and the estimation error $\epsilon$ shown in Figures~5 and~6 of the original paper, for noiseless circuits and circuits with depolarizing noise at different decay rates $\gamma$, respectively. With schedules adapted to the target error and noise level, estimation at $\epsilon\gg\gamma$ retains approximately the quantum scaling $N\propto\epsilon^{-1}$. At finer precision with $\epsilon \ll \gamma$, the measurement contrast is suppressed by $e^{-\gamma q}$ after $q$ oracle calls. More measurements are then needed, and the cost approaches the classical scaling $N\propto\epsilon^{-2}$. The crossover error $\epsilon_C$ marks the transition between these scaling regimes and is expected to be of order $\gamma$. DSH reproduced the noiseless relation but obtained a noisy crossover at $\epsilon_C\approx0.05\gamma$, well below this scale, as shown in Fig.~\ref{fig:si-ra-paper35}\textbf{c}. Although it documented this discrepancy, DSH incorrectly declared the noisy reproduction successful. The judge agent marked the noisy target figure as failed because the crossover fell outside the prescribed evaluation interval $[0.3\gamma,3\gamma]$.

\smallskip
\noindent\textit{Critic-guided issue resolution.}
At the initial stage, \RA also underestimated the cost in the noise-dominated regime. The executor attributed the discrepancy to the published curve, but the critic checked the recorded circuit depths against the original method. In the power-law algorithm, these depths depend on $\beta$. For a requested error $\epsilon<\gamma$, the requirement $\beta\geq1-\log\gamma/\log(\epsilon/c)$ holds, where $c$ is a constant. In the initial simulation, the executor selected the circuit sequence requiring the fewest oracle calls among candidates designed for other error values, without enforcing this bound at the requested precision. The recorded parameters confirmed violations of the bound, and exponential schedules were retained below the expected crossover. The critic therefore identified the circuit-selection rule as a methodological defect and returned the affected sub-plan to the executor for local repair.

Following this review, the executor rebuilt each noisy-regime schedule with $\beta$ at the bound for the requested $(\epsilon,\gamma)$, keeping $c=1$ throughout, and recomputed the affected results. 
The corrected schedule rule was recorded as a revised shared methodological interpretation, and the critic reviewed the recomputed results before accepting the affected sub-plan.
As shown in Fig.~\ref{fig:si-ra-paper35}\textbf{b}, the correction increased the estimated cost by roughly an order of magnitude near $\epsilon\sim10^{-4}$, while the costs at large target errors remained unchanged. Across all three noise strengths, the schedule switches in Fig.~\ref{fig:si-ra-paper35}\textbf{c} moved from below the evaluation interval to within it. The judge agent evaluated the completed reproduction and marked both target figures as passed.

\medskip
\noindent \textbf{Case study 3 (Paper~\#29): Experiment-driven diagnosis.} 
This case illustrates how critic corrects failures in individual subplans by first analyzing their objectives and intermediate results to formulate diagnostic hypotheses, then deploying probing agents to test these hypotheses through small-scale experiments and guide targeted corrections.

\begin{figure}[!htbp]
    \centering
    \includegraphics[width=0.98\textwidth]{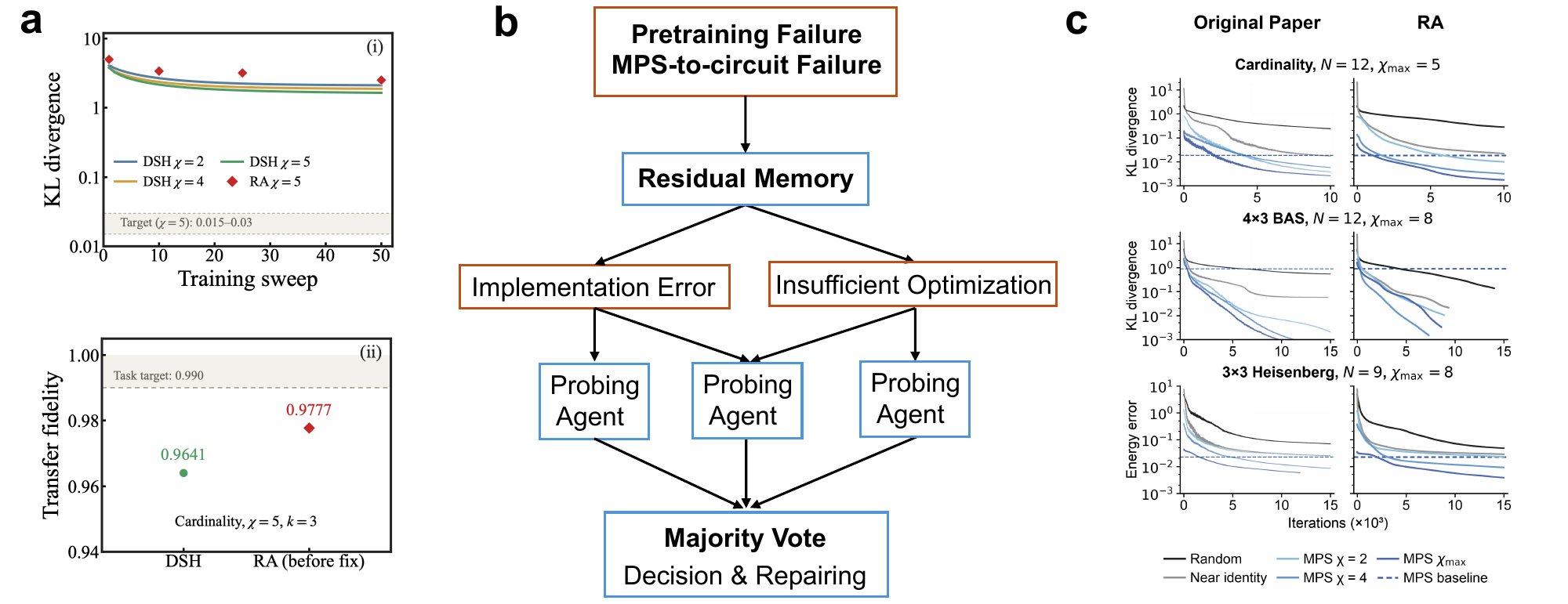}
    \caption{\textbf{Experiment-driven diagnosis and correction for Paper~\#29.}
    \textbf{a}. Initial reproduction failures of DSH and \RA{} before correction. \textbf{(i)}, KL divergence during MPS pretraining with target accuracy. \textbf{(ii)}, MPS-to-circuit transfer fidelities of DSH (0.9641) and \RA{} (0.9777) with target accuracy. \textbf{b}. Workflow of experiment-driven diagnosis in Paper~\#29. Knowledge retrieved from residual memory guides the identification of potential implementation errors and insufficient optimization. Three probing agents conduct small-scale experiments to test these hypotheses and inform subsequent corrections by major voting. \textbf{c}. Comparison between the results reported in the original paper (left) and the corrected reproduction by \RA{} (right) for different tasks. The corrected reproduction successfully recovers the improved training performance.}
    \label{fig:si-experiment-driven-correction}
\end{figure}

\smallskip 
\noindent \textit{Background and baseline failure.}
Paper~\#29~\cite{rudolph2023synergistic} aims to improve the performance of parameterized quantum circuits (PQCs) by classically pretraining a matrix product state (MPS) and converting it into a quantum circuit as the initial ansatz for subsequent optimization. More concretely, the proposed method first uses MPS pretraining to obtain an MPS that closely approximates the target state and then performs MPS-to-circuit conversion to transfer the pretrained state into a PQC with minimal fidelity loss. As Fig.~\ref{fig:si-experiment-driven-correction}\textbf{a} shows, both DSH with DeepSeek V4.1 Flash and \RA{} with the same base LLM failed at MPS pretraining and MPS-to-circuit conversion in their initial attempts. Specifically, the KL divergence between the pretrained MPS and the target state for both methods remained above $1$ after $50$ training sweeps, far exceeding the target range of $[0.015,0.03]$.  Similarly, the MPS-to-circuit conversion achieved transfer fidelities of only $0.9641$ for DSH and $0.9777$ for \RA, both below the required 0.990. These discrepancies ultimately prevented successful reproduction of the PQC training results reported in the original paper.

\smallskip
\noindent \textit{Experiment-driven correction.}
By retrieving relevant knowledge from its residual memory, the critic hypothesized that the failures in both MPS pretraining and MPS-to-circuit conversion could arise from implementation errors or insufficient variational optimization. To investigate these possibilities, the critic deployed three probing agents for each subplan to test each hypothesis from small-scale experiments, as shown in Fig.~\ref{fig:si-experiment-driven-correction}\textbf{b}. For MPS pretraining, two probing agents independently identified errors in the tensor contractions used for gradient computation by comparing independently computed reference gradients with those produced by the original code. For MPS-to-circuit conversion, one probing agent found that the original implementation failed to correctly follow the truncation-and-disentangling procedure for constructing circuit layers prescribed in the paper. Based on majority voting on these findings, \RA{} corrected the errors in both subplans and finally reproduced the performance curves reported in the original paper correctly, as illustrated in Fig.~\ref{fig:si-experiment-driven-correction}\textbf{c}.

\end{document}